\documentclass[11pt,english]{article}
\usepackage{lmodern}
\usepackage[T1]{fontenc}
\usepackage[latin9]{inputenc}
\usepackage{array}
\usepackage{float}
\usepackage{booktabs}
\usepackage{url}
\usepackage{multirow}
\usepackage{amsmath}
\usepackage{amssymb}
\usepackage{cancel}
\usepackage{stackrel}
\usepackage{graphicx}
\usepackage{setspace}
\usepackage{wasysym}
\makeatletter

\providecommand{\tabularnewline}{\\}

\usepackage{jheppub}% for details on the use of the package, please
\usepackage{subfig}
\usepackage[compat=1.1.0]{tikz-feynman}
\usepackage{geometry}
\title{\boldmath $B$-anomalies in a twin Pati-Salam theory of flavour including the 2022 LHCb $R_{K^{(*)}}$ analysis}

\author[]{Mario Fern\'andez Navarro}
\author[]{and Stephen F. King}
\affiliation[]{School of Physics \& Astronomy, University of Southampton, Southampton SO17 1BJ, UK}

\emailAdd{M.F.Navarro@soton.ac.uk}
\emailAdd{S.F.King@soton.ac.uk}

\abstract{We perform a comprehensive phenomenological analysis of the 
twin Pati-Salam theory of flavour, focussing on the parameter space 
relevant for interpreting the $B$-anomalies via vector leptoquark $U_1$ exchange. 
This model provides a very predictive framework in which the $U_1$ 
couplings and the Yukawa couplings find a common origin via mixing of chiral quarks 
and leptons with vector-like fermions, providing a direct link between the $B$-anomalies 
and the fermion masses and mixing. We propose and study a simplified model with three vector-like 
fermion families, in the massless first family approximation, and show that the second and third 
family charged fermion masses and mixings and the $B$-anomalies can be simultaneously explained and 
related. The model has the proper flavour structure to be compatible with all low-energy observables, 
and leads to predictions in promising observables such as $\tau\rightarrow3\mu$, $\tau\rightarrow\mu\gamma$ 
and $B\rightarrow K^{(*)}\nu\bar{\nu}$ at Belle II and LHCb. The model also predicts a rich spectrum of TeV 
scale gauge bosons and vector-like fermions, all accessible to the LHC.  In this updated version we have 
included an extended analysis considering the new 2022 LHCb data on $R_{K^{(*)}}$, which has slightly shifted 
the preferred parameter space with respect to the 2021 case. The model can still explain the $R_{D^{(*)}}$ anomalies 
at 1$\sigma$ in a narrow window, however we expect small deviations from the SM 
on the $R_{K^{(*)}}$ ratios, to be tested in the future via more precise measurements by the LHCb collaboration. 
We also predict $\Delta R_{D}= \Delta R_{D^{*}}$, with future measurements shifting the world averages to slightly smaller central values.}

\@ifundefined{showcaptionsetup}{}{%
 \PassOptionsToPackage{caption=false}{subfig}}
\usepackage{subfig}
\makeatother

\usepackage{babel}
\begin{document}
\makeatletter
\gdef\@fpheader{version accepted in JHEP}
\makeatother

\maketitle \flushbottom

\clearpage
\setcounter{page}{1}
\pagenumbering{arabic}
\numberwithin{equation}{section}
\numberwithin{figure}{section}

\section{Introduction}

Fundamental fermions in the Standard Model (SM) come in three copies,
denoted as ``flavours'', which share universal gauge interactions
but have different masses and mixings, also known as flavour parameters.
The origin of flavour in the SM remains as a complete mystery, as
it lacks of any dynamical explanation to the high number of flavour
parameters and their hierarchical patterns. A further theory of flavour
beyond the SM should provide a solution to the long-lasting ``flavour
puzzle''.

Simultaneously, the non-universal structure of such a theory of flavour
could leave its imprints in flavour physics observables, which are
becoming accessible up to a high precision level in the current generation
of colliders and meson factories. Given the prolific history of flavour
physics anticipating the discovery of new physics, searching for the
origin of flavour in flavour physics is well motivated. In this direction,
a conspicuous series of anomalies in flavour observables emerged in
the last years.

Back in 2021, when this project was started, the $R_{K^{(*)}}$ ratios
had been measured by LHCb to be smaller than 1 \cite{LHCb:2017avl,LHCb:2021trn},
in good agreement with other anomalies in $b\rightarrow s\mu\mu$
data which were hinting for flavourful new physics (NP) affecting
muons rather than electrons. In particular, $R_{K}^{[1.1,6]}$ was
alone in more than 3$\sigma$ tension with the SM prediction. The
breaking of SM lepton flavour universality (LFU) was not only suggested
by $R_{K^{(*)}}$, but also the ratios $R_{D^{(*)}}$ had been measured
to show discrepancy with the SM (see the world averages in \cite{HFLAV:2022pwe}),
hinting for flavourful NP affecting tau leptons. Although no single
measurement of $R_{D^{(*)}}$ is very significant, the combination
of all of them hints for a consistent deviation from the SM prediction
with more than 3$\sigma$ significance. Both LFU ratios together gave
rise to a very consistent picture of hierarchical anomalies, where
strong NP mainly coupled to the third family interfere with a SM charged
current tree-level effect, while weaker NP couple to the much lighter
muons, interfering with 1-loop and GIM-suppressed SM neutral currents.

This picture of ``$B$-anomalies'' led to important model building
efforts by the community during the last 8 years, in order to interpret
these anomalies as a low energy signal of a consistent NP model. A
massive, electrically neutral $Z'$ vector was identified as a possible
explanation of the $R_{K^{(*)}}$ anomalies (see e.g.~\cite{Crivellin:2015mga,Crivellin:2015lwa,Chiang:2017hlj,King:2017anf,Falkowski:2018dsl,Navarro:2021sfb}),
while different leptoquarks were proposed to address either $R_{K^{(*)}}$
or $R_{D^{(*)}}$ separately (see e.g.~\cite{Becirevic:2017jtw,deMedeirosVarzielas:2018bcy,Angelescu:2021lln,Becirevic:2022tsj}).
Interestingly, the vector leptoquark $U_{1}(\boldsymbol{3},\boldsymbol{1},2/3)$
was identified as the only single mediator capable of addressing both
$B$-anomalies simultaneously \cite{Angelescu:2021lln}. However,
the gauge nature of $U_{1}$ requires to specify a clear ultra-violet
(UV) completion that explains its origin. The original ideas by Pati
and Salam (PS) \cite{Pati:1974yy}, led to tensions with unobserved
processes such as $K_{L}\rightarrow\mu e$. Instead, an interesting
proposal was firstly laid out in the Appendix of \cite{Diaz:2017lit},
and more formally later in \cite{DiLuzio:2017vat}, following the
idea introduced in \cite{Georgi:2016xhm} that color could appear
as a diagonal subgroup of a larger $SU(3+N)\times SU(3)'$ local symmetry
valid at high energies. The particular choice $N=1$ leads to the
so-called ``4321'' gauge symmetry,
\begin{equation}
G_{4321}\equiv SU(4)\times SU(3)_{c}^{'}\times SU(2)_{L}\times U(1)_{Y'}\,,
\end{equation}
which can be broken at the TeV scale while satisfying the experimental
bounds \cite{DiLuzio:2017vat,DiLuzio:2018zxy,Cornella:2019hct,Cornella:2021sby},
provided that at least the first and second families of SM fermions
are singlets under $SU(4)$. This breaking leads to a rich gauge boson
spectrum at the TeV scale, containing the vector leptoquark $U_{1}$
along with a massive colour octet $g'(\boldsymbol{8},\boldsymbol{1},0)$
and a massive $Z'\left(\boldsymbol{1},\boldsymbol{1},0\right)$ with
suppressed couplings to light SM fermions. Vector-like (VL) fermions
need to be introduced in order to obtain effective couplings of (at
least) second family fermions to $U_{1}$. The model, even if not
minimal, is very predictive and leads to a rich phenomenology in both
low-energy and high-$p_{T}$ searches. However, the flavour structure
of the model was rather ad-hoc, and it was hinted that the 4321 gauge
group could be the TeV scale effective field theory of a complete
model addressing more open questions of the SM. In particular, the
4321 model seemed to be a nice playground to connect the picture of
$B$-anomalies with the flavour puzzle of the SM.

\begin{figure}[t]
\begin{centering}
\includegraphics[scale=0.23]{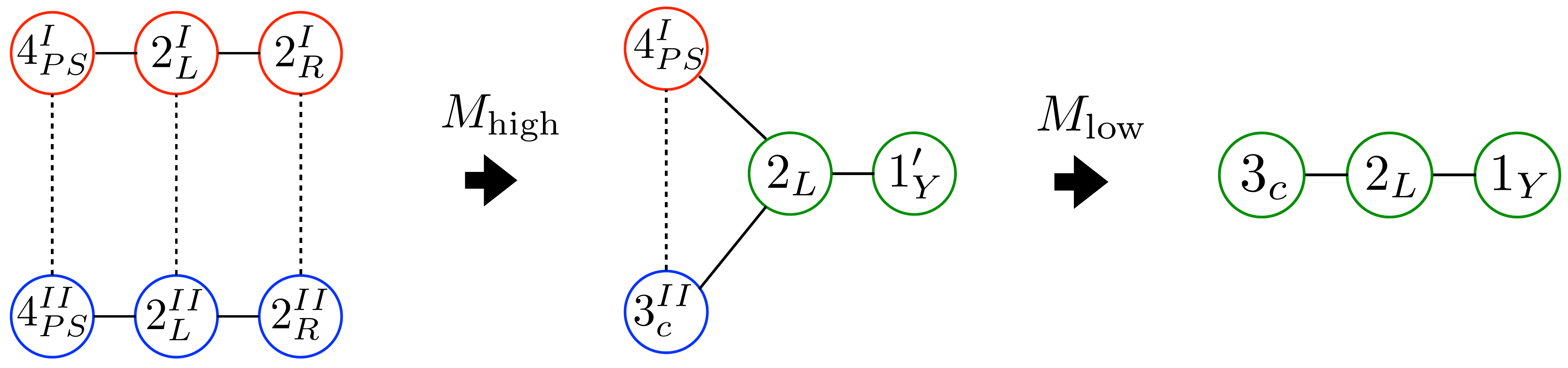}
\par\end{centering}
\caption{The model is based on two copies of the Pati-Salam gauge group $SU(4)_{PS}\times SU(2)_{L}\times SU(2)_{R}$.
The circles represent the gauge groups with the indicated symmetry
breaking. The twin Pati-Salam symmetry is broken down to the 4321
symmetry at high energies $M_{\mathrm{High}}\apprge1\,\mathrm{PeV}$,
then the 4321 group is further broken to the SM at the TeV scale $M_{\mathrm{low}}\sim\mathcal{O}(\mathrm{TeV}).$\label{fig:Model_Diagram}}
\end{figure}

Motivated by the desire to link the origin of the $B$-anomalies with
the origin of Yukawa couplings in the SM, one of us proposed a theory
of flavour involving a twin Pati-Salam group \cite{King:2021jeo}.
Unlike the other models already present in the market \cite{Bordone:2017bld,Bordone:2018nbg,Fuentes-Martin:2022xnb},
the twin PS treats all three fermion families in the same way. The
basic idea is that all three families of SM chiral fermions transform
under one PS group, while families of vector-like fermions transform
under the other one. The first PS group, broken at a high scale, provides
Pati-Salam unification of all SM quarks and leptons, while a fourth
family of vector-like fermions transforms under a second PS group,
broken at the TeV scale to the SM, as in Fig.~\ref{fig:Model_Diagram}.
The full twin Pati-Salam symmetry, together with the absence of a
standard Higgs electroweak (EW) doublet, forbids the usual Yukawa
couplings for the SM fermions. Instead, effective Yukawa couplings
arise through the mixing between SM fermions and vector-like partners.
The same mixing leads to $U_{1}$ couplings for SM fermions which
could address the $B$-anomalies. This way, $B$-anomalies and the
flavour puzzle are dynamically and parametrically connected. Furthermore,
the twin PS model predicts dominantly left-handed (LH) $U_{1}$ currents
that were preferred by the 2021 picture of $B$-anomalies \cite{Geng:2021nhg,Angelescu:2021lln,Altmannshofer:2021qrr},
while the other proposals \cite{Bordone:2017bld,Bordone:2018nbg,Fuentes-Martin:2022xnb}
predict large couplings for right-handed (RH) third family fermion,
which lead to tight constraints from high-$p_{T}$ searches.

In this paper, we studied the phenomenology of the simplified twin
PS model presented in \cite{King:2021jeo}, which turned out to be
incompatible with low-energy data. Afterwards, we performed further
model building and presented an extended version of the model that
can explain the 2021 picture of $B$-anomalies and address charged
fermion masses and mixings, while being compatible with all existing
data. However, during the peer-review process of this paper, the LHCb
collaboration presented a reanalysis of the LFU ratios $R_{K^{(*)}}$,
with the new measurements in the central $q^{2}$ shown below (where
$q^{2}$ denotes the dilepton invariant-mass squared) \cite{LHCb:2022qnv}

\begin{equation}
R_{K}^{[1.1,6]}=\frac{\mathrm{Br}\left(B\rightarrow K\mu^{+}\mu^{-}\right)}{\mathrm{Br}\left(B\rightarrow Ke^{+}e^{-}\right)}=0.949_{-0.046}^{+0.047}\,,\qquad R_{K^{*}}^{[1.1,6]}=\frac{\mathrm{Br}\left(B\rightarrow K^{*}\mu^{+}\mu^{-}\right)}{\mathrm{Br}\left(B\rightarrow K^{*}e^{+}e^{-}\right)}=1.027_{-0.073}^{+0.077}\,,\label{eq:RK}
\end{equation}
with correlation $\rho=-0.017$. Unexpectedly, the updated results
are in good agreement with the SM predictions of $R_{K^{(*)}}^{[1.1,6]}=1\pm0.01$
\cite{Bordone:2016gaq}, as a result of backgrounds in the electron
channel which were misidentified in all the previous analyses. Although
our model was originally built to explain large deviations from the
SM in both $R_{K^{(*)}}$ and $R_{D^{(*)}}$, the new experimental
data offers the opportunity to further test the model as a legitimate
theory of flavour addressing the origin of quark and lepton masses
and mixings. Therefore, we present here an updated analysis which
includes the new 2022 LHCb data on $R_{K^{(*)}}$, and we confront
the new results versus the previous 2021 picture for which the model
was intended. Beyond the $R_{K^{(*)}}$ ratios, a new combined measurement
of $R_{D^{(*)}}$ was presented by LHCb in late 2022 \cite{LHCb:2023zxo}.
This measurement is in line with the previous experimental data on
$R_{D^{(*)}}$, and does not significantly modify the HFLAV average
\cite{HFLAV:2022pwe},
\begin{equation}
R_{D}=\left.\frac{\mathrm{Br}\left(B\rightarrow D\tau\nu\right)}{\mathrm{Br}\left(B\rightarrow D\ell\nu\right)}\right|_{\ell\epsilon\,\left\{ e,\mu\right\} }=0.358\pm0.028\,,\qquad R_{D^{*}}=\left.\frac{\mathrm{Br}\left(B\rightarrow D^{*}\tau\nu\right)}{\mathrm{Br}\left(B\rightarrow D^{*}\ell\nu\right)}\right|_{\ell\epsilon\,\left\{ e,\mu\right\} }=0.285\pm0.013\,.\label{eq:R_D}
\end{equation}
which remains at roughly 3$\sigma$ discrepancy with the SM predictions.

The layout of the remainder of the paper is as follows. In Section
\ref{sec:Twin-Pati-Salam-Theory_1VL} we introduce the simplified
twin Pati-Salam model as presented in \cite{King:2022sxb}, featuring
only one vector-like family, and show that it is unable to explain
$R_{D^{(*)}}$ in a natural way, while being compatible with the stringent
constraints from $B_{s}-\bar{B}_{s}$ mixing. Instead, in Section~\ref{sec:Twin-Pati-Salam-Theory_3VL}
we present a new, extended version of the twin Pati-Salam model including
three vector-like families and a discrete flavour symmetry, which
is compatible with low-energy data and high-$p_{T}$ searches. Section
\ref{subsec:Effective_Yukawa_3VL} shows how effective Yukawa couplings
for the SM fermions arise in the model, addressing charged fermions
masses and mixings. Similarly, Section~\ref{subsec:Couplings_3VL}
shows the origin of effective couplings between SM fermions and the
exotic gauge bosons. Section~\ref{subsec:Low-energy-phenomenology}
shows the phenomenological analysis and the discussion of the results,
including promising signals to test the model in low-energy observables
and high-$p_{T}$ searches, along with a study of the perturbativity
of the model. Section~\ref{sec:Comparison_models} includes a comparison
of our predictions with other models in the market. Finally, we conclude
the paper in Section~\ref{sec:Conclusions}.

\section{Simplified twin Pati-Salam theory of flavour\label{sec:Twin-Pati-Salam-Theory_1VL}}

\subsection{The High Energy Model}

In the traditional PS theory, the chiral quarks and leptons are unified
into $SU(4)_{PS}$ multiplets with leptons as the fourth colour (red,
blue, green, lepton) \cite{Pati:1974yy},
\begin{equation}
\psi_{i}(4,2,1)=\left(\begin{array}{cccc}
u_{r} & u_{b} & u_{g} & \nu\\
d_{r} & d_{b} & d_{g} & e
\end{array}\right)_{i}\equiv\left(Q_{i},L_{i}\right)\,,\quad\psi_{j}^{c}(\overline{4},1,\overline{2})=\left(\begin{array}{cccc}
u_{r}^{c} & u_{b}^{c} & u_{g}^{c} & \nu^{c}\\
d_{r}^{c} & d_{b}^{c} & d_{g}^{c} & e^{c}
\end{array}\right)_{j}\equiv\left(u_{j}^{c},d_{j}^{c},\nu_{j}^{c},e_{j}^{c}\right)\,,
\end{equation}
where $\psi_{i}$ contains the left-handed quark and leptons while
$\psi_{j}^{c}$ contains the CP-conjugated right-handed (RH) quarks
and lepton (so that they become LH), and $i,j=1,2,3$ are family indices.
We consider here two copies of the Pati-Salam symmetry \cite{King:2021jeo},
\begin{equation}
G_{422}^{I}\times G_{422}^{II}=\left(SU(4)_{PS}^{I}\times SU(2)_{L}^{I}\times SU(2)_{R}^{I}\right)\times\left(SU(4)_{PS}^{II}\times SU(2)_{L}^{II}\times SU(2)_{R}^{II}\right)\,.\label{eq:TwinPS_symmetry}
\end{equation}

The matter content and the quantum numbers of each field are displayed
in Table~\ref{tab:Field_content_TwinPS}. The usual three chiral
fermion families, SM-like, originate from the second PS group $G_{422}^{II}$,
broken at a high scale, and transform under Eq.~\eqref{eq:TwinPS_symmetry}
as
\begin{equation}
\psi_{1,2,3}(1,1,1;4,2,1)\,,\quad\psi_{1,2,3}^{c}(1,1,1;\overline{4},1,\overline{2})\,.
\end{equation}
This simplified version of the theory includes one vector-like family
of fermions which originates under the first PS group, whose $SU(4)^{I}$
is broken at the TeV scale, and transforms under Eq.~\eqref{eq:TwinPS_symmetry}
as
\begin{equation}
\psi_{4}(4,2,1;1,1,1)\,,\quad\overline{\psi_{4}}(\overline{4},\overline{2},1;1,1,1)\,,\psi_{4}^{c}(\overline{4},1,\overline{2};1,1,1)\,,\overline{\psi_{4}^{c}}(4,1,2;1,1,1)\,.
\end{equation}

On the other hand, according to the matter content in Table \ref{tab:Field_content_TwinPS},
there are no standard Higgs fields which transform as ($1,\overline{2},2$)
under $G_{422}^{II}$, hence the standard Yukawa couplings involving
the chiral fermions are forbidden by the twin PS symmetry. These will
be generated effectively via mixing with the fourth family of vector-like
fermions which only have quantum numbers under the first PS group,
$G_{422}^{I}$. This mixing is facilitated by the non-standard Higgs
scalar doublets contained in $\phi$, $\overline{\phi}$, $H$, $\overline{H}$
in Table \ref{tab:Field_content_TwinPS}, via the couplings,
\begin{table}[t]
\begin{centering}
\begin{tabular}{lcccccc}
\toprule 
Field & $SU(4)_{PS}^{I}$ & $SU(2)_{L}^{I}$ & $SU(2)_{R}^{I}$ & $SU(4)_{PS}^{II}$ & $SU(2)_{L}^{II}$ & $SU(2)_{R}^{II}$\tabularnewline
\midrule
\midrule 
$\psi_{1,2,3}$ & $\mathbf{1}$ & $\mathbf{1}$ & $\mathbf{1}$ & $\mathbf{4}$ & $\mathbf{2}$ & $\mathbf{1}$\tabularnewline
$\psi_{1,2,3}^{c}$ & $\mathbf{1}$ & $\mathbf{1}$ & $\mathbf{1}$ & $\mathbf{\overline{4}}$ & $\mathbf{1}$ & $\mathbf{\overline{2}}$\tabularnewline
\midrule 
$\psi_{4}$ & $\mathbf{4}$ & $\mathbf{2}$ & $\mathbf{1}$ & $\mathbf{1}$ & $\mathbf{1}$ & $\mathbf{1}$\tabularnewline
$\overline{\psi_{4}}$ & $\mathbf{\overline{4}}$ & $\mathbf{\overline{2}}$ & $\mathbf{1}$ & $\mathbf{1}$ & $\mathbf{1}$ & $\mathbf{1}$\tabularnewline
$\psi_{4}^{c}$ & $\mathbf{\overline{4}}$ & $\mathbf{1}$ & $\mathbf{\overline{2}}$ & $\mathbf{1}$ & $\mathbf{1}$ & $\mathbf{1}$\tabularnewline
$\overline{\psi_{4}^{c}}$ & $\mathbf{4}$ & $\mathbf{1}$ & $\mathbf{2}$ & $\mathbf{1}$ & $\mathbf{1}$ & $\mathbf{1}$\tabularnewline
\midrule 
$\phi$ & $\mathbf{4}$ & $\mathbf{2}$ & $\mathbf{1}$ & $\mathbf{\overline{4}}$ & $\mathbf{\overline{2}}$ & $\mathbf{1}$\tabularnewline
$\overline{\phi}$ & $\mathbf{\overline{4}}$ & $\mathbf{1}$ & $\mathbf{\overline{2}}$ & $\mathbf{4}$ & $\mathbf{1}$ & $\mathbf{2}$\tabularnewline
\midrule
$H$ & $\mathbf{\overline{4}}$ & $\mathbf{\overline{2}}$ & $\mathbf{1}$ & $\mathbf{4}$ & $\mathbf{1}$ & $\mathbf{2}$\tabularnewline
$\overline{H}$ & $\mathbf{4}$ & $\mathbf{1}$ & $\mathbf{2}$ & $\mathbf{\overline{4}}$ & $\mathbf{\overline{2}}$ & $\mathbf{1}$\tabularnewline
\midrule
$H'$ & $\mathbf{1}$ & $\mathbf{1}$ & $\mathbf{1}$ & $\mathbf{4}$ & $\mathbf{1}$ & $\mathbf{2}$\tabularnewline
$\Phi$ & $\mathbf{1}$ & $\mathbf{2}$ & $\mathbf{1}$ & $\mathbf{1}$ & $\mathbf{\overline{2}}$ & $\mathbf{1}$\tabularnewline
$\overline{\Phi}$ & $\mathbf{1}$ & $\mathbf{1}$ & $\mathbf{\overline{2}}$ & $\mathbf{1}$ & $\mathbf{1}$ & $\mathbf{2}$\tabularnewline
\bottomrule
\end{tabular}
\par\end{centering}
\caption{The field content under $G_{422}^{I}\times G_{422}^{II}$, see the
main text for details. \label{tab:Field_content_TwinPS}}
\end{table}
\begin{equation}
\mathcal{L}_{\mathrm{mass}}^{ren}=y_{i4}^{\psi}\overline{H}\psi_{i}\psi_{4}^{c}+y_{4i}^{\psi}H\psi_{4}\psi_{i}^{c}+x_{i4}^{\psi}\phi\psi_{i}\overline{\psi_{4}}+x_{4i}^{\psi^{c}}\overline{\psi_{4}^{c}\phi}\psi_{i}^{c}+M_{4}^{\psi}\psi_{4}\overline{\psi_{4}}+M_{4}^{\psi^{c}}\psi_{4}^{c}\overline{\psi_{4}^{c}}\,,\label{eq:Lren_4thVL-1}
\end{equation}
plus h.c., where $i=1,2,3$; $x,y$ are dimensionless universal coupling
constants and $M_{4}^{\psi,\psi^{c}}$ are the VL mass terms. These
couplings mix the chiral fermions with the VL fermions, and will be
responsible for generating effective Yukawa couplings for the second
and third families. Moreover, the same mixing leads to effective couplings
to TeV scale $SU(4)^{I}$ gauge bosons which violate lepton universality
between the second and third families, as we shall see.

\subsection{High scale symmetry breaking\label{subsec:High-scale-symmetry}}

The twin Pati-Salam symmetry displayed in Eq.~\eqref{eq:TwinPS_symmetry}
is spontaneously broken to the ``4321'' symmetry at the high scale
$M_{\mathrm{High}}\apprge1\,\mathrm{PeV}$ (the latter bound due to
the non-observation of $K_{L}\rightarrow\mu e$ \cite{Valencia:1994cj}),
\begin{equation}
G_{422}^{I}\times G_{422}^{II}\rightarrow G_{4321}\equiv SU(4)_{PS}^{I}\times SU(3)_{c}^{II}\times SU(2)_{L}^{I+II}\times U(1)_{Y'}\,.
\end{equation}
We can think of this as a two part symmetry breaking:\\
(i) The two pairs of left-right groups break down to their diagonal
left-right subgroup, via the VEVs $\left\langle \Phi\right\rangle \sim v_{\Phi}$
and $\langle\overline{\Phi}\rangle\sim v_{\overline{\Phi}}$, leading
to the symmetry breaking,
\begin{equation}
SU(2)_{L}^{I}\times SU(2)_{L}^{II}\rightarrow SU(2)_{L}^{I+II}\,,\qquad SU(2)_{R}^{I}\times SU(2)_{R}^{II}\rightarrow SU(2)_{R}^{I+II}\,.
\end{equation}
Since the two $SU(4)_{PS}$ groups remain intact, the above symmetry
breaking corresponds to
\begin{equation}
G_{422}^{I}\times G_{422}^{II}\rightarrow G_{4422}\equiv SU(4)_{PS}^{I}\times SU(4)_{PS}^{II}\times SU(2)_{L}^{I+II}\times SU(2)_{R}^{I+II}\,.
\end{equation}
(ii) Then we assume the second PS group is broken at a high scale
via the Higgs $H'$ in Table~\ref{tab:Field_content_TwinPS}, which
under $G_{4422}$ transform as
\begin{equation}
H'(1,4,1,2)=\left(\begin{array}{cccc}
u_{H'}^{r} & u_{H'}^{b} & u_{H'}^{g} & \nu_{H'}\\
d_{H'}^{r} & d_{H'}^{b} & d_{H'}^{g} & e_{H'}
\end{array}\right)\,,
\end{equation}
and develops VEV in its right-handed neutrino component\footnote{This VEV is also responsible for heavy right-handed neutrino masses
leading to a seesaw mechanism with naturally light neutrinos as discussed
in \cite{King:2021jeo}. In the present paper we shall ignore such
small neutrino masses which play no role in the phenomenological analysis.},$\left\langle \nu_{H'}\right\rangle \apprge1\,\mathrm{PeV}$, leading
to the symmetry breaking
\begin{equation}
G_{4422}\rightarrow G_{4321}\equiv SU(4)_{PS}^{I}\times SU(3)_{c}^{II}\times SU(2)_{L}^{I+II}\times U(1)_{Y'}\,,
\end{equation}
where $SU(4)_{PS}^{II}$ is broken to $SU(3)_{c}^{II}\times U(1)_{B-L}^{II}$
(at the level of fermion representations, chiral quarks and leptons
are split $\mathbf{4}^{II}\rightarrow(\mathbf{3},1/6)^{II}\oplus(\mathbf{1},-1/2)^{II}$),
while $SU(2)_{R}^{I+II}$ is broken to $U(1)_{T_{3R}}^{I+II}$ and
the abelian generators are broken to $U(1)_{Y'}$, where $Y'=T_{B-L}^{II}+T_{3R}^{I+II}$.
The broken generators of $SU(4)_{PS}^{II}$ are associated with PeV-scale
gauge bosons that will mediate processes at acceptable rates, beyond
the sensitivity of current experiments and colliders. Instead, the
further symmetry breaking of $G_{4321}$ will lead to a rich phenomenology
at the TeV scale, as we shall see. We anticipate that $SU(2)_{L}^{I+II}$
is already the $SU(2)_{L}$ of the SM gauge group, while SM color
and hypercharge are embedded in $SU(4)_{PS}^{I}\times SU(3)_{c}^{II}\times U(1)_{Y'}$.

On the other hand, the Yukon scalars $\phi$ and $\overline{\phi}$
in Table~\ref{tab:Field_content_TwinPS} decompose under $G_{422}^{I}\times G_{422}^{II}\rightarrow G_{4422}\rightarrow G_{4321}$
as 
\begin{equation}
\begin{array}{c}
\phi(\mathbf{4,2,1;\overline{4},\overline{2},1})\rightarrow\phi(\mathbf{4,\overline{4},1\oplus3,1})\rightarrow\phi_{3}(\mathbf{4,\overline{3},1\oplus3},-1/6)\oplus\phi_{1}(\mathbf{4,1,1\oplus3},1/2)\,,\\
\,\\
\overline{\phi}(\mathbf{\overline{4},1,\overline{2};4,1,2})\rightarrow\overline{\phi}(\mathbf{\overline{4},4,1,1\oplus3})\rightarrow\overline{\phi_{3}}(\mathbf{\overline{4},3,1},1/6)\oplus\overline{\phi_{1}}(\mathbf{\overline{4},1,1},-1/2)\,,
\end{array}\label{eq:phi_decomposition}
\end{equation}
plus extra $\overline{\phi_{3}}$ and $\overline{\phi_{1}}$ with
different values of $Y'$ associated to the breaking of the $SU(2)_{R}^{I+II}$
triplet, that we ignore because they do not couple to fermions. The
decomposition above is of phenomenological interest, as the Yukons
$\phi_{3}$, $\overline{\phi_{3}}$ will couple to quarks while $\phi_{1}$,
$\overline{\phi_{1}}$ will couple to leptons, allowing non-trivial
mixing between SM fermions and VL fermions. They will also lead to
a non-trivial breaking of $G_{4321}$ down to the SM.

The Higgs scalars $H$ and $\overline{H}$ in Table~\ref{tab:Field_content_TwinPS}
decompose under $G_{422}^{I}\times G_{422}^{II}\rightarrow G_{4321}$
as (we skip the $G_{4422}$ decomposition here for simplicity) 
\begin{equation}
H(\mathbf{\overline{4},\overline{2},1;4,1,2})\rightarrow H_{t}(\mathbf{\overline{4},3,\overline{2}},2/3),\:H_{b}(\mathbf{\overline{4},3,\overline{2}},-1/3),\,H_{\tau}(\mathbf{\overline{4},1,\overline{2}},-1),\,H_{\nu_{\tau}}(\mathbf{\overline{4},1,\overline{2}},0)\,,\label{eq:H_Higgs}
\end{equation}
\begin{equation}
\overline{H}(\mathbf{4,1,2;\overline{4},\overline{2},1})\rightarrow H_{c}(\mathbf{4,\overline{3},\overline{2}},1/3),\:H_{s}(\mathbf{4,\overline{3},\overline{2}},-2/3),\,H_{\mu}(\mathbf{\overline{4},1,\overline{2}},0),\,H_{\nu_{\mu}}(\mathbf{\overline{4},1,\overline{2}},1)\,,\label{eq:Hbar_Higgs}
\end{equation}
where the notation anticipates that a separate personal Higgs doublet
contributes to each of the second and third family quark and lepton
masses, as we shall see. Models with multiple light Higgs doublets
face the phenomenological challenge of FCNCs arising from tree-level
exchange of the scalar doublets in the Higgs basis. Therefore we assume
that only one pair of Higgs doublets, $H_{u}$ and $H_{d}$ are light,
given by linear combinations of the personal Higgs,
\begin{equation}
\begin{array}{c}
H_{u}=\widetilde{\alpha}_{u}H_{t}+\widetilde{\beta}_{u}H_{c}+\widetilde{\gamma}_{u}H_{\nu_{\tau}}+\widetilde{\delta}_{u}H_{\nu_{\mu}}\,,\quad H_{d}=\widetilde{\alpha}_{d}H_{b}+\widetilde{\beta}_{d}H_{s}+\widetilde{\gamma}_{d}H_{\tau}+\widetilde{\delta}_{d}H_{\mu}\,,\end{array}\label{eq:Light_Higgses}
\end{equation}
where $\widetilde{\alpha}_{u,d}$, $\widetilde{\beta}_{u,d}$, $\widetilde{\gamma}_{u,d}$,
$\widetilde{\delta}_{u,d}$ are complex elements of two unitary Higgs
mixing matrices. The orthogonal linear combinations are assumed to
be very heavy, well above the TeV scale in order to sufficiently suppress
the FCNCs. We will further assume that only the light Higgs doublet
states get VEVs in order to perform EW symmetry breaking,
\begin{equation}
\left\langle H_{u}\right\rangle =v_{u},\quad\left\langle H_{d}\right\rangle =v_{d},\label{eq:VeVs_2HDM}
\end{equation}
while the heavy linear combinations do not, i.e.~we assume that in
the Higgs basis the linear combinations which do not get VEVs are
very heavy. The discussion of such Higgs potential is beyond the scope
of this paper, for the interested reader a deeper discussion was made
in Section 3.4 of \cite{King:2021jeo}. In any case, we shall invert
the unitary transformations in Eq.~\eqref{eq:Light_Higgses} to express
each of the personal Higgs doublets in terms of the light doublets
$H_{u}$, $H_{d}$,
\begin{equation}
\begin{array}{c}
H_{t}=\alpha_{u}H_{u}+...\,,\quad H_{b}=\alpha_{d}H_{d}+...\,,\quad H_{\tau}=\gamma_{d}H_{d}+...\,,\quad H_{\nu_{\tau}}=\gamma_{u}H_{u}+...\,,\\
\,\\
H_{c}=\beta_{u}H_{u}+...\,,\quad H_{s}=\beta_{d}H_{d}+...\,,\quad H_{\mu}=\delta_{d}H_{d}+...\,,\quad\,H_{\nu_{\mu}}=\delta_{u}H_{u}+...\,,
\end{array}\label{eq:Higgs_matching}
\end{equation}
ignoring the heavy states indicated by dots. When the light Higgs
$H_{u}$, $H_{d}$ gain their VEVs in Eq.~\eqref{eq:VeVs_2HDM},
the personal Higgs in the original basis can be thought of as gaining
effective VEVs $\left\langle H_{t}\right\rangle =\alpha_{u}v_{u}$,
etc... This approach will be used in the next section, when constructing
the low-energy quark and lepton mass matrices.

\subsection{Effective Yukawa couplings and fermion masses \label{subsec:Effective-Yukawa-couplings}}

We have already remarked that the usual Yukawa couplings involving
purely chiral fermions are absent in the twin PS model. In this subsection,
we show how they may be generated effectively via mixing with the
vector-like fermions. 

We may write the mass terms and couplings in Eq.~\eqref{eq:Lren_4thVL-1}
as a $5\times5$ matrix in flavour space (we also define 5-dimensional
vectors as $\psi_{\alpha}^{\mathrm{T}}$ and $\psi_{\beta}^{c}$),

\begin{equation}
\mathcal{L}_{\mathrm{mass}}^{ren}=\psi_{\alpha}^{\mathrm{T}}M^{\psi}\psi_{\beta}^{c}+\mathrm{h.c.}\,,
\end{equation}
\begin{equation}
\psi_{\alpha}^{\mathrm{T}}\equiv\left(\begin{array}{ccccc}
\psi_{1} & \psi_{2} & \psi_{3} & \psi_{4} & \overline{\psi_{4}^{c}}\end{array}\right)\,,\qquad\psi_{\beta}^{c}\equiv\left(\begin{array}{ccccc}
\psi_{1}^{c} & \psi_{2}^{c} & \psi_{3}^{c} & \psi_{4}^{c} & \overline{\psi_{4}}\end{array}\right)^{\mathrm{T}}\,,
\end{equation}
\begin{equation}
M^{\psi}=\left(
\global\long\def\arraystretch{1.3}%
\begin{array}{@{}llcccc@{}}
 & \multicolumn{1}{c@{}}{\psi_{1}^{c}} & \psi_{2}^{c} & \psi_{3}^{c} & \psi_{4}^{c} & \overline{\psi_{4}}\\
\cmidrule(l){2-6}\left.\psi_{1}\right| & 0 & 0 & 0 & 0 & 0\\
\left.\psi_{2}\right| & 0 & 0 & 0 & y_{24}^{\psi}\overline{H} & 0\\
\left.\psi_{3}\right| & 0 & 0 & 0 & y_{34}^{\psi}\overline{H} & x_{34}^{\psi}\phi\\
\left.\psi_{4}\right| & 0 & 0 & y_{43}^{\psi}H & 0 & M_{4}^{\psi}\\
\left.\overline{\psi_{4}^{c}}\right| & 0 & x_{42}^{\psi^{c}}\overline{\phi} & x_{43}^{\psi^{c}}\overline{\phi} & M_{4}^{\psi^{c}} & 0
\end{array}\right)\,.\label{eq:MassMatrix_4thVL}
\end{equation}
where extra zeroes had been achieved via suitable rotations that leave
unchanged the upper $3\times3$ blocks. There are several distinct
mass scales in this matrix: the Higgs VEVs $\left\langle H\right\rangle $
and $\langle\overline{H}\rangle$, the Yukon VEVs $\left\langle \phi\right\rangle $
and $\langle\overline{\phi}\rangle$ and the VL fourth family masses
$M_{4}^{\psi}$, $M_{4}^{\psi^{c}}$. Assuming the latter are heavier
than all the scalars VEVs, we may integrate out the fourth family,
to generate effective Yukawa couplings for chiral quarks and leptons
which originate from the diagrams in Fig.~\ref{fig: mass_insertion_4thVL}.
This is denoted as the mass insertion approximation.

As anticipated in \cite{King:2021jeo}, the heavy top mass requires
$\left\langle \phi\right\rangle /M_{4}^{\psi}\sim1$ and thus it is
necessary to go beyond the mass insertion approximation, where the
large mixing angle formalism introduced in Appendix \ref{sec:Mixing-angle-formalism}
applies. We shall block-diagonalise the mass matrix in Eq.~(\ref{eq:MassMatrix_4thVL})
in order to obtain the SM Yukawa couplings for the chiral families,
\begin{equation}
M^{\psi'}=\left(
\global\long\def\arraystretch{1.3}%
\begin{array}{@{}llcccc@{}}
 & \multicolumn{1}{c@{}}{\psi'{}_{1}^{c}} & \psi'{}_{2}^{c} & \psi'{}_{3}^{c} & \psi'{}_{4}^{c} & \overline{\psi'_{4}}\\
\cmidrule(l){2-6}\left.\psi'_{1}\right| &  &  &  &  & 0\\
\left.\psi'_{2}\right| &  &  &  &  & 0\\
\left.\psi'_{3}\right| &  &  & \widetilde{y}{}_{\alpha\beta}^{\psi'} &  & 0\\
\left.\psi'_{4}\right| &  &  &  &  & \widetilde{M}_{4}^{\psi}\\
\left.\overline{\psi_{4}^{c'}}\right| & 0 & 0 & 0 & \widetilde{M}_{4}^{\psi^{c}} & 0
\end{array}\right)\,,\label{MassMatrix_4thVL_decoupling}
\end{equation}
where $\widetilde{y}{}_{\alpha\beta}^{\psi'}$ are the upper $4\times4$
block of the mass matrices in this basis. The key feature of Eq.~\eqref{MassMatrix_4thVL_decoupling}
is the zeros in the fifth row and column which are achieved by rotating
the four families by the unitary $4\times4$ transformations,
\begin{equation}
V_{\psi}=V_{34}^{\psi}=\left(\begin{array}{cccc}
1 & 0 & 0 & 0\\
0 & 1 & 0 & 0\\
0 & 0 & c_{34}^{\psi} & s_{34}^{\psi}\\
0 & 0 & -s_{34}^{\psi} & c_{34}^{\psi}
\end{array}\right)\,,\quad V_{\psi^{c}}=V_{34}^{\psi^{c}}V_{24}^{\psi^{c}}=\left(\begin{array}{cccc}
1 & 0 & 0 & 0\\
0 & 1 & 0 & 0\\
0 & 0 & c_{34}^{\psi^{c}} & s_{34}^{\psi^{c}}\\
0 & 0 & -s_{34}^{\psi^{c}} & c_{34}^{\psi^{c}}
\end{array}\right)\left(\begin{array}{cccc}
1 & 0 & 0 & 0\\
0 & c_{24}^{\psi^{c}} & 0 & s_{24}^{\psi^{c}}\\
0 & 0 & 1 & 0\\
0 & -s_{24}^{\psi^{c}} & 0 & c_{24}^{\psi^{c}}
\end{array}\right)\,,\label{eq:34mixing}
\end{equation}
where the mixing angles are given in Appendix~\ref{sec:Mixing-angle-formalism},
we define $s_{i\alpha}^{\psi^{(c)}}\equiv\sin\theta_{i\alpha}^{\psi^{(c)}}$,~$c_{i\alpha}^{\psi^{(c)}}\equiv\cos\theta_{i\alpha}^{\psi^{(c)}}$.
\begin{flalign}
 & s_{34}^{\psi}=\frac{x_{34}^{\psi}\left\langle \phi\right\rangle }{\sqrt{\left(x_{34}^{\psi}\left\langle \phi\right\rangle \right)^{2}+\left(M_{4}^{\psi}\right)^{2}}}\,, &  & s_{24}^{\psi^{c}}=\frac{x_{42}^{\psi^{c}}\langle\overline{\phi}\rangle}{\sqrt{\left(x_{42}^{\psi^{c}}\langle\overline{\phi}\rangle\right)^{2}+\left(M_{4}^{\psi^{c}}\right)^{2}}}\,,\label{eq:34_mixing_extended-1}\\
 & s_{34}^{\psi^{c}}=\frac{x_{43}^{\psi^{c}}\langle\overline{\phi}\rangle}{\sqrt{\left(x_{42}^{\psi^{c}}\langle\overline{\phi}\rangle\right)^{2}+\left(x_{43}^{\psi^{c}}\langle\overline{\phi}\rangle\right)^{2}+\left(M_{4}^{\psi^{c}}\right)^{2}}}\,, &  & \tilde{M}_{4}^{\psi}=\sqrt{\left(x_{34}^{\psi}\left\langle \phi\right\rangle \right)^{2}+\left(M_{4}^{\psi}\right)^{2}}\,,\label{eq:sqc24_mixing-1}\\
 & \widetilde{M}_{4}^{\psi^{c}}=\sqrt{\left(x_{42}^{\psi^{c}}\langle\overline{\phi}\rangle\right)^{2}+\left(x_{43}^{\psi^{c}}\langle\overline{\phi}\rangle\right)^{2}+\left(M_{4}^{\psi^{c}}\right)^{2}}\,,\label{eq:sqc34_mixing-1}
\end{flalign}

Now we apply the transformations in Eq.~\eqref{eq:34mixing} to the
upper $4\times4$ block of \eqref{eq:MassMatrix_4thVL}, obtaining
effective Yukawa couplings for the chiral fermions as the upper $3\times3$
block of the mass matrix in the new basis,
\begin{figure}[t]
\subfloat[]{\begin{centering}
\begin{tikzpicture}
	\begin{feynman}
		\vertex (a) {\(\psi_{3}\)};
		\vertex [right=18mm of a] (b);
		\vertex [right=of b] (c) [label={ [xshift=0.1cm, yshift=0.1cm] \small $M^{\psi}_{4}$}];
		\vertex [right=of c] (d);
		\vertex [right=of d] (e) {\(\psi^{c}_{3}\)};
		\vertex [above=of b] (f1) {\(\phi\)};
		\vertex [above=of d] (f2) {\(H\)};
		\diagram* {
			(a) -- [fermion] (b) -- [charged scalar] (f1),
			(b) -- [edge label'=\(\overline{\psi_{4}}\)] (c),
			(c) -- [edge label'=\(\psi_{4}\), inner sep=6pt, insertion=0] (d) -- [charged scalar] (f2),
			(d) -- [fermion] (e),
	};
	\end{feynman}
\end{tikzpicture}
\par\end{centering}
}$\quad$\subfloat[]{\begin{centering}
\begin{tikzpicture}
	\begin{feynman}
		\vertex (a) {\(\psi_{i}\)};
		\vertex [right=18mm of a] (b);
		\vertex [right=of b] (c) [label={ [xshift=0.1cm, yshift=0.1cm] \small $M^{\psi^{c}}_{4}$}];
		\vertex [right=of c] (d);
		\vertex [right=of d] (e) {\(\psi^{c}_{j}\)};
		\vertex [above=of b] (f1) {\(\overline{H}\)};
		\vertex [above=of d] (f2) {\(\overline{\phi}\)};
		\diagram* {
			(a) -- [fermion] (b) -- [charged scalar] (f1),
			(b) -- [edge label'=\(\overline{\psi^{c}_{4}}\)] (c),
			(c) -- [edge label'=\(\psi^{c}_{4}\), inner sep=6pt, insertion=0] (d) -- [charged scalar] (f2),
			(d) -- [fermion] (e),
	};
	\end{feynman}
\end{tikzpicture}
\par\end{centering}
}\caption{Diagrams in the model which lead to the effective Yukawa couplings
in the mass insertion approximation, $i,j=2,3$. \label{fig: mass_insertion_4thVL}}
\end{figure}
\begin{equation}
\mathcal{L}_{eff}^{Yuk3\times3}=\psi'{}_{i}^{\mathrm{T}}V_{\psi}y_{\alpha\beta}^{\psi}V_{\psi^{c}}^{\dagger}\psi'{}_{j}^{c}+\mathrm{h.c.}\,,\label{eq:Primed_basis}
\end{equation}
\begin{equation}
\psi'{}_{\alpha}^{\mathrm{T}}=\psi_{\alpha}^{\mathrm{T}}V_{\psi}^{\dagger}\,,\qquad\psi'{}_{\alpha}^{c}=V_{\psi^{c}}\psi_{\alpha}^{c}\,,
\end{equation}
where $i,j=1,2,3$. We obtain
\begin{equation}
\mathcal{L}_{eff}^{Yuk,3\times3}=\left(
\global\long\def\arraystretch{0.7}%
\begin{array}{@{}llcc@{}}
 & \multicolumn{1}{c@{}}{\psi'{}_{1}^{c}} & \psi'{}_{2}^{c} & \psi'{}_{3}^{c}\\
\cmidrule(l){2-4}\left.\psi'_{1}\right| & 0 & 0 & 0\\
\left.\psi'_{2}\right| & 0 & 0 & 0\\
\left.\psi'_{3}\right| & 0 & 0 & c_{34}^{\psi^{c}}s_{34}^{\psi}y_{43}^{\psi}
\end{array}\right)H+\left(
\global\long\def\arraystretch{0.7}%
\begin{array}{@{}llcc@{}}
 & \multicolumn{1}{c@{}}{\psi'{}_{1}^{c}} & \psi'{}_{2}^{c} & \psi'{}_{3}^{c}\\
\cmidrule(l){2-4}\left.\psi'_{1}\right| & 0 & 0 & 0\\
\left.\psi'_{2}\right| & 0 & s_{24}^{\psi^{c}}y_{24}^{\psi} & c_{24}^{\psi^{c}}s_{34}^{\psi^{c}}y_{24}^{\psi}\\
\left.\psi'_{3}\right| & 0 & c_{34}^{\psi}s_{24}^{\psi^{c}}y_{34}^{\psi} & c_{34}^{\psi}c_{24}^{\psi^{c}}s_{34}^{\psi^{c}}y_{34}^{\psi}
\end{array}\right)\overline{H}+\mathrm{h.c.}\label{eq:MassMatrix_4thVL_effective-1-1}
\end{equation}
Until the breaking of the twin PS symmetry, the matrix above is Pati-Salam
universal, so all fermions of the same flavour share the same effective
Yukawa $y_{\mathrm{eff}}^{\psi}$. If we assume a hierarchy of scales
for the VL masses
\begin{equation}
M_{4}^{\psi}\ll M_{4}^{\psi^{c}}\,,\label{eq:hierarchy_scalesVL}
\end{equation}
then the first matrix in Eq.~\eqref{MassMatrix_4thVL_decoupling}
generates larger effective third family Yukawa couplings, while the
second matrix generates suppressed second family Yukawa couplings
and mixings. This way, the hierarchy of quark and lepton masses in
the SM Yukawa couplings is re-expressed as the hierarchy of scales
in Eq.~\eqref{eq:hierarchy_scalesVL}. Remarkably, the hierarchical
relation in Eq.~\eqref{eq:hierarchy_scalesVL} will lead to small
couplings of $\psi^{c}$ chiral fermions (or SM EW singlets) to $SU(4)^{I}$
gauge bosons, hence obtaining dominantly left-handed $U_{1}$ couplings.
The couplings to RH fermions will be suppressed, connected to the
origin of second family fermion masses, and this way the tight high-$p_{T}$
constraints that afflict other 4321 models can be relaxed (see Section~\ref{subsec:Colliders}).

On the other hand, since the sum of the two matrices in Eq.~\eqref{MassMatrix_4thVL_decoupling}
has rank 1, the first family will be massless. The masses of first
family fermions can arise via the mechanism presented in \cite{King:2021jeo},
however it leads to no connections with $B$-physics and the relevant
phenomenology discussed here. Therefore, for the phenomenological
purposes of this manuscript, we can safely assume the first family
to remain massless.

After the symmetry breaking of the twin PS group to $G_{4321}$, the
Yukawa couplings $x_{34}^{\psi}$, $x_{42,43}^{\psi^{c}}$ and VL
masses $M_{4}^{\psi},$ $M_{4}^{\psi^{c}}$ remain universal up to
small renormalisation group evolution (RGE) effects, however the Yukons
decompose in a different way for lepton and quarks as per Eq.~\eqref{eq:phi_decomposition}.
Due to this decomposition, the mixing angles in Eq.~\eqref{eq:MassMatrix_4thVL_effective-1-1}
are now different for quark and leptons. The VEVs of the Yukons break
the $SU(4)$ symmetry relating quarks and leptons, but an accidental
$SU(2)_{q^{c}}$ symmetry relating $\psi^{c}$ quarks remains. Hence,
the mixing angles $(s_{i4}^{u^{c}}=s_{i4}^{d^{c}})$ are the same
for up and down quarks, and we define $q^{c}=u^{c},\,d^{c}$. On the
other hand, the Higgs fields $H$, $\overline{H}$ decompose as personal
Higgs doublets for the second and third fermion families as per \eqref{eq:H_Higgs}
and \eqref{eq:Hbar_Higgs}. The personal Higgses are introduced in
order to break the accidental symmetry $SU(2)_{q_{c}}$, otherwise
the mass matrices in the up and down sector would remain identical.
A similar discussion applies to charged leptons and neutrinos, and
personal Higgses apply in the same way. Mass terms for second and
third family fermions will be obtained after the personal Higgses
develop a VEV, see Section \ref{subsec:High-scale-symmetry}. This
way, Eq.~\eqref{eq:MassMatrix_4thVL_effective-1-1} decomposes for
each charged sector as the following effective mass matrices,
\begin{equation}
M_{\mathrm{eff}}^{u}=\left(
\global\long\def\arraystretch{0.7}%
\begin{array}{@{}llcc@{}}
 & \multicolumn{1}{c@{}}{\phantom{\!\,}u'{}_{1}^{c}} & \phantom{\!\,}u'{}_{2}^{c} & \phantom{\!\,}u'{}_{3}^{c}\\
\cmidrule(l){2-4}\left.Q'_{1}\right| & 0 & 0 & 0\\
\left.Q'_{2}\right| & 0 & 0 & 0\\
\left.Q'_{3}\right| & 0 & 0 & s_{34}^{Q}y_{43}^{\psi}
\end{array}\right)\left\langle H_{t}\right\rangle +\left(
\global\long\def\arraystretch{0.7}%
\begin{array}{@{}llcc@{}}
 & \multicolumn{1}{c@{}}{\phantom{\!\,}u'{}_{1}^{c}} & \phantom{\!\,}u'{}_{2}^{c} & \phantom{\!\,}u'{}_{3}^{c}\\
\cmidrule(l){2-4}\left.Q'_{1}\right| & 0 & 0 & 0\\
\left.Q'_{2}\right| & 0 & s_{24}^{q^{c}}y_{24}^{\psi} & s_{34}^{q^{c}}y_{24}^{\psi}\\
\left.Q'_{3}\right| & 0 & c_{34}^{Q}s_{24}^{q^{c}}y_{34}^{\psi} & c_{34}^{Q}s_{34}^{q^{c}}y_{34}^{\psi}
\end{array}\right)\left\langle H_{c}\right\rangle +\mathrm{h.c.}\,,\label{eq:MassMatrix_4thVL_effective_up}
\end{equation}
\begin{equation}
M_{\mathrm{eff}}^{d}=\left(
\global\long\def\arraystretch{0.7}%
\begin{array}{@{}llcc@{}}
 & \multicolumn{1}{c@{}}{\phantom{\!\,}d'{}_{1}^{c}} & \phantom{\!\,}d'{}_{2}^{c} & \phantom{\!\,}d'{}_{3}^{c}\\
\cmidrule(l){2-4}\left.Q'_{1}\right| & 0 & 0 & 0\\
\left.Q'_{2}\right| & 0 & 0 & 0\\
\left.Q'_{3}\right| & 0 & 0 & s_{34}^{Q}y_{43}^{\psi}
\end{array}\right)\left\langle H_{b}\right\rangle +\left(
\global\long\def\arraystretch{0.7}%
\begin{array}{@{}llcc@{}}
 & \multicolumn{1}{c@{}}{\phantom{\!\,}d'{}_{1}^{c}} & \phantom{\!\,}d'{}_{2}^{c} & \phantom{\!\,}d'{}_{3}^{c}\\
\cmidrule(l){2-4}\left.Q'_{1}\right| & 0 & 0 & 0\\
\left.Q'_{2}\right| & 0 & s_{24}^{q^{c}}y_{24}^{\psi} & s_{34}^{q^{c}}y_{24}^{\psi}\\
\left.Q'_{3}\right| & 0 & c_{34}^{Q}s_{24}^{q^{c}}y_{34}^{\psi} & c_{34}^{Q}s_{34}^{q^{c}}y_{34}^{\psi}
\end{array}\right)\left\langle H_{s}\right\rangle +\mathrm{h.c.}\,,\label{eq:MassMatrix_4thVL_effective_down}
\end{equation}
\begin{equation}
M_{\mathrm{eff}}^{e}=\left(
\global\long\def\arraystretch{0.7}%
\begin{array}{@{}llcc@{}}
 & \multicolumn{1}{c@{}}{\phantom{\!\,}e'{}_{1}^{c}} & \phantom{\!\,}e'{}_{2}^{c} & \phantom{\!\,}e'{}_{3}^{c}\\
\cmidrule(l){2-4}\left.L'_{1}\right| & 0 & 0 & 0\\
\left.L'_{2}\right| & 0 & 0 & 0\\
\left.L'_{3}\right| & 0 & 0 & s_{34}^{L}y_{43}^{\psi}
\end{array}\right)\left\langle H_{\tau}\right\rangle +\left(
\global\long\def\arraystretch{0.7}%
\begin{array}{@{}llcc@{}}
 & \multicolumn{1}{c@{}}{\phantom{\!\,}e'{}_{1}^{c}} & \phantom{\!\,}e'{}_{2}^{c} & \phantom{\!\,}e'{}_{3}^{c}\\
\cmidrule(l){2-4}\left.L'_{1}\right| & 0 & 0 & 0\\
\left.L'_{2}\right| & 0 & s_{24}^{e^{c}}y_{24}^{\psi} & s_{34}^{e^{c}}y_{24}^{\psi}\\
\left.L'_{3}\right| & 0 & c_{34}^{L}s_{24}^{e^{c}}y_{34}^{\psi} & c_{34}^{L}s_{34}^{e^{c}}y_{34}^{\psi}
\end{array}\right)\left\langle H_{\mu}\right\rangle +\mathrm{h.c.}\,,\label{eq:MassMatrix_4thVL_effective_leptons}
\end{equation}
where the Yukawas $y_{43}^{\psi}$ and $y_{24,34}^{\psi}$ are Pati-Salam
universal, and we have approximated all cosines related to $\psi^{c}$
fields to be 1 due to the hierarchy of VL masses in Eq.~\eqref{eq:hierarchy_scalesVL}.
We obtain a similar Dirac-like matrix for neutrinos. In the complete
version of the model presented in \cite{King:2021jeo}, a further
Majorana matrix for the singlet neutrinos $\nu^{c}$ is obtained,
and all neutrino masses and mixings are accommodated via a type I
seesaw mechanism (see full discussion in Section 4.2 of \cite{King:2021jeo}).
However, for the sake of simplicity, we will consider massless neutrinos
in this simplified framework, as they are of subleading importance
for the $B$-anomalies and for the phenomenological analysis intended
for this article.

Due to the fact that VL fermions are much heavier than SM fermions,
the fourth row and column, that we have intentionally ignored when
writing Eqs.~\eqref{eq:MassMatrix_4thVL_effective_up},~\eqref{eq:MassMatrix_4thVL_effective_down},~\eqref{eq:MassMatrix_4thVL_effective_leptons},
can be decoupled from the $3\times3$ upper blocks, which we can diagonalise
via independent 2-3 transformations for each charged sector $V_{23}^{u}$,
$V_{23}^{d}$ and $V_{23}^{e}$. Similar transformations apply for
EW singlet fermions $u^{c}$, $d^{c}$, $e^{c}$, in such a way that
the mass matrices in Eqs.~\eqref{eq:MassMatrix_4thVL_effective_up},~\eqref{eq:MassMatrix_4thVL_effective_down},~\eqref{eq:MassMatrix_4thVL_effective_leptons}
are diagonalised as
\begin{equation}
V_{23}^{u}M_{\mathrm{eff}}^{u}V_{23}^{u^{c}\dagger}=\mathrm{diag}(0,m_{c},m_{t})\,,\:V_{23}^{d}M_{\mathrm{eff}}^{d}V_{23}^{d^{c}\dagger}=\mathrm{diag}(0,m_{s},m_{b})\,,\:V_{23}^{e}M_{\mathrm{eff}}^{e}V_{23}^{e^{c}\dagger}=\mathrm{diag}(0,m_{\mu},m_{\tau})\,.
\end{equation}
The CKM matrix is then predicted as
\begin{equation}
V_{\mathrm{CKM}}=V_{23}^{u}V_{23}^{d\dagger}=\left(\begin{array}{ccc}
1 & 0 & 0\\
0 & c_{23}^{u}c_{23}^{d}+s_{23}^{u}s_{23}^{d} & c_{23}^{d}s_{23}^{u}-c_{23}^{u}s_{23}^{d}\\
0 & -\left(c_{23}^{d}s_{23}^{u}-c_{23}^{u}s_{23}^{d}\right) & c_{23}^{u}c_{23}^{d}+s_{23}^{u}s_{23}^{d}
\end{array}\right)\approx\left(\begin{array}{ccc}
1 & 0 & 0\\
0 & V_{cs} & V_{cb}\\
0 & V_{ts} & V_{tb}
\end{array}\right).\label{eq:CKM_matrix}
\end{equation}
We do not address the mixing involving the first family since we are
assuming massless first family fermions, as previously discussed.
We are however required to preserve $V_{cb}$ as \cite{PDG:2022ynf}
\begin{equation}
V_{cb}=\left(41.0\pm1.4\right)\times10^{-3}\approx s_{23}^{u}-s_{23}^{d}\,,\label{eq:Vcb_model}
\end{equation}
positive in our parameterisation, where in the last step we have approximated
the cosines to be 1. We will not fit $V_{ts}$, $V_{tb}$, $V_{cs}$
up to the experimental precision, as corrections related to the first
family mixing (and CPV phase) are required.

In the following we explore the parameters in the mass matrices of
Eqs.~\eqref{eq:MassMatrix_4thVL_effective_up},~\eqref{eq:MassMatrix_4thVL_effective_down},~\eqref{eq:MassMatrix_4thVL_effective_leptons},
and its impact over the diagonalisation of the mass matrices:
\begin{itemize}
\item In very good approximation, the mass of the top quark is given by
the (3,3) entry in the first matrix of Eq.~\eqref{eq:MassMatrix_4thVL_effective_up},
i.e.
\begin{equation}
m_{t}\approx s_{34}^{Q}y_{43}^{\psi}\left\langle H_{t}\right\rangle =s_{34}^{Q}y_{43}^{\psi}\alpha_{u}\frac{1}{\sqrt{1+\tan^{-2}\beta}}\frac{v_{\mathrm{SM}}}{\sqrt{2}}\,,
\end{equation}
where $v_{\mathrm{SM}}=246\,\mathrm{GeV}$ and we have applied $\left\langle H_{t}\right\rangle =\alpha_{u}v_{u}$
as per Eq.~\eqref{eq:Higgs_matching}, where
\begin{equation}
v_{u}=\sin\beta\frac{v_{\mathrm{SM}}}{\sqrt{2}}=\frac{1}{\sqrt{1+\tan^{-2}\beta}}\frac{v_{\mathrm{SM}}}{\sqrt{2}}\,,
\end{equation}
as in usual 2HDM. If we consider $\tan\beta\approx10$ and $\alpha_{u}\approx1$,
then we obtain
\begin{equation}
m_{t}\approx s_{34}^{Q}y_{43}^{\psi}\frac{v_{\mathrm{SM}}}{\sqrt{2}}\equiv y_{t}\frac{v_{\mathrm{SM}}}{\sqrt{2}}\,.\label{eq:Top_Effective Yukawa}
\end{equation}
From the expression above, it is clear that very large or maximal
$s_{34}^{Q}\approx1$ is required in order to preserve a natural $y_{43}^{\psi}$,
and to avoid perturbativity issues. Moreover, we will see that maximal
values for $s_{34}^{Q}$ are also well motivated by the $R_{D^{(*)}}$
anomaly, leading to a clear connection between the $B$-physics and
the flavour puzzle only present in this model.
\item In the bullet point above, the effective top Yukawa coupling in the
Higgs basis has been estimated as $y_{t}\approx1$. By following the
same procedure, we can see that all fermion masses can be accommodated
with natural parameters. Remarkably, we obtain that all the effective
Yukawa couplings are SM-like in the Higgs basis, explaining the observed
pattern of SM Yukawa couplings at low-energy.
\item The mixing between left-handed quark fields arise mainly from the
off-diagonal (2,3) entry in the quark mass matrices, which is controlled
by $s_{34}^{q^{c}}$. This mixing can be estimated for each sector
by the ratio of the (2,3) entry over the (3,3) entry, i.e.
\begin{equation}
\theta_{23}^{u}\approx\frac{s_{34}^{q^{c}}y_{24}^{\psi}\left\langle H_{c}\right\rangle }{s_{34}^{Q}y_{43}^{\psi}\left\langle H_{t}\right\rangle }\approx\frac{m_{c}}{m_{t}}\simeq\mathcal{O}(0.1V_{cb})\,,\qquad\theta_{23}^{d}\approx\frac{s_{34}^{q^{c}}y_{24}^{\psi}\left\langle H_{s}\right\rangle }{s_{34}^{Q}y_{43}^{\psi}\left\langle H_{b}\right\rangle }\approx\frac{m_{s}}{m_{b}}\simeq\mathcal{O}(V_{cb})\,,\label{eq:up_mixing_4thVL}
\end{equation}
obtained under the approximation $s_{34}^{q^{c}}\approx s_{24}^{q^{c}}$.
Therefore, the model predicts that $V_{cb}$ originates mainly from
the down sector, while the mixing in the up sector is small, suppressed
by the heavy top mass. The specific values of the mixing angles can
be different if we relax $s_{24}^{q^{c}}\approx s_{34}^{q^{c}}$,
but the CKM remains down-dominated in any case.
\item The lepton sector follows a similar discussion as that of the quark
sector. However, the phenomenological relation $\left\langle \phi_{3}\right\rangle \gg\left\langle \phi_{1}\right\rangle $
will lead to smaller angles than those of quarks, since the Yukawa
couplings $x_{34}^{\psi}$, $x_{42,43}^{\psi^{c}}$ and VL masses
$M_{4}^{\psi},$ $M_{4}^{\psi^{c}}$are universal. If $s_{34}^{Q}\approx1$,
then $s_{34}^{L}$ is expected to be large as well and we obtain $\left\langle H_{\tau}\right\rangle \approx m_{\tau}$.
Under the assumption $s_{24}^{e^{c}}\approx s_{34}^{e^{c}}$, the
charge lepton mixing is predicted as
\begin{equation}
\theta_{23}^{e}\approx\frac{s_{34}^{e^{c}}y_{24}^{\psi}\left\langle H_{\mu}\right\rangle }{s_{34}^{L}y_{43}^{\psi}\left\langle H_{\tau}\right\rangle }\approx\frac{m_{\mu}}{m_{\tau}}\simeq0.06\,.
\end{equation}
A particularly interesting situation arises when $s_{34}^{e^{c}}>s_{24}^{e^{c}}$,
where a larger $\theta_{23}^{e}$ contributing to large atmospheric
neutrino mixing is obtained. In this scenario, interesting signals
in lepton flavour-violating (LFV) processes such as $\tau\rightarrow3\mu$
or $\tau\rightarrow\mu\gamma$ arise, mediated at tree-level by $SU(4)^{I}$
gauge bosons. This is obtained if $x_{43}^{\psi^{c}}>x_{42}^{\psi^{c}}$,
without the need of any tuning.
\item Unlike private Higgs models \cite{Porto:2007ed,Porto:2008hb,BenTov:2012cx,Rodejohann:2019izm},
the personal Higgs VEVs are not hierarchical, all of order 1-10 GeV,
with the exception of the top one whose VEV is approximately that
of the SM Higgs doublet, as discussed above. The reason is that the
fermion mass hierarchies arise from the hierarchies $s_{34}^{\psi}\gg s_{24}^{\psi^{c}},s_{34}^{\psi^{c}}$,
which find their natural origin in the hierarchy of VL masses $M_{4}^{\psi}\ll M_{4}^{\psi^{c}}$
in Eq.~\eqref{eq:hierarchy_scalesVL}. The latter simultaneously
leads to dominantly left-handed leptoquark currents, as mentioned
before.
\end{itemize}

\subsection{The low-energy theory $G_{4321}$}

In this section we shall discuss the $G_{4321}$ theory that breaks
to the SM symmetry group at low energies $G_{4321}\rightarrow G_{\mathrm{SM}}$,
which is achieved via the scalars $\phi_{3}(4,\bar{3},1+3,-1/6)$
and $\phi_{1}(4,1,1+3,1/2)$ developing the VEVs
\begin{equation}
\left\langle \phi_{3}\right\rangle =\left(\begin{array}{ccc}
\frac{v_{3}}{\sqrt{2}} & 0 & 0\\
0 & \frac{v_{3}}{\sqrt{2}} & 0\\
0 & 0 & \frac{v_{3}}{\sqrt{2}}\\
0 & 0 & 0
\end{array}\right)\,,\quad\left\langle \phi_{1}\right\rangle =\left(\begin{array}{c}
0\\
0\\
0\\
\frac{v_{1}}{\sqrt{2}}
\end{array}\right)\,,
\end{equation}
where $v_{1},\,v_{3}\lesssim1\,\mathrm{TeV}$, and analogously for
$\overline{\phi}_{3}$ and $\overline{\phi}_{1}$ developing VEVs
$\overline{v}_{3}$ and $\overline{v}_{1}$, leading to the symmetry
breaking of $G_{4321}$ down to the SM gauge group,
\begin{equation}
SU(4)_{PS}^{I}\times SU(3)_{c}^{II}\times SU(2)_{L}^{I+II}\times U(1)_{Y'}\rightarrow SU(3)_{c}\times SU(2)_{L}\times U(1)_{Y}\,.\label{eq:4321_breaking}
\end{equation}
Here the $SU(4)_{PS}^{I}$ is broken to $SU(3)_{c}^{I}\times U(1)_{B-L}^{I}$($4\rightarrow3_{1/6}+1_{-1/2}$),
with $SU(3)_{c}^{I}\times SU(3)_{c}^{II}$ further broken to the diagonal
subgroup $SU(3)_{c}^{I+II}$, identified as SM QCD. On the other hand,
$SU(2)_{L}^{I+II}$ remains as the SM $SU(2)_{L}$. The Abelian generators
are broken to SM hypercharge $U(1)_{Y}$ where $Y=T_{B-L}^{I}+Y'=T_{B-L}^{I}+T_{B-L}^{II}+T_{3R}$.
The physical massive scalar spectrum includes a real colour octet,
three SM singlets and a complex scalar transforming as (3,1,2/3).
The heavy gauge boson spectrum includes a vector leptoquark $U_{1}^{\mu}=(3,1,2/3)$,
a colour octet $g'_{\mu}=(8,1,0)$ also identified as coloron, and
$Z'_{\mu}=(1,1,0)$. The heavy gauge bosons arise from the different
steps of the symmetry breaking,
\begin{alignat}{2}
 & SU(4)_{PS}^{I}\rightarrow SU(3)_{c}^{I}\times U(1)_{B-L}^{I} & \quad & \Rightarrow U_{1}^{\mu}(3,1,2/3)\,,\\
 & SU(3)_{c}^{I}\times SU(3)_{c}^{II}\rightarrow SU(3)_{c}^{I+II} & \quad & \Rightarrow g'_{\mu}(8,1,0)\,,\\
 & U(1)_{B-L}^{I}\times U(1)_{Y'}\rightarrow U(1)_{Y} & \quad & \Rightarrow Z'_{\mu}(1,1,0)\,.
\end{alignat}
The gauge boson masses resulting from the symmetry breaking in Eq.~\eqref{eq:hierarchy_scalesVL}
are a generalisation of the results in \cite{Diaz:2017lit,DiLuzio:2017vat},
\begin{equation}
M_{U_{1}}=\frac{1}{\sqrt{2}}g_{4}\sqrt{v_{1}^{2}+v_{3}^{2}}\,,\quad M_{g'}=\sqrt{g_{4}^{2}+g_{3}^{2}}v_{3}\,,\quad M_{Z'}=\frac{\sqrt{3}}{2}\sqrt{g_{4}^{2}+\frac{2}{3}g_{1}^{2}}\sqrt{v_{1}^{2}+\frac{1}{3}v_{3}^{2}}\,,\label{eq:MU1}
\end{equation}
where we have assumed $\overline{v}_{3}\approx v_{3}$ and $\overline{v}_{1}\approx v_{1}$
for simplicity. The mass of the coloron depends only on $v_{3}$,
and the scenario $v_{3}\gg v_{1}$ leads to the approximated relation
$M_{g'}\approx\sqrt{2}M_{U_{1}}$. This way, the coloron can be slightly
heavier than the vector leptoquark, which can help to pass the stringent
bounds from high-$p_{T}$ searches.

In the original gauge basis, the heavy gauge bosons couple to the
EW doublets (including the EW doublets formed by fourth family VL
fermions) via the left-handed interactions
\begin{equation}
\frac{g_{4}}{\sqrt{2}}\left(Q_{4}^{\dagger}\gamma^{\mu}L_{4}+\mathrm{h.c.}\right)U_{1\mu}+\mathrm{h.c.}\,,\label{eq:U1couplings_4th}
\end{equation}
\begin{equation}
\frac{g_{4}g_{s}}{g_{3}}\left(Q_{4}^{\dagger}\gamma^{\mu}T^{a}Q_{4}-\frac{g_{3}^{2}}{g_{4}^{2}}Q_{i}^{\dagger}\gamma^{\mu}T^{a}Q_{i}\right)g'{}_{\mu}^{a}\,,\label{eq:ColoronCouplings_4th}
\end{equation}
\begin{equation}
\frac{\sqrt{3}}{\sqrt{2}}\frac{g_{4}g_{Y}}{g_{1}}\left(\frac{1}{6}Q_{4}^{\dagger}\gamma^{\mu}Q_{4}-\frac{1}{2}L_{4}^{\dagger}\gamma^{\mu}L_{4}-\frac{g_{1}^{2}}{9g_{4}^{2}}Q_{i}^{\dagger}\gamma^{\mu}Q_{i}+\frac{g_{1}^{2}}{3g_{4}^{2}}L_{i}^{\dagger}\gamma^{\mu}L_{i}\right)Z'_{\mu}\,.\label{eq:ZprimeCouplings_4th}
\end{equation}
and also to the EW singlets, although these couplings are suppressed
by small mixing angles connected to the origin of second family fermion
masses. Therefore, they can be safely neglected\footnote{Although flavour universal terms similar to those in Eqs.~\eqref{eq:ColoronCouplings_4th}-\eqref{eq:ZprimeCouplings_4th}
can be relevant for direct production at high-$p_{T}$.}. This way, the $U_{1}$ couplings will be purely left-handed, which
can alleviate the stringent bounds from high-$p_{T}$. Similar couplings
are obtained for the VL partners in the conjugated representations,
however those couplings are irrelevant for the phenomenology since
the conjugated partners do not mix with the SM fermions. The expressions
above can be readily written from CP-conjugated notation to L, R notation
via the formulae of Appendix~\ref{sec:From-CP-conjugated-notation}.

The gauge couplings of $SU(3)_{c}$ and $U(1)_{Y}$ are given by
\begin{equation}
g_{s}=\frac{g_{4}g_{3}}{\sqrt{g_{4}^{2}+g_{3}^{2}}}\,,\qquad g_{Y}=\frac{g_{4}g_{1}}{\sqrt{g_{4}^{2}+\frac{2}{3}g_{1}^{2}}}\,,\label{eq:SM_gauge_couplings}
\end{equation}
where $g_{4,3,2,1}$ are the gauge couplings of $G_{4321}$. The scenario
$g_{4}\gg g_{3,2,1}$ is well motivated from the phenomenological
point of view, since here the flavour-universal couplings of light
fermions to the heavy $Z'$ and $g'$ are suppressed by the ratios
$g_{1}/g_{4}$ and $g_{3}/g_{4}$, which will inhibit the direct production
of these states at the LHC. In this scenario, the relations above
yield the simple expressions $g_{s}\approx g_{3}$ and $g_{Y}\approx g_{1}$
for the SM gauge couplings.

A key feature of the gauge boson couplings in Eqs.~(\ref{eq:U1couplings_4th}-\ref{eq:ZprimeCouplings_4th})
is that, while the coloron $g'_{\mu}$ and the $Z'_{\mu}$ couple
to all chiral and VL quarks and leptons, the vector leptoquark $U_{1}^{\mu}$
only couples to the fourth family VL fermions. However, the couplings
in Eqs.~(\ref{eq:U1couplings_4th}-\ref{eq:ZprimeCouplings_4th})
are written in the original gauge basis. We shall perform the transformation
to the decoupling basis (primed) as per Eq.~\eqref{eq:34mixing},
\begin{equation}
\mathcal{L}_{U_{1}}^{\mathrm{gauge}}=\frac{g_{4}}{\sqrt{2}}Q'{}_{\alpha}^{\dagger}V_{34}^{Q}\gamma_{\mu}\mathrm{diag}\left(0,0,0,1\right)V_{34}^{L\dagger}L'_{\beta}U_{1}^{\mu}+\mathrm{h.c.}\,,
\end{equation}
where $\alpha,\beta=1,..,4$ and the indexes of the matrices are implicit.
We obtain an effective coupling for the third family due to mixing
with the fourth family,
\begin{equation}
\mathcal{L}_{U_{1}}^{\mathrm{gauge}}=\frac{g_{4}}{\sqrt{2}}Q'{}_{i}^{\dagger}\gamma_{\mu}\mathrm{diag}(0,0,s_{34}^{Q}s_{34}^{L})L'_{j}U_{1}^{\mu}+\mathrm{h.c.}\,,\label{eq:U1_couplings_decoupling_4th}
\end{equation}
where we have omitted the fourth column and row for simplicity. The
diagrams in Fig.~\ref{fig:U1_couplings_mass_insertion} are illustrative,
however it must be remembered that the mass insertion approximation
is not accurate here due to the heavy top mass, instead we have to
work in the large mixing angle formalism. In principle, the couplings
in \eqref{eq:U1_couplings_decoupling_4th} can simultaneously contribute
to both LFU ratios $R_{K^{(*)}}$ and $R_{D^{(*)}}$ once the further
2-3 transformations required to diagonalise the quark and lepton mass
matrices are taken into account. Such transformations split the $SU(2)_{L}$
doublets and lead to different couplings for the different chiral
fermions, included in Appendix~\ref{sec:Vector-fermion-interactions}.

In Eq.~\eqref{eq:U1couplings_u_4th} it is shown how the leptoquark
couplings that contribute to LFU ratios arise due to the same mixing
effects which diagonalise the mass matrices of the model, yielding
mass terms for the SM fermions. Therefore, the flavour puzzle and
the $B$-physics anomalies are dynamically and parametrically connected
in this model, leading to a predictive framework.
\begin{figure}[t]
\begin{centering}
\begin{tikzpicture}
	\begin{feynman}
		\vertex (a) {\(Q_{3}\)};
		\vertex [right=13mm of a] (b);
		\vertex [right=11mm of b] (c) [label={ [xshift=0.1cm, yshift=0.1cm] \small $M^{Q}_{4}$}];
		\vertex [right=11mm of c] (d);
		\vertex [right=11mm of d] (e) [label={ [xshift=0.1cm, yshift=0.1cm] \small $M^{L}_{4}$}];
		\vertex [right=11mm of e] (f);
		\vertex [right=11mm of f] (g) {\(L_{3}\)};
		\vertex [above=14mm of b] (f1) {\(\phi_{3}\)};
		\vertex [above=14mm of d] (f2) {\(U_{1}\)};
		\vertex [above=14mm of f] (f3) {\(\phi_{1}\)};
		\diagram* {
			(a) -- [fermion] (b) -- [charged scalar] (f1),
			(b) -- [edge label'=\(\overline{Q}_{4}\)] (c),
			(c) -- [edge label'=\(Q_{4}\), inner sep=6pt, insertion=0] (d) -- [boson, blue] (f2),
			(d) -- [edge label'=\(L_{4}\), inner sep=6pt] (e),
			(e) -- [edge label'=\(\overline{L}_{4}\), insertion=0] (f) -- [charged scalar] (f3),
			(f) -- [fermion] (g),
	};
	\end{feynman}
\end{tikzpicture}
\par\end{centering}
\caption{Diagram in the model which leads to the effective $U_{1}$ couplings
in the mass insertion approximation. \label{fig:U1_couplings_mass_insertion}}
\end{figure}

Following the same methodology, we obtain the coloron and $Z'$ couplings
in the basis of mass eigenstates, which can be found in Appendix \ref{sec:Vector-fermion-interactions}.

The flavour-violating couplings of $U_{1}$ in Eq.~\eqref{eq:U1couplings_u_4th}
are all proportional to mixing between chiral fermions. In principle,
such mixing is of order $V_{cb}$ in the down sector, and of order
$0.1V_{cb}$ in the up sector (see discussion in Section \ref{subsec:Effective-Yukawa-couplings}).
The small mixing in the up sector leads to a small $U_{1}$ 2-3 coupling,
possibly too small for $R_{D^{(*)}}$, however a deeper analysis was
required and we will perform such analysis in the next section. Moreover,
flavour-violating couplings involving the coloron and $Z'$ could
be sizable in the down 2-3 sector, since the CKM is predicted to be
originated from the down sector in this model. We shall study whether
this is compatible or not with the stringent constraints coming from
$B_{s}-\overline{B}_{s}$ meson mixing.

\subsubsection{$R_{K^{(*)}}$ and $R_{D^{(*)}}$\label{subsec:RK_RD}}

New contributions to the $R_{D^{(*)}}$ and $R_{K^{(*)}}$ ratios
arise in our model via tree-level contributions mediated by the $U_{1}$
vector leptoquark, see the formulae in Appendix \ref{subsec:b_c_tau_nu_Appendix}
and \ref{subsec:b_s_ll_Appendix}. After integrating out $U_{1}$,
we obtain the following scaling
\begin{equation}
\left|\Delta R_{D^{(*)}}\right|\propto\left(s_{34}^{L}s_{34}^{Q}\right)^{2}s_{23}^{u}c_{23}^{u}\,,\label{eq:RD_U1}
\end{equation}

\begin{equation}
\left|\Delta R_{K^{(*)}}\right|\propto\left(s_{34}^{L}s_{34}^{Q}\right)^{2}\left(s_{23}^{e}\right)^{2}s_{23}^{d}c_{23}^{d}\,.\label{eq:RK_U1}
\end{equation}
From Eq.~\eqref{eq:RD_U1} it can be seen that our contribution to
$R_{D^{(*)}}$ is proportional to the mixing angle $\theta_{23}^{u}$.
Such angle is naturally small in this model, roughly $\mathcal{O}(0.1V_{cb})$
as per Eq.~\eqref{eq:up_mixing_4thVL}, due to the fact that the
CKM mixing is originated from the down sector. As a consequence, the
contribution to $R_{D^{(*)}}$ is suppressed. On the other hand, the
contribution of $U_{1}$ to $R_{K^{(*)}}$ is further suppressed by
the $\mathcal{O}(V_{cb})$ mixing angles $\theta_{23}^{d}$ and $\theta_{23}^{e}$,
for a total suppression of $\mathcal{O}(V_{cb}^{3})$.

\subsubsection{$B_{s}-\bar{B}_{s}$ mixing \label{subsec:Bs_Mixing}}

Flavour-violating couplings involving the coloron and $Z'$ could
be sizable in the 2-3 down sector, since the CKM is predicted to be
originated from the down sector in this model. The formulae, the treatment
and the bounds obtained from $B_{s}-\bar{B}_{s}$ mixing are derived
in Appendix \ref{subsec:Bs_Mixing_Appendix}.

The bounds are highly constraining over this model because both the
coloron and $Z'$ mediate tree-level contributions to $\Delta M_{s}$,
which interfere positively with the SM prediction, while the latter
are already larger than the experimental result. We estimate that,
in order to satisfy the bound $\Delta M_{s}^{\mathrm{NP}}/\Delta M_{s}^{\mathrm{SM}}<0.11$,
the 2-3 down-quark mixing needs to satisfy $\left|s_{23}^{d}\right|\apprle0.1V_{cb}$
if the 3-4 mixing is maximal $s_{34}^{Q}\approx1$.

\subsubsection{Results in the simplified model}

\begin{figure}[t]
\subfloat[\label{fig:sQ34_sd_plane}]{\includegraphics[scale=0.5]{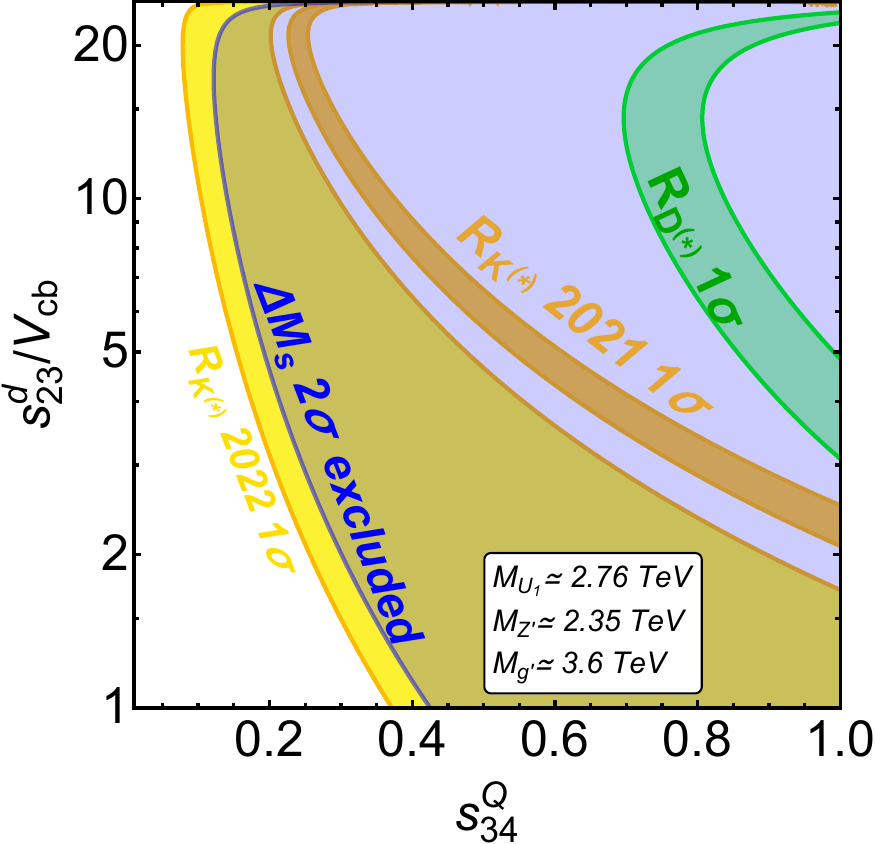}

}$\qquad$\subfloat[\label{fig:v3_v1}]{\includegraphics[scale=0.51]{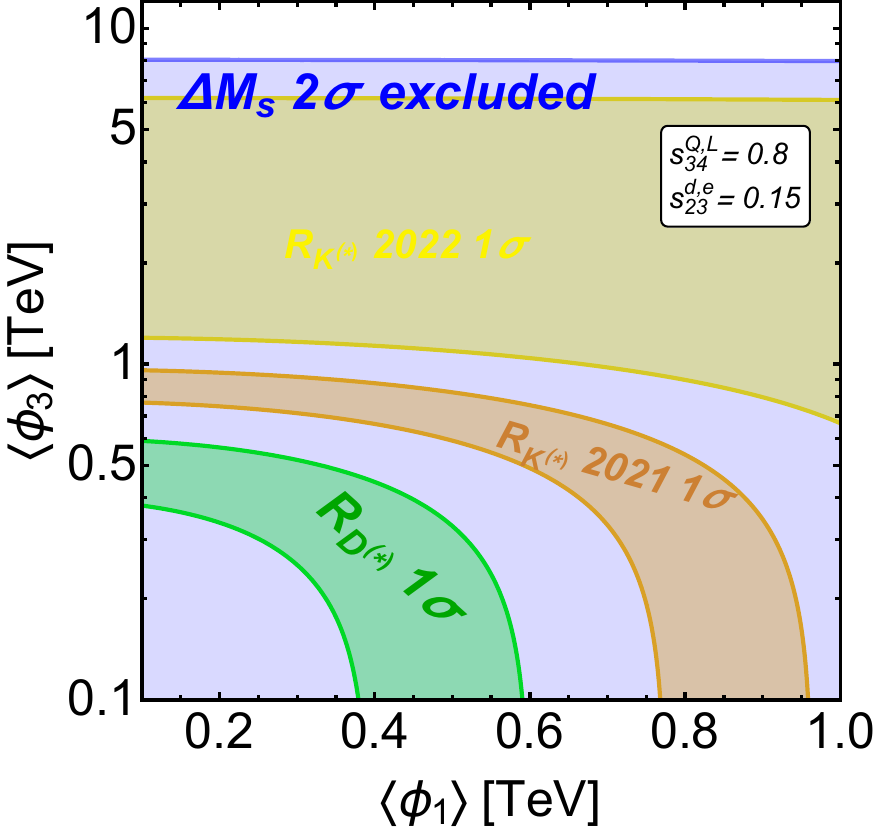}

}

\caption{\textbf{\textit{(Left)}} Regions compatible with $R_{D^{(*)}}$ and
$R_{K^{(*)}}$ (2022 and 2021 data) in the plane ($s_{34}^{Q}$, $s_{23}^{d}$),
the heavy gauge boson masses are fixed as depicted in the panel. \textbf{\textit{(Right)}}
Regions compatible with $R_{D^{(*)}}$ and $R_{K^{(*)}}$ (2022 and
2021 data) in the plane ($\left\langle \phi_{1}\right\rangle $,$\left\langle \phi_{3}\right\rangle $),
which allows to explore the spectrum of heavy gauge boson masses.
The mixing angles are fixed as depicted in the plot. In both panels
the blue regions are excluded by the $\Delta M_{s}$ bound, see Eq.~\eqref{eq:DeltaMs_bound}.}
\end{figure}

As anticipated in the previous sections, the contribution of the leptoquark
to the $R_{D^{(*)}}$ anomaly is strongly suppressed by a naturally
small mixing angle $\theta_{23}^{u}\approx m_{c}/m_{t}$, leading
to a suppression of $\mathcal{O}(0.1V_{cb})$. In Fig.~\ref{fig:sQ34_sd_plane}
it can be seen that for a typical benchmark mass $M_{U_{1}}=3\,\mathrm{TeV}$,
a larger $s_{23}^{u}\apprge4V_{cb}$ is needed in order to address
the $R_{D^{(*)}}$ anomaly, provided that the 3-4 mixing is maximal.

The contribution to $R_{K^{(*)}}$ also suffers from an overall suppression
of $\mathcal{O}(V_{cb}^{3})$. We can go beyond the natural value
of $\theta_{23}^{u}$ by increasing the mixing angle $s_{34}^{q_{c}}$
(i.e. increasing the fundamental Yukawa $x_{34}^{\psi^{c}}$, or reducing
the VL mass $M_{4}^{\psi^{c}}$), which controls the overall size
of the off-diagonal (2,3) entry in the effective mass matrices of
Eqs.~\eqref{eq:MassMatrix_4thVL_effective_up} and \eqref{eq:MassMatrix_4thVL_effective_down}.
This way, we can explore the parameter space of larger 2-3 mixing
angles, provided that the experimental value of $V_{cb}$ is preserved
through Eq.~\eqref{eq:Vcb_model}, which entangles both quark mixings
$\theta_{23}^{u}$ and $\theta_{23}^{d}$. We further assume $s_{34}^{Q}=s_{34}^{L}$
and $s_{23}^{d}=s_{23}^{e}$ to simplify the parameter space. Both
assumptions are well motivated, the former due to universality of
the Yukawa $x_{34}^{\psi}$ and VL mass $M_{4}^{\psi}$, the latter
due to both mixing angles being proportional to similar parameters,
with the mass matrices having the same mass scale. 

Our results are depicted in Fig.~\ref{fig:sQ34_sd_plane} for a spectrum
of heavy gauge boson masses compatible with high-$p_{T}$ searches
(see Section \ref{subsec:Colliders}). We find that for the given
benchmark, a small region of the parameter space is compatible with
the 2022 data of $R_{K^{(*)}}$, however the 1$\sigma$ region of
$R_{D^{(*)}}$ is not compatible with $\Delta M_{s}$. This version
of the model was already unable to explain the 2021 data of both LFU
anomalies due to the large constraints from tree-level $Z'$ and coloron
contributions to $\Delta M_{s}$.

In Fig.~\ref{fig:v3_v1} we have varied the VEVs of $\phi_{3}$ and
$\phi_{1}$, effectively exploring the parameter space of gauge boson
masses in line with Eq.~\eqref{eq:MU1}. However, we find that the
stringent constraints from $\Delta M_{s}$ are only alleviated when
$\left\langle \phi_{3}\right\rangle \apprge8\,\mathrm{TeV}$, which
corresponds to a coloron with mass $M_{g'}\gtrsim50\,\mathrm{TeV}$
and a vector leptoquark with mass $M_{U_{1}}\apprge34\,\mathrm{TeV}$,
too heavy to address $R_{D^{(*)}}$.

We conclude that the model in this simplified version is over-constrained
by large tree-level contributions to $\Delta M_{s}$ mediated by the
coloron and $Z'$. Such FCNCs arise due to the 2-3 CKM mixing having
its origin in the down sector. Moreover, the same small 2-3 mixing
angles suppress the contribution of the model to $R_{D^{(*)}}$. However,
we shall show that the proper flavour structure to be compatible with
all data is achieved in the extended version of the model presented
in Section \ref{sec:Twin-Pati-Salam-Theory_3VL}.

\section{Extending the simplified twin Pati-Salam theory of flavour\label{sec:Twin-Pati-Salam-Theory_3VL}}

In this section we present an extended version of the simplified twin
Pati-Salam model, featuring extra matter content and a discrete flavour
symmetry. This new version can achieve the proper flavour structure
required to be compatible with all data, solving the problems of the
simplified twin Pati-Salam model discussed in Section~\ref{sec:Twin-Pati-Salam-Theory_1VL}.
Firstly, we will introduce the extended version of the model. Secondly,
we will revisit the diagonalisation of the mass matrix, leading to
the fermion masses and to the new couplings with the heavy gauge bosons.
Finally, we will study the phenomenology, showing that the model is
compatible with all data while predicting promising signals in flavour-violating
observables, rare $B$-decays and high-$p_{T}$ searches.

\subsection{New matter content and discrete flavour symmetry}

As identified in \cite{DiLuzio:2018zxy}, when one considers a 4321
model with all chiral fermions transforming as $SU(4)$ singlets (fermiophobic
framework), three vector-like fermion families can achieve the proper
flavour structure to explain the $B$-anomalies. Such flavour structure
can provide a GIM-like suppression of FCNCs, along with large leptoquark
couplings that can contribute to the LFU ratios. Hence, as depicted
in Table~\ref{tab:The-field-content_ExtendedModel}, we extend now
the simplified model by two extra vector-like families, to a total
of three,
\begin{table}[t]
\begin{centering}
\begin{tabular}{lccccccc}
\toprule 
Field & $SU(4)_{PS}^{I}$ & $SU(2)_{L}^{I}$ & $SU(2)_{R}^{I}$ & $SU(4)_{PS}^{II}$ & $SU(2)_{L}^{II}$ & $SU(2)_{R}^{II}$ & $Z_{4}$\tabularnewline
\midrule
\midrule 
$\psi_{1,2,3}$ & $\mathbf{1}$ & $\mathbf{1}$ & $\mathbf{1}$ & $\mathbf{4}$ & $\mathbf{2}$ & $\mathbf{1}$ & $\alpha$, 1, 1\tabularnewline
$\psi_{1,2,3}^{c}$ & $\mathbf{1}$ & $\mathbf{1}$ & $\mathbf{1}$ & $\mathbf{\overline{4}}$ & $\mathbf{1}$ & $\mathbf{\overline{2}}$ & $\alpha$, $\alpha^{2}$, 1\tabularnewline
\midrule 
$\psi_{4,5,6}$ & $\mathbf{4}$ & $\mathbf{2}$ & $\mathbf{1}$ & $\mathbf{1}$ & $\mathbf{1}$ & $\mathbf{1}$ & 1, 1, $\alpha$\tabularnewline
$\overline{\psi}_{4,5,6}$ & $\mathbf{\overline{4}}$ & $\mathbf{\overline{2}}$ & $\mathbf{1}$ & $\mathbf{1}$ & $\mathbf{1}$ & $\mathbf{1}$ & 1, 1, $\alpha^{3}$\tabularnewline
$\psi_{4,5,6}^{c}$ & $\mathbf{\overline{4}}$ & $\mathbf{1}$ & $\mathbf{\overline{2}}$ & $\mathbf{1}$ & $\mathbf{1}$ & $\mathbf{1}$ & 1, 1, $\alpha$\tabularnewline
$\overline{\psi^{c}}_{4,5,6}$ & $\mathbf{4}$ & $\mathbf{1}$ & $\mathbf{2}$ & $\mathbf{1}$ & $\mathbf{1}$ & $\mathbf{1}$ & 1, 1, $\alpha^{3}$\tabularnewline
\midrule 
$\phi$ & $\mathbf{4}$ & $\mathbf{2}$ & $\mathbf{1}$ & $\mathbf{\overline{4}}$ & $\mathbf{\overline{2}}$ & $\mathbf{1}$ & 1\tabularnewline
$\overline{\phi}$,$\,\overline{\phi'}$ & $\mathbf{\overline{4}}$ & $\mathbf{1}$ & $\mathbf{\overline{2}}$ & $\mathbf{4}$ & $\mathbf{1}$ & $\mathbf{2}$ & 1, $\alpha^{2}$\tabularnewline
\midrule
$H$ & $\mathbf{\overline{4}}$ & $\mathbf{\overline{2}}$ & $\mathbf{1}$ & $\mathbf{4}$ & $\mathbf{1}$ & $\mathbf{2}$ & 1\tabularnewline
$\overline{H}$ & $\mathbf{4}$ & $\mathbf{1}$ & $\mathbf{2}$ & $\mathbf{\overline{4}}$ & $\mathbf{\overline{2}}$ & $\mathbf{1}$ & 1\tabularnewline
\midrule
$H'$ & $\mathbf{1}$ & $\mathbf{1}$ & $\mathbf{1}$ & $\mathbf{4}$ & $\mathbf{1}$ & $\mathbf{2}$ & 1\tabularnewline
$\overline{H}'$ & $\mathbf{1}$ & $\mathbf{1}$ & $\mathbf{1}$ & $\mathbf{\overline{4}}$ & $\mathbf{1}$ & $\mathbf{\overline{2}}$ & 1\tabularnewline
$\Phi$ & $\mathbf{1}$ & $\mathbf{2}$ & $\mathbf{1}$ & $\mathbf{1}$ & $\mathbf{\overline{2}}$ & $\mathbf{1}$ & 1\tabularnewline
$\overline{\Phi}$ & $\mathbf{1}$ & $\mathbf{1}$ & $\mathbf{\overline{2}}$ & $\mathbf{1}$ & $\mathbf{1}$ & $\mathbf{2}$ & 1\tabularnewline
\midrule
$\Omega_{15}$ & $\mathbf{15}$ & $\mathbf{1}$ & $\mathbf{1}$ & $\mathbf{1}$ & $\mathbf{1}$ & $\mathbf{1}$ & 1\tabularnewline
\bottomrule
\end{tabular}
\par\end{centering}
\caption{The field content under $G_{422}^{I}\times G_{422}^{II}\times Z_{4}$,
see the main text for details. \label{tab:The-field-content_ExtendedModel}}
\end{table}
\begin{equation}
\begin{array}{c}
\psi_{4,5,6}(4,2,1;1,1,1)_{(1,1,\alpha)}\,,\overline{\psi}_{4,5,6}(\overline{4},\overline{2},1;1,1,1)_{(1,1,\alpha^{3})}\,,\\
\,\\
\psi_{4,5,6}^{c}(\overline{4},1,\overline{2};1,1,1)_{(1,1,\alpha)}\,,\overline{\psi^{c}}_{4,5,6}(4,1,2;1,1,1)_{(1,1,\alpha^{3})}\,,
\end{array}
\end{equation}
where it can be seen that all VL families originate from the first
Pati-Salam group, being singlets under the second. They are indistinguishable
under the twin Pati-Salam symmetry in Eq.~\eqref{eq:TwinPS_symmetry},
however a newly introduced $Z_{4}$ flavour symmetry discriminates
the sixth family from the fourth and fifth, via different powers of
the $Z_{4}$ charge $\alpha=e^{i\pi/2}$. This way, the total symmetry
group of the high energy model is extended to
\begin{equation}
G_{422}^{I}\times G_{422}^{II}\times Z_{4}\,.
\end{equation}
The new $Z_{4}$ discrete symmetry is introduced for phenomenological
purposes, as it will prevent fine-tuning, reduce the total number
of parameters of the model and protect from FCNCs involving the first
family of SM-like chiral fermions. Moreover, $Z_{4}$ will simplify
the diagonalisation of the full mass matrices and preserve the effective
Yukawa couplings for SM fermions given in Section \ref{subsec:Effective-Yukawa-couplings},
with specific modifications. The origin of the chiral fermion families
is still the second Pati-Salam group, however now they transform in
a non-trivial way under $Z_{4}$,
\begin{equation}
\psi_{1,2,3}(1,1,1;4,2,1)_{(\alpha,1,1)}\,,\quad\psi_{1,2,3}^{c}(1,1,1;\overline{4},1,\overline{2})_{(\alpha,\alpha^{2},1)}\,.
\end{equation}
Finally, the scalar content is extended by an additional scalar $\Omega_{15}$
which transforms in the adjoint representation of $SU(4)^{I}$, whose
VEV $\left\langle \Omega_{15}\right\rangle =T_{15}v_{15}$ splits
the vector-like masses, and

\begin{equation}
T_{15}=\frac{1}{2\sqrt{6}}\mathrm{diag}(1,1,1,-3)\,.
\end{equation}
We also include an additional copy of the Yukon $\overline{\phi}$,
denoted as $\overline{\phi'}$, featuring $\alpha^{2}$ charge under
$Z_{4}$. The simplified Lagrangian in Eq.~\eqref{eq:Lren_4thVL-1}
is extended by the new matter content to

\begin{align}
\begin{aligned}\mathcal{L}_{\mathrm{mass}}^{ren} & =y_{ia}^{\psi}\overline{H}\psi_{i}\psi_{a}^{c}+y_{a3}^{\psi}H\psi_{a}\psi_{3}^{c}+x_{ia}^{\psi}\phi\psi_{i}\overline{\psi}_{a}+x_{a2}^{\psi^{c}}\overline{\psi_{a}^{c}\phi'}\psi_{2}^{c}+x_{a3}^{\psi^{c}}\overline{\psi_{a}^{c}\phi}\psi_{3}^{c}+x_{16}^{\psi}\phi\psi_{1}\overline{\psi}_{6}+x_{61}^{\psi^{c}}\overline{\psi_{6}^{c}\phi}\psi_{1}^{c} & {}\\
{} & +M_{ab}^{\psi}\psi_{a}\overline{\psi_{b}}+M_{ab}^{\psi^{c}}\psi_{a}^{c}\overline{\psi_{b}^{c}}+M_{66}^{\psi}\psi_{6}\overline{\psi_{6}}+M_{66}^{\psi^{c}}\psi_{6}^{c}\overline{\psi_{6}^{c}} & {}\\
{} & +\lambda_{15}^{aa}\Omega_{15}\psi_{a}\overline{\psi_{a}}+\lambda_{15}^{66}\Omega_{15}\psi_{6}\overline{\psi_{6}}+\bar{\lambda}_{15}^{aa}\Omega_{15}\psi_{a}^{c}\overline{\psi_{a}^{c}}+\bar{\lambda}_{15}^{66}\Omega_{15}\psi_{6}^{c}\overline{\psi_{6}^{c}}+\mathrm{h.c.}\,, & {}
\end{aligned}
\label{eq: full_lagrangian}
\end{align}
where $i=2,3$ and $a,b=4,5$ (terms $i=1$ and $a,b=6$ forbidden
by $Z_{4}$). The symmetry breaking and the decomposition of the different
fields proceeds just like in the simplified model, see Section~\ref{subsec:High-scale-symmetry},
however the VEVs of the additional scalars $\overline{\phi'}$ and
$\Omega_{15}$ play a role in the spontaneous breaking of the 4321
symmetry, and the corresponding gauge boson masses become (assuming
$v_{1,3}\approx\overline{v}_{1,3}\approx\overline{v'}_{1,3}$ for
simplicity)
\begin{equation}
M_{U_{1}}=\frac{1}{2}g_{4}\sqrt{3v_{1}^{2}+3v_{3}^{2}+\frac{4}{3}v_{15}^{2}}\,,\quad M_{g'}=\frac{\sqrt{3}}{\sqrt{2}}\sqrt{g_{4}^{2}+g_{3}^{2}}v_{3}\,,\quad M_{Z'}=\frac{1}{2}\sqrt{\frac{3}{2}}\sqrt{g_{4}^{2}+\frac{2}{3}g_{1}^{2}}\sqrt{3v_{1}^{2}+v_{3}^{2}}\,.
\end{equation}

\subsection{Effective Yukawa couplings revisited\label{subsec:Effective_Yukawa_3VL}}

In this section, we diagonalise the full mass matrix of the extended
model, following the same procedure as in Section \ref{subsec:Effective-Yukawa-couplings},
but including the extra matter content of the extended model. We may
write the mass terms and couplings in Eq.~\eqref{eq: full_lagrangian}
as a $9\times9$ matrix in flavour space (we also define 9-dimensional
vectors as $\psi_{\alpha}$ and $\psi_{\beta}^{c}$ below),
\begin{equation}
\mathcal{L}_{4,5,6}^{ren}=\psi_{\alpha}^{\mathrm{T}}M^{\psi}\psi_{\beta}^{c}+\mathrm{h.c.}\,,
\end{equation}
\begin{equation}
\psi_{\alpha}=\left(\begin{array}{ccccccccc}
\psi_{1} & \psi_{2} & \psi_{3} & \psi_{4} & \psi_{5} & \psi_{6} & \overline{\psi_{4}^{c}} & \overline{\psi_{5}^{c}} & \overline{\psi_{6}^{c}}\end{array}\right)^{\mathrm{T}}\,,\quad\psi_{\beta}^{c}=\left(\begin{array}{ccccccccc}
\psi_{1}^{c} & \psi_{2}^{c} & \psi_{3}^{c} & \psi_{4}^{c} & \psi_{5}^{c} & \psi_{6}^{c} & \overline{\psi_{4}} & \overline{\psi_{5}} & \overline{\psi_{6}}\end{array}\right)^{\mathrm{T}}\,,
\end{equation}
\begin{equation}
M^{\psi}=\left(
\global\long\def\arraystretch{1.3}%
\begin{array}{@{}llcccccccc@{}}
 & \multicolumn{1}{c@{}}{\psi_{1}^{c}} & \psi_{2}^{c} & \psi_{3}^{c} & \psi_{4}^{c} & \psi_{5}^{c} & \psi_{6}^{c} & \overline{\psi_{4}} & \overline{\psi_{5}} & \overline{\psi_{6}}\\
\cmidrule(l){2-10}\left.\psi_{1}\right| & 0 & 0 & 0 & 0 & 0 & 0 & 0 & 0 & x_{16}^{\psi}\phi\\
\left.\psi_{2}\right| & 0 & 0 & 0 & y_{24}^{\psi}\overline{H} & y_{25}^{\psi}\overline{H} & 0 & 0 & x_{25}^{\psi}\phi & 0\\
\left.\psi_{3}\right| & 0 & 0 & 0 & y_{34}^{\psi}\overline{H} & y_{35}^{\psi}\overline{H} & 0 & x_{34}^{\psi}\phi & x_{35}^{\psi}\phi & 0\\
\left.\psi_{4}\right| & 0 & 0 & y_{43}^{\psi}H & 0 & 0 & 0 & M_{44}^{Q,L} & M_{45}^{\psi} & 0\\
\left.\psi_{5}\right| & 0 & 0 & y_{53}^{\psi}H & 0 & 0 & 0 & M_{54}^{\psi} & M_{55}^{Q,L} & 0\\
\left.\psi_{6}\right| & 0 & 0 & 0 & 0 & 0 & 0 & 0 & 0 & M_{66}^{Q,L}\\
\left.\overline{\psi_{4}^{c}}\right| & 0 & x_{42}^{\psi^{c}}\overline{\phi'} & x_{43}^{\psi^{c}}\overline{\phi} & M_{44}^{\psi^{c}} & M_{45}^{\psi^{c}} & 0 & 0 & 0 & 0\\
\left.\overline{\psi_{5}^{c}}\right| & 0 & x_{52}^{\psi^{c}}\overline{\phi'} & x_{53}^{\psi^{c}}\overline{\phi} & M_{54}^{\psi^{c}} & M_{55}^{\psi^{c}} & 0 & 0 & 0 & 0\\
\left.\overline{\psi_{6}^{c}}\right| & x_{61}^{\psi^{c}}\overline{\phi} & 0 & 0 & 0 & 0 & M_{66}^{\psi^{c}} & 0 & 0 & 0
\end{array}\right)\,,\label{eq:9x9_mass_matrix}
\end{equation}
where the diagonal mass parameters $M_{44,55,66}^{Q,L}$ are splitted
for quarks and leptons due to the VEV of $\Omega_{15}$,

\begin{equation}
M_{aa}^{Q}\equiv M_{aa}^{\psi}+\frac{\lambda_{15}^{aa}\left\langle \Omega_{15}\right\rangle }{2\sqrt{6}}\,,\quad M_{aa}^{L}\equiv M_{aa}^{\psi}-3\frac{\lambda_{15}^{aa}\left\langle \Omega_{15}\right\rangle }{2\sqrt{6}}\,,
\end{equation}
where $a=4,5,6$. Similar equations are obtained for the $\psi^{c}$
sector, however in the $\psi^{c}$ sector the mass splitting is minimal
due to $\left\langle \Omega_{15}\right\rangle $ being of order a
few hundreds GeV while $M_{aa}^{\psi^{c}}$ are much heavier due to
a generalisation of the hierarchy in Eq.~\eqref{eq:hierarchy_scalesVL-1}.
In Eq.~\eqref{eq:9x9_mass_matrix} we have achieved an extra zero
in the (2,7) entry by rotating $\psi_{2}$ and $\psi_{3}$, without
loss of generality thanks to the zeros in the upper $3\times3$ block
(see Section \ref{subsec:Effective-Yukawa-couplings}).

The matrix in Eq.~\eqref{eq:9x9_mass_matrix} features three different
mass scales, the Higgs VEVs $\left\langle H\right\rangle $ and $\langle\overline{H}\rangle$,
the Yukon VEVs $\left\langle \phi\right\rangle $, $\langle\overline{\phi}\rangle$,
$\langle\overline{\phi'}\rangle$ and the VL mass terms $M_{ab}^{\psi}$
and $M_{ab}^{\psi^{c}}$. We can block diagonalise the matrix above
by taking advantage of the different mass scales. Firstly, we diagonalise
the $2\times2$ sub-blocks containing the heavy masses $M_{ab}^{\psi}$
and $M_{ab}^{\psi^{c}}$, 

\begin{equation}
\left(\begin{array}{cc}
M_{4}^{Q} & 0\\
0 & M_{5}^{Q}
\end{array}\right)=V_{45}^{Q}\left(\begin{array}{cc}
M_{44}^{Q} & M_{45}^{\psi}\\
M_{54}^{\psi} & M_{55}^{Q}
\end{array}\right)V_{45}^{\bar{Q}\dagger}\,,\qquad\left(\begin{array}{cc}
M_{4}^{L} & 0\\
0 & M_{5}^{L}
\end{array}\right)=V_{45}^{L}\left(\begin{array}{cc}
M_{44}^{L} & M_{45}^{\psi}\\
M_{54}^{\psi} & M_{55}^{L}
\end{array}\right)V_{45}^{\bar{L}\dagger}\,,\label{eq:VL_splitting_quark}
\end{equation}
and similarly in the $\psi^{c}$ sector. The 4-5 rotations above just
redefine the elements in the 4th, 5th, 7th and 8th rows and columns
of the full mass matrix, leaving the upper $3\times3$ blocks unchanged
(plus we reintroduce the zero in the (2,7) entry by another rotation
of $\psi_{2}$ and $\psi_{3}$). Then we perform a further sequence
of rotations to go to the decoupling basis, where no large elements
appear apart from the diagonal heavy masses (i.e.~those terms in
the seventh, eighth and ninth rows and columns involving the fields
$\phi$ and $\overline{\phi}$ are all absorbed into a redefinition
of the heavy masses), and we obtain a block-diagonal matrix similar
to that of Eq.~\eqref{MassMatrix_4thVL_decoupling} but enlarged
with the fifth and sixth VL families. The total set of unitary transformations
is given by

\begin{equation}
\begin{array}{c}
V_{\psi}=V_{16}^{\psi}V_{35}^{\psi}V_{25}^{\psi}V_{34}^{\psi}V_{45}^{\psi}V_{45}^{\overline{\psi^{c}}}\,,\end{array}\label{eq:mixing_matrix_psi}
\end{equation}

\begin{equation}
V_{\psi^{c}}=V_{16}^{\psi^{c}}V_{35}^{\psi^{c}}V_{25}^{\psi^{c}}V_{34}^{\psi^{c}}V_{24}^{\psi^{c}}V_{45}^{\psi^{c}}V_{45}^{\bar{\psi}}\approx V_{34}^{\psi^{c}}V_{24}^{\psi^{c}}\,.\label{eq:mixing_matrix_psic}
\end{equation}
The mixing angles controlling the unitary transformations in Eq.~\eqref{eq:mixing_matrix_psi}
are given in Appendix~\ref{sec:Mixing-angle-formalism}. The transformations
in the $\psi^{c}$ sector of \eqref{eq:mixing_matrix_psic} can be
described by $V_{34}^{\psi^{c}}V_{24}^{\psi^{c}}$ in good approximation,
whose mixing angles are given by Eqs.~\eqref{eq:sqc24_mixing} and
\eqref{eq:sqc34_mixing}. This approximation is accurate as far as
the mixing involving the 5th and 6th $\psi^{c}$ fields is further
suppressed by a generalisation of the hierarchy in Eq.~\eqref{eq:hierarchy_scalesVL}
to three vector-like families, namely

\begin{equation}
M_{44}^{Q,L}\ll M_{55}^{Q,L}\sim M_{66}^{Q,L}\ll M_{44}^{\psi^{c}}\ll M_{55}^{\psi^{c}},M_{66}^{\psi^{c}}\,.\label{eq:hierarchy_scalesVL-1}
\end{equation}
The hierarchy above will preserve most features of the basic simplified
model, such as large third family Yukawa couplings arising from mixing
with $\psi_{4}$ fermions, and small second family Yukawa couplings
arising from mixing with $\psi_{4}^{c}$. The couplings of $U_{1}$
to chiral fermions will remain dominantly left-handed, since the couplings
to $\psi^{c}$ chiral fermions (or equivalently right-handed fermions)
will remain suppressed by small mixing angles. On the other hand,
the hierarchy $M_{44}^{Q,L}\ll M_{55}^{Q,L}$ will provide hierarchical
couplings of $U_{1}$ to third family and second family fermions,
so we anticipate a small contribution to $R_{K^{(*)}}$ and a large
contribution to $R_{D^{(*)}}$.

We obtain the effective Yukawa couplings for SM fermions by applying
the set of unitary transformations in Eqs.~\eqref{eq:mixing_matrix_psi}
and \eqref{eq:mixing_matrix_psic} to the upper $6\times6$ block
of \eqref{eq:MassMatrix_4thVL}, in the same way as in Eq.~\eqref{eq:Primed_basis}.
In this basis (primed), the mass matrix for each charged sector reads
(assuming a small $x_{35}^{\psi}$, see Section \ref{subsec:BsMixing_revisited}
for the motivation, and approximating cosines in the $\psi^{c}$ sector
to be 1),

\begin{equation}
M_{\mathrm{eff}}^{u}=\left(
\global\long\def\arraystretch{0.7}%
\begin{array}{@{}llcc@{}}
 & \multicolumn{1}{c@{}}{\phantom{\!\,}u'{}_{1}^{c}} & \phantom{\!\,}u'{}_{2}^{c} & \phantom{\!\,}u'{}_{3}^{c}\\
\cmidrule(l){2-4}\left.Q'_{1}\right| & 0 & 0 & 0\\
\left.Q'_{2}\right| & 0 & 0 & s_{25}^{Q}y_{53}^{\psi}\\
\left.Q'_{3}\right| & 0 & 0 & s_{34}^{Q}y_{43}^{\psi}
\end{array}\right)\left\langle H_{t}\right\rangle +\left(
\global\long\def\arraystretch{0.7}%
\begin{array}{@{}llcc@{}}
 & \multicolumn{1}{c@{}}{\phantom{\!\,}u'{}_{1}^{c}} & \phantom{\!\,}u'{}_{2}^{c} & \phantom{\!\,}u'{}_{3}^{c}\\
\cmidrule(l){2-4}\left.Q'_{1}\right| & 0 & 0 & 0\\
\left.Q'_{2}\right| & 0 & c_{25}^{Q}s_{24}^{q^{c}}y_{24}^{\psi} & c_{25}^{Q}s_{34}^{q^{c}}y_{24}^{\psi}\\
\left.Q'_{3}\right| & 0 & c_{34}^{Q}s_{24}^{q^{c}}y_{34}^{\psi} & c_{34}^{Q}s_{34}^{q^{c}}y_{34}^{\psi}
\end{array}\right)\left\langle H_{c}\right\rangle +\mathrm{h.c.}\,,\label{eq:MassMatrix_4thVL_effective_up-1}
\end{equation}
\begin{equation}
M_{\mathrm{eff}}^{d}=\left(
\global\long\def\arraystretch{0.7}%
\begin{array}{@{}llcc@{}}
 & \multicolumn{1}{c@{}}{\phantom{\!\,}d'{}_{1}^{c}} & \phantom{\!\,}d'{}_{2}^{c} & \phantom{\!\,}d'{}_{3}^{c}\\
\cmidrule(l){2-4}\left.Q'_{1}\right| & 0 & 0 & 0\\
\left.Q'_{2}\right| & 0 & 0 & s_{25}^{Q}y_{53}^{\psi}\\
\left.Q'_{3}\right| & 0 & 0 & s_{34}^{Q}y_{43}^{\psi}
\end{array}\right)\left\langle H_{b}\right\rangle +\left(
\global\long\def\arraystretch{0.7}%
\begin{array}{@{}llcc@{}}
 & \multicolumn{1}{c@{}}{\phantom{\!\,}d'{}_{1}^{c}} & \phantom{\!\,}d'{}_{2}^{c} & \phantom{\!\,}d'{}_{3}^{c}\\
\cmidrule(l){2-4}\left.Q'_{1}\right| & 0 & 0 & 0\\
\left.Q'_{2}\right| & 0 & c_{25}^{Q}s_{24}^{q^{c}}y_{24}^{\psi} & c_{25}^{Q}s_{34}^{q^{c}}y_{24}^{\psi}\\
\left.Q'_{3}\right| & 0 & c_{34}^{Q}s_{24}^{q^{c}}y_{34}^{\psi} & c_{34}^{Q}s_{34}^{q^{c}}y_{34}^{\psi}
\end{array}\right)\left\langle H_{s}\right\rangle +\mathrm{h.c.}\,,\label{eq:MassMatrix_4thVL_effective_down-1}
\end{equation}
\begin{equation}
M_{\mathrm{eff}}^{e}=\left(
\global\long\def\arraystretch{0.7}%
\begin{array}{@{}llcc@{}}
 & \multicolumn{1}{c@{}}{\phantom{\!\,}e'{}_{1}^{c}} & \phantom{\!\,}e'{}_{2}^{c} & \phantom{\!\,}e'{}_{3}^{c}\\
\cmidrule(l){2-4}\left.L'_{1}\right| & 0 & 0 & 0\\
\left.L'_{2}\right| & 0 & 0 & s_{25}^{L}y_{53}^{\psi}\\
\left.L'_{3}\right| & 0 & 0 & s_{34}^{L}y_{43}^{\psi}
\end{array}\right)\left\langle H_{\tau}\right\rangle +\left(
\global\long\def\arraystretch{0.7}%
\begin{array}{@{}llcc@{}}
 & \multicolumn{1}{c@{}}{\phantom{\!\,}e'{}_{1}^{c}} & \phantom{\!\,}e'{}_{2}^{c} & \phantom{\!\,}e'{}_{3}^{c}\\
\cmidrule(l){2-4}\left.L'_{1}\right| & 0 & 0 & 0\\
\left.L'_{2}\right| & 0 & c_{25}^{L}s_{24}^{e^{c}}y_{24}^{\psi} & c_{25}^{L}s_{34}^{e^{c}}y_{24}^{\psi}\\
\left.L'_{3}\right| & 0 & c_{34}^{L}s_{24}^{e^{c}}y_{34}^{\psi} & c_{34}^{L}s_{34}^{e^{c}}y_{34}^{\psi}
\end{array}\right)\left\langle H_{\mu}\right\rangle +\mathrm{h.c.}\,,\label{eq:MassMatrix_4thVL_effective_leptons-1}
\end{equation}
which are diagonalised by 2-3 rotations, and the CKM matrix is obtained
via Eq.~\eqref{eq:CKM_matrix}. The mass matrices above are of similar
form to Eqs.~\eqref{eq:MassMatrix_4thVL_effective_up},~\eqref{eq:MassMatrix_4thVL_effective_down},~\eqref{eq:MassMatrix_4thVL_effective_leptons},
just featuring an extra off-diagonal component in the (2,3) entry
of the first matrix in each sector, arising from mixing with the 5th
family. This new term can be used to partially cancel the down 2-3
mixing while simultaneously enhancing up mixing to preserve the CKM,
involving a mild tuning:
\begin{itemize}
\item Let us impose that the total (2,3) entry in the down quark mass matrix
is small, i.e.
\begin{equation}
-s_{25}^{Q}\left|y_{53}^{\psi}\right|\left\langle H_{b}\right\rangle +c_{25}^{Q}s_{34}^{q^{c}}y_{24}^{\psi}\left\langle H_{s}\right\rangle \approx0\,.
\end{equation}
Following the discussion of Section \ref{subsec:Effective-Yukawa-couplings},
a natural benchmark is $\left\langle H_{b}\right\rangle \approx m_{b}$
and $s_{34}^{q^{c}}y_{24}^{\psi}\left\langle H_{s}\right\rangle \approx m_{s}$,
hence
\begin{equation}
-s_{25}^{Q}\left|y_{53}^{\psi}\right|m_{b}+m_{s}\approx0\Longleftrightarrow\left|y_{53}^{\psi}\right|=\frac{m_{s}}{s_{25}^{Q}m_{b}}\,.
\end{equation}
On the other hand, the mixing angle $s_{25}^{Q}$ is very relevant
for the $B$-decays and related phenomenology, and we obtain the typical
value $s_{25}^{Q}\approx0.2$ in Section \ref{subsec:Low-energy-phenomenology},
featuring another connection between the flavour puzzle and $B$-physics
in our model. With this input, we obtain 
\begin{equation}
\left|y_{53}^{\psi}\right|\approx\mathcal{O}(0.1).
\end{equation}
In particular, the benchmark in Table \ref{tab:BP} suppresses the
down mixing with the choice $y_{53}^{\psi}=-0.3$, obtaining $s_{23}^{d}\approx\mathcal{O}(10^{-3})$
which is enough to control the stringent constraints from $B_{s}-\bar{B}_{s}$
meson mixing (see Section \ref{subsec:Bs_Mixing}).
\item At the same time that $y_{53}^{\psi}$ partially cancels the down
mixing, it leads to large up mixing which preserves the CKM. Let us
now estimate the 2-3 mixing in the up sector as the ratio of the (2,3)
entry over the (3,3) entry in the up effective mass matrix,
\begin{equation}
\frac{-s_{25}^{Q}\left|y_{53}^{\psi}\right|\left\langle H_{t}\right\rangle +c_{25}^{Q}s_{34}^{q^{c}}y_{24}^{\psi}\left\langle H_{c}\right\rangle }{s_{34}^{Q}y_{43}^{\psi}\left\langle H_{t}\right\rangle }\approx\frac{s_{25}^{Q}\left|y_{53}^{\psi}\right|\left\langle H_{t}\right\rangle +m_{c}}{m_{t}}\approx s_{25}^{Q}\left|y_{53}^{\psi}\right|\approx\mathcal{O}(V_{cb})\,,
\end{equation}
where we have considered $y_{43}^{\psi}=1$, $s_{34}^{Q}\approx1$,
as required to explain the top mass (see the discussion in the first
bullet point of Section \ref{subsec:Effective-Yukawa-couplings})
and we have neglected the (2,3) term proportional to the smaller energy
scale $\left\langle H_{c}\right\rangle $ when compared with the heavier
$\left\langle H_{t}\right\rangle $. This way, we have taken advantage
of the new contribution via the 5th family (and of the different hierarchies
$m_{c}/m_{t}$ and $m_{s}/m_{b}$) to cancel the dangerous down mixing
while preserving the CKM via up mixing.
\item The situation in the lepton sector is similar due to Pati-Salam universality
of the parameters, i.e.
\begin{equation}
\frac{-s_{25}^{L}\left|y_{53}^{\psi}\right|\left\langle H_{\tau}\right\rangle +c_{25}^{L}s_{34}^{e^{c}}y_{24}^{\psi}\left\langle H_{\mu}\right\rangle }{s_{34}^{L}y_{43}^{\psi}\left\langle H_{\tau}\right\rangle }\approx\frac{-s_{25}^{L}\left|y_{53}^{\psi}\right|\left\langle H_{\tau}\right\rangle +m_{\mu}}{m_{\tau}}\approx s_{25}^{L}\left|y_{53}^{\psi}\right|\approx\mathcal{O}(V_{cb})\,.
\end{equation}
However, the leptonic mixing angles $s_{24}^{e^{c}}$ and $s_{34}^{e^{c}}$
are smaller than the quark ones due to the phenomenological relation
$\left\langle \phi_{3}\right\rangle \gg\left\langle \phi_{1}\right\rangle $.
This leads to $\left\langle H_{\mu}\right\rangle $ being above the
scale of the muon mass, which predicts a quick growth of lepton mixing
in the scenario $s_{34}^{e^{c}}>s_{24}^{e^{c}}$. This can be easily
achieved in realistic benchmarks. In this scenario, interesting signals
arise in LFV processes such as $\tau\rightarrow3\mu$ or $\tau\rightarrow\mu\gamma$,
mediated at tree-level by the $Z'$ boson, see Section \ref{subsec:LFV_processes}.
\end{itemize}
Other than the bullet points above, the mass matrices in Eqs.~\eqref{eq:MassMatrix_4thVL_effective_up-1},~\eqref{eq:MassMatrix_4thVL_effective_down-1},~\eqref{eq:MassMatrix_4thVL_effective_leptons-1}
lead to similar predictions as those of the simplified model in Section
\ref{subsec:Effective-Yukawa-couplings}.

\subsection{Vector-fermion interactions in the extended model\label{subsec:Couplings_3VL}}

In this section we shall compute the vector-fermion couplings involving
the heavy gauge bosons $U_{1}$, $g'$, $Z'$. The complete formulae
can be found in Appendix~\ref{sec:Vector-fermion-interactions}.
We omit the couplings of the vector-like partners in the conjugate
representations $\overline{\psi}_{\alpha}$ and $\overline{\psi_{\alpha}^{c}}$,
since they do not mix with SM fermions. 

\subsubsection{$U_{1}$ couplings}

In the original gauge basis, the vector leptoquark couples to the
heavy EW doublets via the left-handed interactions,

\begin{equation}
\mathcal{L}_{U_{1}}^{\mathrm{gauge}}=\frac{g_{4}}{\sqrt{2}}\left(Q_{4}^{\dagger}\gamma_{\mu}L_{4}+Q_{5}^{\dagger}\gamma_{\mu}L_{5}+Q_{6}^{\dagger}\gamma_{\mu}L_{6}+\mathrm{h.c.}\right)U_{1}^{\mu}\,,
\end{equation}
where similar couplings to the heavy EW singlets $\psi^{c}$ are also
present, however they lead to suppressed couplings to SM fermions
due to the hierarchy in Eq.~\eqref{eq:hierarchy_scalesVL-1}. This
way, we obtain purely left-handed $U_{1}$ couplings in good approximation.
Now we shall apply the unitary transformations in Eq.~\eqref{eq:mixing_matrix_psi}
to rotate the fields from the original gauge basis to the decoupling
basis (primed),

\begin{equation}
\mathcal{L}_{U_{1}}^{\mathrm{gauge}}=\frac{g_{4}}{\sqrt{2}}Q'{}_{\alpha}^{\dagger}\gamma_{\mu}V_{Q}\gamma_{\mu}\mathrm{diag}(0,0,0,1,1,1)V_{L}^{\dagger}L'_{\beta}U_{1}^{\mu}+\mathrm{h.c.}\,,\label{eq:rotation_decoupling_U1_6VL}
\end{equation}
where

\begin{equation}
V_{Q}=V_{16}^{Q}V_{35}^{Q}V_{25}^{Q}V_{34}^{Q}V_{45}^{Q}\,,\qquad V_{L}=V_{16}^{L}V_{35}^{L}V_{25}^{L}V_{34}^{L}V_{45}^{L}\,.\label{eq:mixing_matrix_Q}
\end{equation}
The 4-5 rotations are different for quarks and leptons due to $\left\langle \Omega_{15}\right\rangle $
splitting the mass terms of the VL fermions. They lead to a non-trivial
CKM-like matrix for the $U_{1}$ couplings,

\begin{equation}
W_{LQ}=V_{45}^{Q}V_{45}^{L\dagger}=\left(\begin{array}{ccc}
c_{\theta_{LQ}} & -s_{\theta_{LQ}} & 0\\
s_{\theta_{LQ}} & c_{\theta_{LQ}} & 0\\
0 & 0 & 1
\end{array}\right)\,,
\end{equation}
where $s_{\theta_{LQ}}$ depends on the angles $s_{45}^{Q}$ and $s_{45}^{L}$,
obtained from the diagonalisation in Eq.~\eqref{eq:VL_splitting_quark}.
The unitary matrix $W_{LQ}$ can be regarded as a generalisation of
the CKM matrix to $SU(4)$ or quark-lepton space. Similarly to the
CKM case, the $W_{LQ}$ matrix is the only source of flavour-changing
transitions among $SU(4)^{I}$ states, and it appears only in interactions
mediated by $U_{1}$. In this sense, the vector leptoquark, $U_{1}$,
is analogous to the SM $W^{\pm}$ bosons. Similarly, the $Z'$, $g'$
are analogous to the SM $Z$ boson, and we will show that their interactions
are $SU(4)^{I}$ flavour-conserving at tree-level. In analogy to the
SM, we will denote $U_{1}$ transitions as charged currents and $Z'$,
$g'$ transitions as neutral currents. As in the SM, flavour-changing
neutral currents (FCNCs) proportional to the $W_{LQ}$ matrix are
generated at loop level. This mechanism was firstly identified in
\cite{DiLuzio:2018zxy} for a similar 4321 framework.

The same mixing that leads to the SM fermion masses and mixings, see
Eq.~\eqref{eq:mixing_matrix_Q}, also leads to effective $U_{1}$
couplings to SM fermions which can contribute to the LFU ratios,

\begin{equation}
\mathcal{L}_{U_{1}}^{\mathrm{gauge}}=\frac{g_{4}}{\sqrt{2}}Q'{}_{i}^{\dagger}\gamma_{\mu}\left(\begin{array}{ccc}
s_{16}^{Q}s_{16}^{L}\epsilon & 0 & 0\\
0 & c_{\theta_{LQ}}s_{25}^{Q}s_{25}^{L} & s_{\theta_{LQ}}s_{25}^{Q}s_{34}^{L}\\
0 & -s_{\theta_{LQ}}s_{34}^{Q}s_{25}^{L} & c_{\theta_{LQ}}s_{34}^{Q}s_{34}^{L}
\end{array}\right)L'_{j}U_{1}^{\mu}+\mathrm{h.c.}\,,\label{eq:LQ_couplings}
\end{equation}
where we have considered that $s_{35}^{Q,L}$ are small, see Section\textbf{s}
\ref{subsec:Effective_Yukawa_3VL} and \ref{subsec:BsMixing_revisited}.
The first family coupling can be diluted via mixing with vector-like
fermions, which is parameterised via the effective parameter $\epsilon$
(see Appendix~\ref{sec:Epsilon_Dilution} for more details\textbf{)}.
The couplings above receive small corrections due to 2-3 fermion mixing
arising after diagonalising the effective mass matrices in Eqs.~\eqref{eq:MassMatrix_4thVL_effective_up-1},~\eqref{eq:MassMatrix_4thVL_effective_down-1},~\eqref{eq:MassMatrix_4thVL_effective_leptons-1}.
It can be seen from Eq.~\eqref{eq:LQ_couplings} that a large coupling
(2,3) coupling $\beta_{c\nu_{\tau}}$ arises now, proportional to
the large sines $s_{\theta_{LQ}}$, $s_{34}^{L}$ and $s_{25}^{Q}$.
This solves one important issue of the simplified model, where the
flavour-violating couplings $\beta_{c\nu_{\tau}}$ and $\beta_{b\mu}$
where connected to small 2-3 mixing angles, suppressing the contributions
of $U_{1}$ to the LFU ratios. In any case, the leptoquark couplings
that contribute to $B$-decays arise due to the same mixing effects
which diagonalise the mass matrices of the model, yielding mass terms
for the SM fermions. This way, the flavour puzzle and the $B$-anomalies
are dynamically and parametrically connected in this model, leading
to a predictive framework.

\subsubsection{Coloron couplings and GIM-like mechanism}

In the original gauge basis, the coloron couplings are flavour diagonal,
featuring the following couplings to EW doublets,
\begin{equation}
\mathcal{L}_{g'}^{\mathrm{gauge}}=\frac{g_{4}g_{s}}{g_{3}}\left(Q_{4}^{\dagger}\gamma^{\mu}T^{a}Q_{4}+Q_{5}^{\dagger}\gamma^{\mu}T^{a}Q_{5}+Q_{6}^{\dagger}\gamma^{\mu}T^{a}Q_{6}-\frac{g_{3}^{2}}{g_{4}^{2}}Q_{i}^{\dagger}\gamma^{\mu}T^{a}Q_{i}\right)g'{}_{\mu}^{a}\,,\label{eq:interaction_basis_g'}
\end{equation}
where $i=1,2,3$. Now we rotate to the decoupling basis by applying
the transformations in Eq.~\eqref{eq:mixing_matrix_Q}, (assuming
small $x_{35}^{\psi}$ as discussed in Section \ref{subsec:BsMixing_revisited})
obtaining

{\small{}
\begin{equation}
\mathcal{L}_{g'}^{\mathrm{gauge}}=\frac{g_{4}g_{s}}{g_{3}}Q'{}_{i}^{\dagger}\gamma^{\mu}T^{a}\left(\begin{array}{ccc}
\left(s_{16}^{Q}\right)^{2}-\left(c_{16}^{Q}\right)^{2}\frac{g_{3}^{2}}{g_{4}^{2}} & 0 & 0\\
0 & \left(s_{25}^{Q}\right)^{2}-\left(c_{25}^{Q}\right)^{2}\frac{g_{3}^{2}}{g_{4}^{2}} & 0\\
0 & 0 & \left(s_{34}^{Q}\right)^{2}-\left(c_{34}^{Q}\right)^{2}\frac{g_{3}^{2}}{g_{4}^{2}}
\end{array}\right)Q'_{j}g'{}_{\mu}^{a}\,,\label{eq:Coloron_couplings_3VL}
\end{equation}
}{\small\par}

Here $V_{45}^{Q}$ cancels due to unitarity and due to the $g'$ couplings
between VL quarks being flavour-universal in the original basis of
\eqref{eq:interaction_basis_g'}. Therefore, as anticipated before,
the CKM-like matrix $W_{LQ}$ does not affect the neutral currents
mediated by $g'$ (and similarly by $Z'$). The coloron couplings
in \ref{eq:Coloron_couplings_3VL} receive small corrections due to
2-3 mixing arising after diagonalising the effective mass matrices
in Eqs.~\eqref{eq:MassMatrix_4thVL_effective_up-1},~\eqref{eq:MassMatrix_4thVL_effective_down-1},~\eqref{eq:MassMatrix_4thVL_effective_leptons-1},
predominantly in the up sector, due to the down-aligned flavour structure
achieved in Section \ref{subsec:Effective_Yukawa_3VL}. We obtain
similar couplings for EW singlets, however their mixing angles are
suppressed by the hierarchy in Eq.~\eqref{eq:hierarchy_scalesVL-1},
and so they remain like in the original gauge basis.

The coloron couplings of Eq.~\eqref{eq:Coloron_couplings_3VL} are
flavour-universal if

\begin{equation}
s_{34}^{Q}=s_{25}^{Q}=s_{16}^{Q}\,,\label{eq:GIM-like_condition}
\end{equation}
leading to a GIM-like protection from tree-level FCNCs mediated by
the coloron. The condition above was already identified in \cite{DiLuzio:2018zxy},
denoted as \textit{full alignment limit}. However, we have seen that
maximal $s_{34}^{Q}\approx1$ is well motivated in our model to protect
the perturbativity of the top Yukawa, by the fit of the $R_{D^{(*)}}$
anomaly, and furthermore it naturally suppresses $s_{35}^{Q}$ via
a small $c_{34}^{Q}$. The caveat is that if the condition in Eq.~\eqref{eq:GIM-like_condition}
is implemented, then $s_{16}^{Q}$ and $s_{25}^{Q}$ would also be
maximal, leading to large couplings to valence quarks which would
blow up the production of the coloron at the LHC. This fact was already
identified in \cite{DiLuzio:2018zxy}, where large $s_{34}^{Q}$ was
also suggested by the $B$-anomalies, and a partial alignment limit
was implemented, 

\begin{equation}
s_{25}^{Q}=s_{16}^{Q}\,,\label{eq:GIM-like_condition-1}
\end{equation}
which suppresses FCNCs between the first and second quark families,
proportional to the largest off-diagonal elements of the CKM matrix.
FCNCs between the second and third families still arise, however we
are protected from the stringent constraints of $B_{s}-\bar{B}_{s}$
meson mixing due to the down-aligned flavour structure achieved in
Section \ref{subsec:Effective_Yukawa_3VL}. Finally, FCNCs between
the first and third families are also under control, as they are proportional
to the smaller elements of the CKM matrix.

The GIM-like condition of Eq.~\eqref{eq:GIM-like_condition-1} translates,
in terms of fundamental parameters of our model, into

\begin{equation}
\frac{x_{25}^{\psi}\left\langle \phi_{3}\right\rangle }{\sqrt{\left(x_{25}^{\psi}\left\langle \phi_{3}\right\rangle \right)^{2}+\left(M_{5}^{Q}\right)^{2}}}=\frac{x_{16}^{\psi}\left\langle \phi_{3}\right\rangle }{\sqrt{\left(x_{16}^{\psi}\left\langle \phi_{3}\right\rangle \right)^{2}+\left(M_{6}^{Q}\right)^{2}}}\,,\label{eq:GIM-like_2}
\end{equation}
which could be naively achieved with natural couplings and $M_{5}^{Q}$,
$M_{6}^{Q}$ being of the same order, as allowed by the messenger
dominance in Eq.~(\ref{eq:hierarchy_scalesVL-1}). The couplings
and vector-like mass terms can also be chosen differently, as far
as Eq.~(\ref{eq:GIM-like_2}) is preserved. At the moment, the GIM-like
mechanism is accidental. However, Eq.~(\ref{eq:GIM-like_2}) suggests
that the sixth and fifth family, and also the first and second families,
might transform as doublets under a global $SU(2)$ symmetry, enforcing
the parametric relations of Eq.~(\ref{eq:GIM-like_2}).

A similar treatment of $Z'$ couplings can be found in Appendix~\ref{sec:Vector-fermion-interactions},
and a similar condition is obtained to suppress LFV between the first
and second lepton families,

\begin{equation}
s_{25}^{L}=s_{16}^{L}\,.\label{eq:GIM-like_leptons-1}
\end{equation}
Remarkably, if the condition of Eq.~\eqref{eq:GIM-like_condition-1}
is fulfilled, then Eq.~\eqref{eq:GIM-like_leptons-1} would also
be fulfilled in good approximation thanks to the underlying twin Pati-Salam
symmetry, the small breaking effects given by the splitting of VL
masses via $\left\langle \Omega_{15}\right\rangle $.

\subsection{Low-energy phenomenology\label{subsec:Low-energy-phenomenology}}

The twin PS model features a fermiophobic low-energy 4321 theory with
a rich phenomenology. Although extensive analyses of general 4321
models have been performed during the last few years, the vast majority
of them have been performed in the framework of non-fermiophobic 4321
models \cite{Cornella:2019hct,Cornella:2021sby,Barbieri:2022ikw,Bordone:2017bld,Bordone:2018nbg}.
Instead, the twin PS model offers a fermiophobic scenario with a different
phenomenology. Being a theory of flavour, extra constraints and correlations
arise via the generation of the SM Yukawa couplings and the prediction
of fermion masses and mixing, including striking signals in LFV processes.
Moreover, the underlying twin Pati-Salam symmetry introduces universality
(and perturbativity) constraints over several parameters, which are
not present in other models. These features motivate a dedicated analysis.
We will highlight key observables for which the intrinsic nature of
the model can be disentangled from all alternative proposals. All
low-energy observables considered are listed in Table~\ref{tab:Observables},
with references to current experimental bounds and links to theory
expressions.
\begin{table}[t]
\noindent \begin{centering}
\begin{tabular}{ccc}
\toprule 
\multicolumn{1}{c}{Observable} & Experiment/constraint & Theory expr.\tabularnewline
\midrule
\midrule 
$\left[C_{\nu edu}^{*}\right]^{3332}$ ($R_{D^{(*)}}$) & $0.07\pm0.02$\cite{Angelescu:2021lln} & \eqref{eq:R_D_WilsonCoefficient}\tabularnewline
\midrule 
$C_{9}^{\mu\mu}=-C_{10}^{\mu\mu}$ ($R_{K^{(*)}}^{2021}$) & $[-0.31,-0.48]$ (68\% CL)\cite{Geng:2021nhg} & \eqref{eq:R_K_WilsonCoefficients}\tabularnewline
\midrule 
$C_{9}^{\mu\mu}=-C_{10}^{\mu\mu}$ ($R_{K^{(*)}}^{2022}$) & $[-0.01,-0.14]$ (68\% CL)\eqref{eq:C9_-C10} & \eqref{eq:R_K_WilsonCoefficients}\tabularnewline
\midrule 
$\delta(\Delta M_{s})$ ($B_{s}-\bar{B}_{s}$ mixing) & $\apprle0.11$ (95\% CL) \cite{DiLuzio:2019jyq} & \eqref{eq:delta_DeltaMs}\tabularnewline
\midrule 
$\mathcal{B}\left(\tau\rightarrow3\mu\right)$ & $<2.1\cdot10^{-8}$ (90\% CL)\cite{Hayasaka:2010np} & \eqref{eq:tau_3mu}\tabularnewline
\midrule 
$\mathcal{B}\left(\tau\rightarrow\mu\gamma\right)$ & $<5.0\cdot10^{-8}$ (90\% CL)\cite{HFLAV:2019otj} & \eqref{eq:tau_mu_photon}\tabularnewline
\midrule 
$\mathcal{B}\left(B_{s}\rightarrow\tau^{\pm}\mu^{\mp}\right)$ & $<3.4\cdot10^{-5}$ (90\% CL)\cite{LHCb:2019ujz} & \eqref{eq:Bs_tau_mu}\tabularnewline
\midrule 
$\mathcal{B}\left(B^{+}\rightarrow K^{+}\tau^{\pm}\mu^{\mp}\right)$ & $<2.8\cdot10^{-5}$ (90\% CL)\cite{BaBar:2012azg} & \eqref{eq:B_Ktaumu}\tabularnewline
\midrule 
$\mathcal{B}\left(\tau\rightarrow\mu\phi\right)$ & $<8.4\cdot10^{-8}$ (90\% CL)\cite{Belle:2011ogy} & \eqref{eq:tau_muphi}\tabularnewline
\midrule 
$\mathcal{B}\left(K_{L}\rightarrow\mu e\right)$ & $<4.7\cdot10^{-12}$ (90\% CL) \cite{PDG:2022ynf} & \eqref{eq:KL_mue}\tabularnewline
\midrule 
$(g_{\tau}/g_{e,\mu})_{\ell+\pi+K}$ & $1.0003\pm0.0014$ (68\% CL)\cite{HFLAV:2022pwe} & \eqref{eq:LFUratios_tau}\tabularnewline
\midrule 
$\mathcal{B}\left(B_{s}\rightarrow\tau^{+}\tau^{-}\right)$ & $<5.2\times10^{-3}$ (90\% CL)\cite{LHCb:2017myy} & \eqref{eq:Bs_tautau}\tabularnewline
\midrule 
$\mathcal{B}\left(B\rightarrow K\tau^{+}\tau^{-}\right)$ & $<2.25\times10^{-3}$ (90\% CL)\cite{BaBar:2016wgb} & \eqref{eq:B_Ktautau}\tabularnewline
\midrule 
$\mathcal{B}\left(B\rightarrow K^{(*)}\nu\bar{\nu}\right)/\mathcal{B}\left(B\rightarrow K^{(*)}\nu\bar{\nu}\right)_{\mathrm{SM}}$ & $<3.6\,(2.7)$ (90\% CL)\cite{BaBar:2013npw,Belle:2017oht} & \eqref{eq:BtoK_nunu}\tabularnewline
\bottomrule
\end{tabular}
\par\end{centering}
\caption{Set of observables explored in the phenomenological analysis, including
current experimental constraints.\label{tab:Observables}}
\end{table}

The benchmark points BP1 and BP2 in Table~\ref{tab:BP} address the
$R_{D^{(*)}}$ anomalies and are compatible with the 2021 and 2022
data on $R_{K^{(*)}}$, respectively, plus all the considered low-energy
observables and high-$p_{T}$ searches. They provide a good starting
point to study the relevant phenomenology, featuring typical configurations
of the model, and allow us to confront the 2021 picture of the model
versus the new situation with LFU preserved in $\mu/e$ ratios. Moreover,
they fit second and third family charged fermion masses and mixings,
featuring a down-aligned flavour structure with $\mathcal{O}(0.1)$
$\mu-\tau$ lepton mixing. The latter is more benchmark dependent,
with the common range being $s_{23}^{e}=[V_{cb},5V_{cb}]$. The case
$s_{23}^{e}\approx0.1$ is interesting because it leads to intriguing
signals in LFV processes, as we shall see. BP1 and BP2 also feature
$x_{25}^{\psi}\approx x_{16}^{\psi}$ and $M_{5}^{Q,L}\approx M_{6}^{Q,L}$,
providing a GIM-like suppression of 1-2 FCNCs.

In the forthcoming sections we will assume the couplings of the fundamental
Lagrangian to be universal, such as $x_{34}^{\psi}$ and $x_{25}^{\psi}$,
however their universality is broken by small RGE effects which we
estimate in Section~\ref{subsec:Pertubativity} to be below 8\%.
We neglect the small RGE effects and preserve universal parameters
for the phenomenological analysis, in order to simplify the exploration
of the parameter space and highlight the underlying twin Pati-Salam
symmetry.

\subsubsection{Model independent analysis of 2022 clean $b\rightarrow s\mu\mu$
data}

This model was originally built to address and relate the 2021 $R_{K^{(*)}}$
and $R_{D^{(*)}}$ LFU anomalies, while connecting their origin to
the origin of Yukawa couplings in the SM. This picture changed completely
after the 2022 LHCb update of the $R_{K^{(*)}}$ ratios \cite{LHCb:2022qnv},
which are now broadly compatible with the SM predictions (see Eq.~\eqref{eq:RK}).
New 2022 data of $\mathcal{B}(B_{s}\rightarrow\mu^{+}\mu^{-})$ by
CMS \cite{Kar:2022tor} is also compatible with the SM, while the
previous measurements were hinting for values smaller than the SM
prediction, including the 2021 measurement by LHCb \cite{LHCb:2021vsc}.
In our analysis, we will consider the global average of $\mathcal{B}(B_{s}\rightarrow\mu^{+}\mu^{-})$
experimental data by Allanach and Davighi \cite{Allanach:2022iod},
\begin{figure}[t]
\begin{centering}
\includegraphics[scale=0.5]{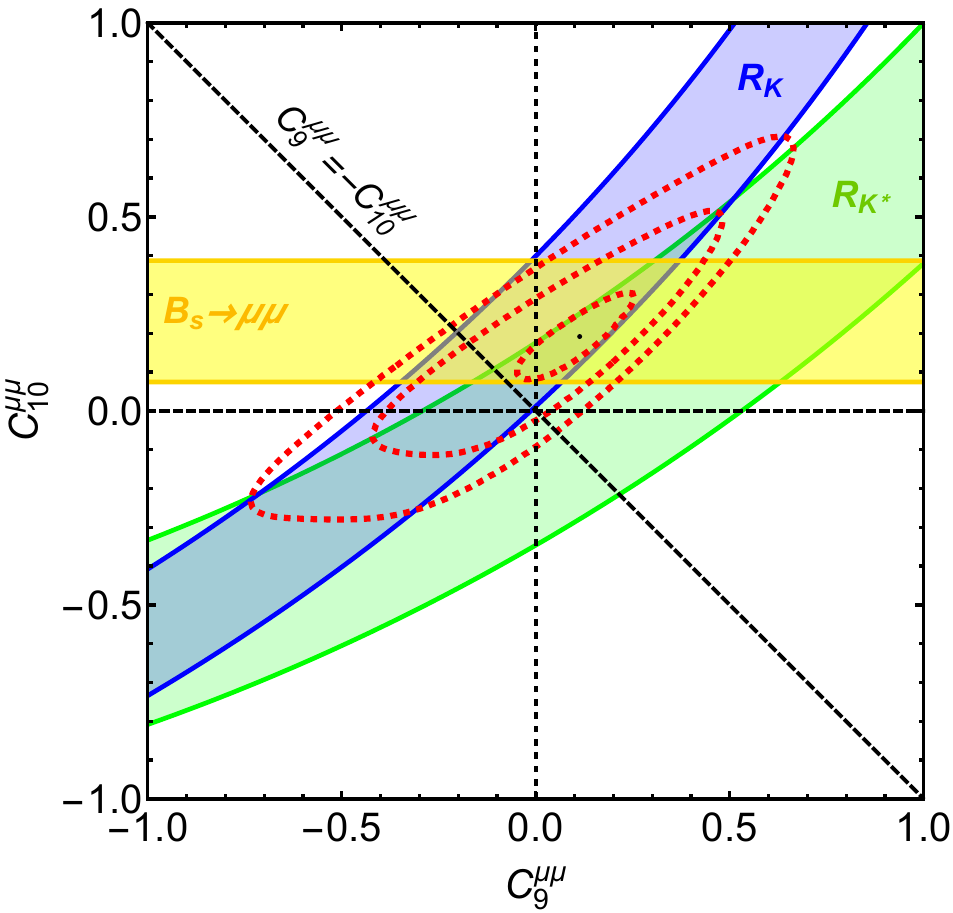}
\par\end{centering}
\caption{Allowed regions in the plane $C_{9}^{\mu\mu}$ vs $C_{10}^{\mu\mu}$
to $1\sigma$ accuracy derived by using 2022 data on $R_{K}$ (blue
region), $R_{K^{*}}$ (green region) and $\mathcal{B}(B_{s}\rightarrow\mu^{+}\mu^{-})$
(yellow region). The red contours denote the $1\sigma$, $2\sigma$
and $3\sigma$ regions of the $\chi^{2}$ fit. The black dot denotes
the best fit point with $\Delta\chi^{2}/\mathrm{dof}\approx0.56$.\label{fig:C9vsC10}}
\end{figure}

\begin{equation}
\mathcal{B}(B_{s}\rightarrow\mu^{+}\mu^{-})=(3.28\pm0.26)\times10^{-9}\,,
\end{equation}
which is roughly $1\sigma$ below the SM prediction $\mathcal{B}(B_{s}\rightarrow\mu^{+}\mu^{-})_{\mathrm{SM}}=(3.67\pm0.15)\times10^{-9}$.
Instead, other observables such as the $B\rightarrow K\mu\mu$ \cite{Gubernari:2022hxn}
branching fraction and the angular observable $P'_{5}$ \cite{LHCb:2020lmf,LHCb:2020gog}
show important tensions with the SM, however such anomalies rely on
important assumptions about the hadronic uncertainties, and their
study is beyond the scope of this manuscript. Nonetheless, an extended
discussion can be found in Section~\ref{subsec:Off-shell-photon-penguin}.

Following common practice, we describe $b\rightarrow s\mu\mu$ transitions
in terms of the low-energy Lagrangian containing the usual $\mathcal{O}_{9}^{23\mu\mu}$
and $\mathcal{O}_{10}^{23\mu\mu}$ operators, defined in Eq.~\eqref{eq:O9_O10_operators}.
More details and all the formulae are included in Appendix~\ref{subsec:b_s_ll_Appendix}.
For the sake of clarity, we further simplify the notation by removing
quark indexes and denote the corresponding NP Wilson coefficients
(WCs) as $C_{9}^{\mu\mu}$ and $C_{10}^{\mu\mu}$. To the best of
our knowledge, no explicit data for the theoretically clean fit of
the WCs considering the new SM-like $R_{K^{(*)}}$ is given in the
literature, motivating our own model independent analysis. The fits
presented in the recent analyses \cite{Greljo:2022jac,Ciuchini:2022wbq}
include observables which are not theoretically clean, and assumptions
about the hadronic uncertainties need to be made. Instead, we need
to explore how the theoretically clean, SM-like observables constrain
the LFUV $C_{9}^{\mu\mu}=-C_{10}^{\mu\mu}$ which unavoidably arises
in our model.

In Fig.~\ref{fig:C9vsC10} we show the parameter space in the plane
($C_{9}^{\mu\mu}$, $C_{10}^{\mu\mu}$) preferred by the 2022 $R_{K^{(*)}}$
ratios (in the central $q^{2}$) and the average of $\mathcal{B}(B_{s}\rightarrow\mu^{+}\mu^{-})$.
We also display the result of a combined $\chi^{2}$ fit to all three
observables as the red ellipses, denoting 1$\sigma$, $2\sigma$ and
$3\sigma$ intervals. Our results show that a small but non-zero value
of $C_{10}^{\mu\mu}$ is still preferred by $\mathcal{B}(B_{s}\rightarrow\mu^{+}\mu^{-})$.
On the other hand, $C_{9}^{\mu\mu}$ is compatible with zero, but
small positive and negative values are still allowed by the new $R_{K^{(*)}}$
ratios at $1\sigma$.

In particular, left-handed NP $C_{9}^{\mu\mu}=-C_{10}^{\mu\mu}$ are
not far away from the $1\sigma$ region, and our 1-dimensional fit
for the latter is

\begin{equation}
C_{9}^{\mu\mu}=-C_{10}^{\mu\mu}=[-0.0111,-0.1425]\:(1\ensuremath{\sigma})\,,\label{eq:C9_-C10}
\end{equation}
with a best fit value of $C_{9}^{\mu\mu}=-C_{10}^{\mu\mu}=-0.0725$
with $\Delta\chi^{2}/\mathrm{dof}\approx0.58$. Although left-handed
NP are still allowed by the new data, the WCs are much smaller than
those preferred by 2021 data (see Table \ref{tab:Observables}). 

In our model, the left-handed WC in Eq.~\eqref{eq:C9_-C10} is obtained
after integrating out the heavy gauge bosons, with the overall contribution
being dominated by $U_{1}$ tree-level exchange. We shall constrain
such contribution to the $\text{1}\sigma$ region in Eq.~\eqref{eq:C9_-C10},
and confront the new results against with the previous picture of
2021 data for which the model was developed.

\subsubsection{$R_{D^{(*)}}$ and $R_{K^{(*)}}$ }

Beyond the contribution to $C_{9}^{\mu\mu}=-C_{10}^{\mu\mu}$ (see
the EFT of the model in Appendix~\ref{subsec:b_s_ll_Appendix}),
our model also generates a contribution to the WC $\left[C_{\nu edu}^{*}\right]^{3332}$
as defined in Appendix \ref{subsec:b_c_tau_nu_Appendix}. The latter
contribution can accommodate existing tensions between the $R_{D^{(*)}}$
ratios and the SM. Namely, in terms of fundamental parameters of the
model, the deviations from the SM of the LFU ratios scale as follows,

\begin{equation}
\left|\Delta R_{D^{(*)}}\right|\propto\left(x_{34}^{\psi}\right)^{3}x_{25}^{\psi}\,,\label{eq:RD_U1-1}
\end{equation}

\begin{equation}
\left|\Delta R_{K^{(*)}}\right|\propto x_{34}^{\psi}\left(x_{25}^{\psi}\right)^{3}\,,\label{eq:RK_U1-1}
\end{equation}
where we have fixed the VL masses and the 4321-breaking VEVs to the
values of our benchmark (Table \ref{tab:BP}). This way, the Yukawa
couplings above control the contributions to most of the relevant
phenomenology, including the LFU ratios. The Pati-Salam universality
of $x_{34}^{\psi}$ and $x_{25}^{\psi}$ provides here a welcome constraint,
not present in other 4321 models. In particular, one can see that
both $R_{D^{(*)}}$ and $R_{K^{(*)}}$ are connected via the same
parameters and deviations in both are expected, while in other 4321
models the equivalent of $x_{i\alpha}^{\psi}$ decompose in different
parameters for quarks and leptons, which decouple $R_{K^{(*)}}$ from
$R_{D^{(*)}}$.

Following from Eqs.~\eqref{eq:RD_U1-1} and \eqref{eq:RK_U1-1},
the cubic dependence of $R_{K^{(*)}}$ on $x_{25}^{\psi}$ anticipates
that we can suppress the contribution to $R_{K^{(*)}}$, while preserving
a large contribution to $R_{D^{(*)}}$ thanks to its linear dependence
on $x_{25}^{\psi}$. As a consequence, the yellow band of parameter
space preferred by 2022 $R_{K^{(*)}}$ is just shifted below the orange
band of 2021 $R_{K^{(*)}}$ in Fig.~\ref{fig:BsMixing_parameter_space}.
The 2022 $R_{K^{(*)}}$ band is compatible with $R_{D^{(*)}}$ at
$1\sigma$ only in a narrow region of the parameter space. This is
encouraging, given the fact that the model was built to address the
2021 tensions in both LFU ratios. However, in order to explain $R_{D^{(*)}}$,
small deviations from the SM in the $R_{K^{(*)}}$ ratios are unavoidable,
to be tested in the future via more precise measurements of LFU by
the LHCb collaboration. Moreover, lower central values for $R_{D^{(*)}}$
are also expected.

Remarkably, the fact that the twin PS model only generates the effective
operator $(\bar{c}_{L}\gamma_{\mu}b_{L})$ $(\bar{\text{\ensuremath{\tau}}}_{L}\gamma^{\mu}\nu_{\tau L})$
implies that both $R_{D}$ and $R_{D^{*}}$ are corrected in the same
direction and with the same size. Instead, non-fermiophobic 4321 models
also predict the scalar operator $(\bar{c}_{L}b_{R})(\bar{\text{\ensuremath{\tau}}}_{R}\nu_{\tau L})$,
which leads to a larger correction of $R_{D}$ than that of $R_{D^{*}}$
(about 5/2 larger for the $\mathrm{PS}^{3}$ model, see Eq.~(27)
in \cite{Bordone:2017bld}). 

\subsubsection{Off-shell photon penguin with tau leptons\label{subsec:Off-shell-photon-penguin}}

The explanation of $R_{D^{(*)}}$ in our model is correlated to new
contributions to $b\rightarrow s\tau\tau$ due to $SU(2)_{L}$ invariance
of the $U_{1}$ couplings, to be explored in detail in Section \ref{subsec:Signals-in-rare-processes}.
Interestingly, the same couplings that contribute to $b\rightarrow s\tau\tau$
also lead to an off-shell photon penguin diagram with tau leptons
running in the loop, which generates a lepton universal contribution
to the operator $\mathcal{O}_{9}^{23\ell\ell}=(\bar{s}_{L}\gamma_{\mu}b_{L})(\bar{\ell}\gamma^{\mu}\ell)$
entering in $b\rightarrow s\ell\ell$ transitions, namely
\begin{figure}[t]
\subfloat[\label{fig:Off-shell penguin}]{\noindent \begin{centering}
\begin{tikzpicture}
	\begin{feynman}
		\vertex (a) {\(b_{L}\)};
		\vertex [right=20mm of a] (b);
		\vertex [below right=16mm of b] (c);
		\vertex [below=16mm of c] (f);
		\vertex [below left=16mm of f] (f1) {\(\ell^{+}\)};
		\vertex [below right=16mm of f] (f2) {\(\ell^{-}\)};
		\vertex [above right=16mm of c] (d);
		\vertex [right=16mm of d] (e) {\(s_{L}\)};
		\diagram* {
			(a) -- [fermion] (b) -- [boson, half left, blue, edge label'=\(U_{1}\)] (d) -- [fermion] (e),
			(b) -- [fermion, quarter right, edge label'={\(\tau_{L},\,E_{L4},\,E_{L5}\)}, inner sep=2pt] (c) -- [boson, blue, edge label'=\(\gamma\)] (f),
			(f1) -- [fermion] (f) -- [fermion] (f2),
			(c) -- [fermion, quarter right, edge label'={\(\tau_{L},\,E_{L4},\,E_{L5}\)}, inner sep=2pt] (d),
	};
	\end{feynman}
\end{tikzpicture}
\par\end{centering}
}$\qquad$\subfloat[\label{fig:C9U}]{\includegraphics[scale=0.36]{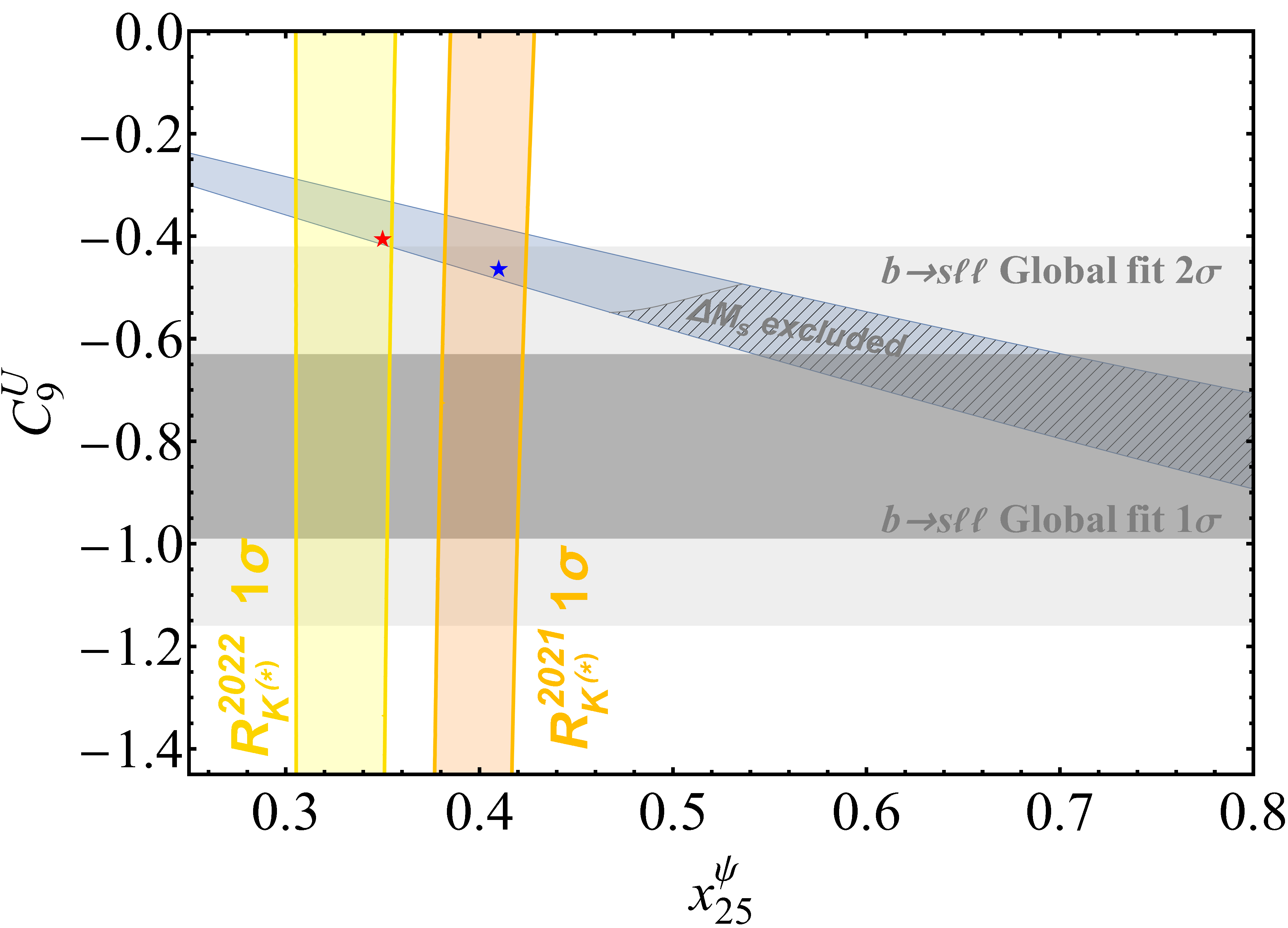}

}

\caption{\textbf{\textit{(Left)}}\textit{ }Off-shell photon penguin with tau
leptons in the loop that generates $C_{9}^{U}$, a contribution to
the lepton universal operator $\mathcal{O}_{9}^{23\ell\ell}$ that
participates in $b\rightarrow s\ell\ell$ transitions. \textbf{\textit{(Right)}}
$C_{9}^{U}$ as a function of $x_{25}^{\psi}$ via Eq.~\eqref{eq:C9U},
with $x_{34}^{\psi}$ varied in the range {[}1, 3.5{]} as preferred
by $R_{D^{(*)}}$ (blue region), with the rest of parameters fixed
as in Table~\ref{tab:BP}. The gray (light gray) region denotes the
1$\sigma$ ($2\sigma$) contour of $C_{9}^{U}$ as preferred by a
global fit to $b\rightarrow s\ell\ell$ data taken from \cite{Alguero:2021anc}.
The yellow (orange) band denotes the $1\sigma$ region preferred by
$R_{K^{(*)}}^{2022}$ ($R_{K^{(*)}}^{2021}$). The blue and red stars
denote BP1 and BP2 respectively.}
\end{figure}

\begin{equation}
C_{9}^{U}=-\frac{v_{\mathrm{SM}}^{2}g_{4}^{2}}{6V_{tb}V_{ts}^{*}M_{U_{\text{1}}}^{2}}\left(\log\left[\frac{2m_{b}^{2}}{g_{4}^{2}M_{U_{1}}^{2}}\right]\beta_{s\tau}\beta_{b\tau}^{*}+\log\left[\frac{2m_{E_{5}}^{2}}{g_{4}^{2}M_{U_{1}}^{2}}\right]\beta_{sE_{5}}\beta_{bE_{5}}^{*}+\log\left[\frac{2m_{E_{4}}^{2}}{g_{4}^{2}M_{U_{1}}^{2}}\right]\beta_{sE_{4}}\beta_{bE_{4}}^{*}\right)\,,\label{eq:C9U}
\end{equation}
which is explicitly correlated to $b\rightarrow s\tau\tau$, as well
as to $R_{D^{(*)}}$ since $SU(2)_{L}$ invariance implies $\beta_{s\tau}\approx\beta_{c\nu_{\tau}}$
for the $U_{1}$ couplings. Therefore, the scaling is $|C_{9}^{U}|\propto(x_{34}^{\psi})^{3}x_{25}^{\psi}$,
just like $R_{D^{(*)}}$. A similar contribution has been studied
in the literature in a model independent framework \cite{Bobeth:2014rda,Capdevila:2017iqn,Crivellin:2018yvo,Alguero:2022wkd},
however in our model we need to add the contributions via the extra
VL charged leptons $E_{4,5}$, see Fig.~\ref{fig:Off-shell penguin}.
Unfortunately, due to the flavour structure of our model, the contributions
via VL leptons interfere negatively with the leading contribution
via the tau loop, and hence our overall contribution to $C_{9}^{U}$
is smaller than in other models. The contribution from $E_{4}$ is
negligible, but the contribution from $E_{5}$ reduces $C_{9}^{U}$
by a 20\% factor of the tau loop contribution.

In our model, $R_{D^{(*)}}$ and $R_{K^{(*)}}$ are correlated, as
can be seen from Eqs.~\eqref{eq:RD_U1-1} and \eqref{eq:RK_U1-1}.
Therefore, $C_{9}^{U}$ is not only correlated with $R_{D^{(*)}}$
but also with $R_{K^{(*)}}$. Given that deviations from 1 in $R_{K^{(*)}}$
are now constrained by the new LHCb measurements, our final contribution
to $C_{9}^{U}$ is constrained to be $C_{9}^{U}\approx-0.4$, as can
be seen in Fig.~\ref{fig:C9U}. However, global fits of $b\rightarrow s\ell\ell$
data (see e.g.~\cite{Alguero:2021anc,Greljo:2022jac,Ciuchini:2022wbq}),
mostly driven by anomalies in $\mathrm{Br}(B\rightarrow K\mu\mu)$,
$\mathrm{Br}(B_{s}\rightarrow\phi\mu\mu)$ and $P'_{5}(B\rightarrow K^{*}\mu\mu)$
(see e.g.~\cite{Gubernari:2022hxn}), prefer a larger value $C_{9}^{U}\approx-0.8$.
Therefore, we conclude that our model is not able to fully address
the anomalies in $b\rightarrow s\ell\ell$ via the off-shell photon
penguin, although our contribution to $C_{9}^{U}$ ameliorates the
tensions. Performing a more ambitious analysis would require to make
assumptions about the hadronic uncertainties afflicting $\mathrm{Br}(B\rightarrow K\mu\mu)$,
$\mathrm{Br}(B_{s}\rightarrow\phi\mu\mu)$ and $P'_{5}(B\rightarrow K^{*}\mu\mu)$,
which is beyond the scope of this paper.

\subsubsection{$B_{s}-\bar{B}_{s}$ mixing \label{subsec:BsMixing_revisited}}

In the twin PS model as presented in Section~\ref{sec:Twin-Pati-Salam-Theory_3VL},
tree-level contributions to $B_{s}-\bar{B}_{s}$ mixing via 2-3 quark
mixing are suppressed due to the down-aligned flavour structure achieved
in Section~\ref{subsec:Effective_Yukawa_3VL}. A further 1-loop contribution
mediated by $U_{1}$ has been studied in the literature \cite{DiLuzio:2018zxy,Cornella:2021sby,Fuentes-Martin:2020hvc}
for other 4321 models, and vector-like charged leptons are known to
play a crucial role. In \cite{DiLuzio:2018zxy} a framework with three
VL charged leptons was considered, however the loop function was generalised
from the SM $W$ box, so the bounds where expected to be slightly
overestimated. Instead, in \cite{Fuentes-Martin:2020hvc} the proper
loop function was derived, but a framework with only one VL charged
lepton was considered. For this work, we have generalised the loop
function of \cite{Fuentes-Martin:2020hvc} to the case of three VL
leptons. The 1-loop contribution mediated by $U_{1}$ reads, 

\begin{equation}
C_{bs}^{\mathrm{NP-loop}}=\frac{g_{4}^{4}}{\left(8\pi M_{U_{1}}\right)^{2}}\sum_{\alpha,\beta}\left(\beta_{s\alpha}^{*}\beta_{b\alpha}\right)\left(\beta_{s\beta}^{*}\beta_{b\beta}\right)F(x_{\alpha},x_{\beta})\,,\label{eq:Cbs_U1}
\end{equation}
where $\alpha,\beta=\mu,\tau,E_{4},E_{5}$ run for all charged leptons,
including the vector-like partners (except for electrons and the sixth
charged lepton which do not couple to the second or third generation),
and $x_{\alpha}=(m_{\alpha}/M_{U_{1}})^{2}$. The contribution corresponds
to the box diagrams in Fig.~\ref{fig:Box_BsMixing}. The proper loop
function for our framework is given in Appendix~\ref{subsec:Loop_functions_BsMixing}.

The product of couplings $\beta_{s\alpha}^{*}\beta_{b\alpha}$ has
the fundamental property

\begin{equation}
\sum_{\alpha}\beta_{s\alpha}^{*}\beta_{b\alpha}=0\,,\label{eq:unitarity_loop}
\end{equation}
which arises trivially from unitarity of the transformations in Eq.~\eqref{eq:mixing_matrix_Q}.
This property, similarly to the GIM mechanism in the SM, is essential
to render the loop finite. However, the property holds as long as
the 2-3 down mixing and $s_{35}^{Q}$ are small. In particular, $s_{35}^{Q}$
is naturally small in the scenario $s_{34}^{Q}\approx1$, as it is
suppressed by the small cosine $c_{34}^{Q}$, see the definition of
$s_{35}^{Q}$ in Eq.~\eqref{eq:35_mixing}. Ultimately, the mixing
angle $s_{35}^{Q}$ is controlled by the fundamental parameter $x_{35}^{\psi}$,
and we obtained that $x_{35}^{\psi}\apprle0.09$ is required to survive
the $\Delta M_{s}$ bound.

The loop function is dominated by the heavy vector-like partners.
In particular, in the motivated scenario with maximal $s_{34}^{L}$,
the couplings with the fourth family $\beta_{sE_{4}}^{*}\beta_{bE_{4}}$
are suppressed by the small cosine $c_{34}^{L}$. This way, the loop
is dominated by $E_{5}$ in good approximation, and we can apply the
property \eqref{eq:unitarity_loop} to obtain
\begin{figure}[t]
\subfloat[\label{fig:ML5_deltaMs}]{\includegraphics[scale=0.4]{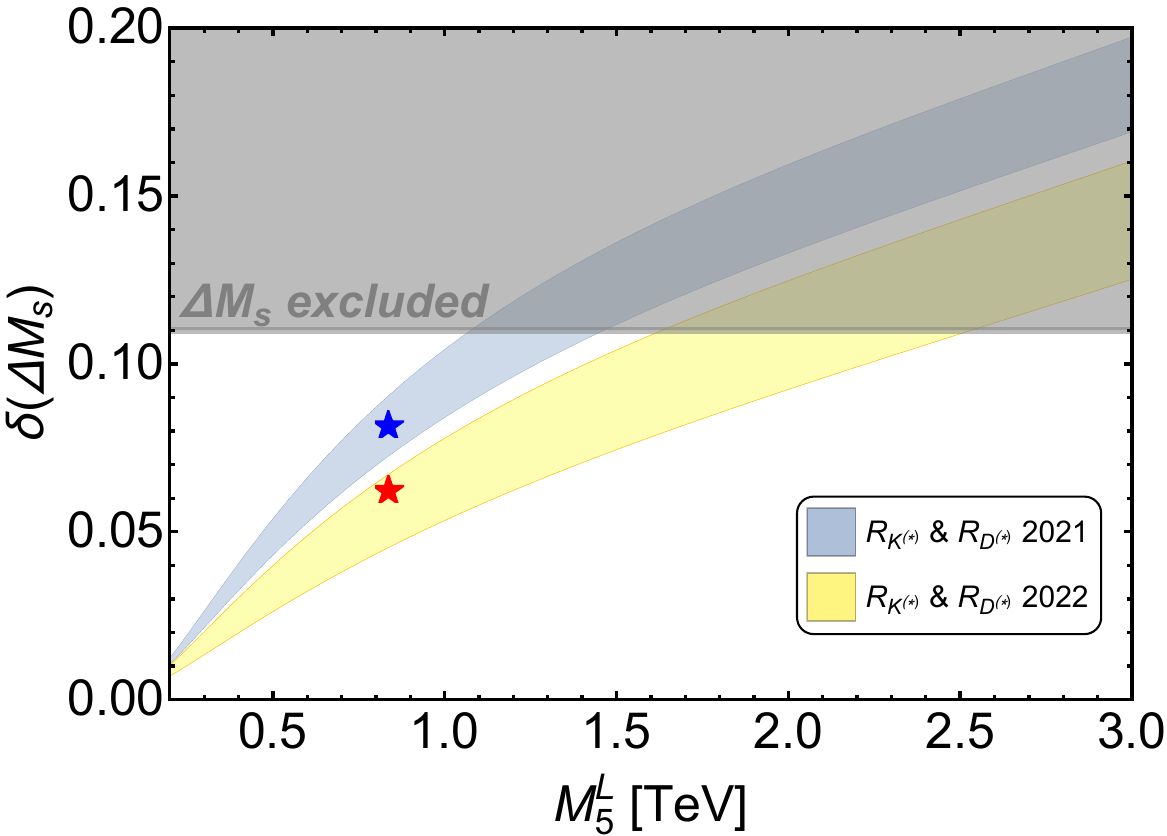}

}$\qquad$\subfloat[\label{fig:BsMixing_parameter_space}]{\includegraphics[scale=0.4]{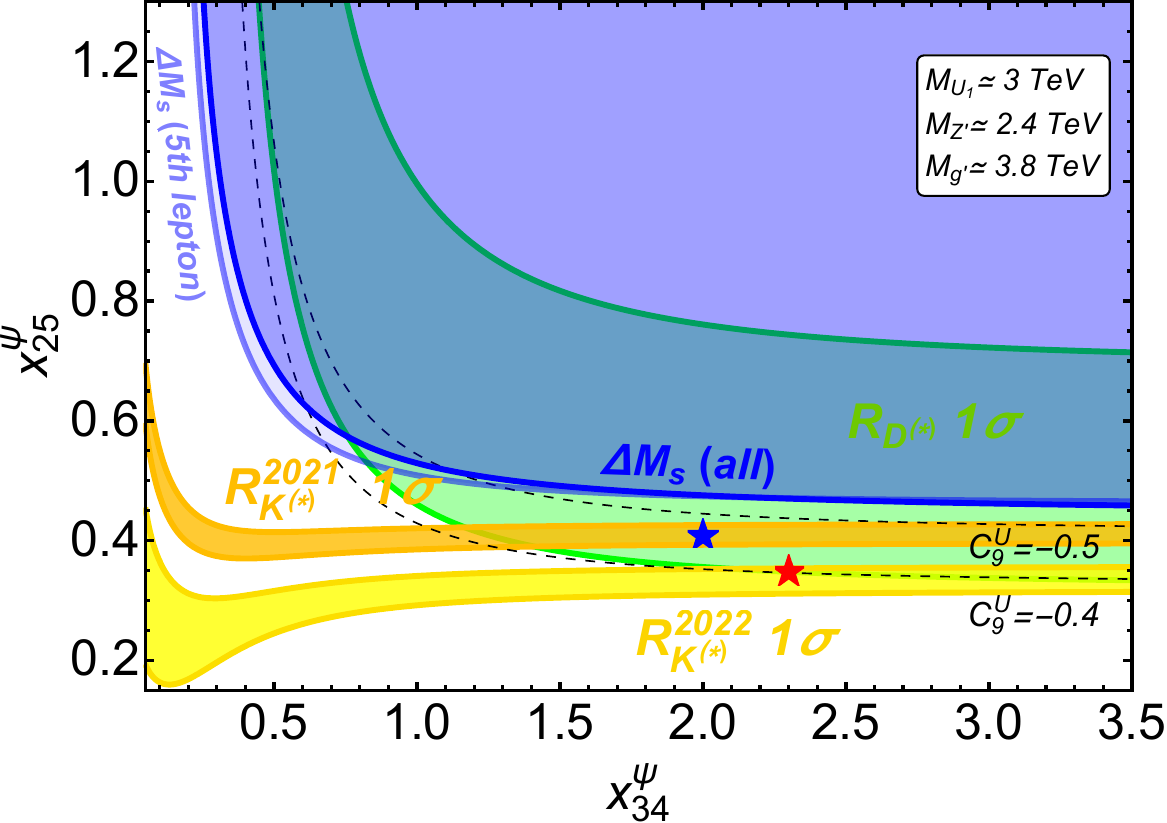}

}

\caption{\textbf{\textit{(Left)}}\textit{ }$\delta(\Delta M_{s})$ (Eq.~\eqref{eq:delta_DeltaMs})
as a function of the 5th vector-like mass term. $x_{25}^{\psi}$ is
varied in the range $x_{25}^{\psi}=[0.3,\,0.35]$ ($[0.4,\,0.45]$)
preferred by $R_{K^{(*)}}^{2022}$ ($R_{K^{(*)}}^{2021}$ ), obtaining
the yellow (blue) band. The gray region is excluded by the $\Delta M_{s}$
bound, see Eq.~\eqref{eq:DeltaMs_bound}. \textbf{\textit{(Right)}}
Parameter space in the plane ($x_{34}^{\psi}$, $x_{25}^{\psi}$)
compatible with $R_{D^{(*)}}$ and $R_{K^{(*)}}$ at 1$\sigma$. The
remaining parameters are fixed as in Table~\ref{tab:BP} for both
panels. The dashed lines show contours of constant $C_{9}^{U}$. The
blue region is excluded by the $\Delta M_{s}$ bound, the region excluded
only due to the contribution via the 5th lepton is also shown in lighter
blue for comparison. The blue and red stars denote BP1 and BP2 respectively.}
\end{figure}

\begin{equation}
C_{bs}^{\mathrm{NP-loop}}=\frac{g_{4}^{4}}{\left(8\pi M_{U_{1}}\right)^{2}}\left(\beta_{sE_{5}}^{*}\beta_{bE_{5}}\right)^{2}\tilde{F}(x_{E_{5}})\,.\label{eq:Cbs_U1_5th}
\end{equation}
The loop function grows with $x_{E_{5}}$ (see Appendix~\ref{subsec:Loop_functions_BsMixing}).
However, in the limit of large bare mass term $M_{5}^{L}$ the effective
coupling $\beta_{sE_{5}}^{*}\propto s_{25}^{Q}$ vanishes (since large
$M_{5}^{L}$ also implies large $M_{5}^{Q}$ due to the Pati-Salam
symmetry), hence both the contribution to $\Delta M_{s}$ and $R_{D^{(*)}}$
go away. In Fig.~\ref{fig:ML5_deltaMs} we plot $\delta(\Delta M_{s})$
defined in Eq.~\eqref{eq:delta_DeltaMs} in terms of $M_{5}^{L}$,
and we vary $x_{25}^{\psi}$ in the ranges compatible with $R_{D^{(*)}}$
and $R_{K^{(*)}}^{2022}$ ($R_{K^{(*)}}^{2021}$). We can see that
the $\Delta M_{s}$ bound requires a vector-like lepton around 1.5-2
TeV in the 2022 case, while 2021 data was pointing to a VL lepton
with a mass around 1 TeV.

In Fig.~\ref{fig:BsMixing_parameter_space} we show that Eq.~\eqref{eq:Cbs_U1_5th}
is indeed a good approximation, up to small interference effects between
the 4th and 5th family contributions in the small $x_{34}^{\psi}$
region, where the fourth lepton is lighter. We also show the parameter
space compatible with $\Delta M_{s}$ and the LFU ratios in our benchmark
scenario. In particular, $\Delta M_{s}$ turns out to be the largest
constraint over the parameter space other than $R_{K^{(*)}}^{2022}$.

\subsubsection{LFV processes\label{subsec:LFV_processes}}

\subsubsection*{$\boldsymbol{\tau\rightarrow3\mu}$}

The partial alignment condition of Eq.~\eqref{eq:GIM-like_leptons-1}
allows for $Z'$-mediated FCNCs in $\tau\mu$ processes, due to the
fact that the model predicts significant mixing between the muon and
tau leptons. This is a crucial prediction of the twin PS theory of
flavour, not present in general 4321 models. Of particular interest
is the process $\tau\rightarrow3\mu$, which receives a tree-level
$Z'$ contribution that grows with the $\tau\mu$ mixing angle $s_{23}^{e}$.
Beyond the latter, $\tau\rightarrow3\mu$ also receives a $U_{1}$-mediated
1-loop contribution

\begin{equation}
C_{\tau\mu\mu\mu}^{U_{1}}=\frac{3g_{4}^{4}}{128\pi^{2}M_{U_{1}}^{2}}\beta_{D_{5}\mu}^{*}\beta_{D_{5}\tau}\left(\beta_{D_{5}\mu}\right)^{2}\tilde{F}(x_{D_{5}})\,.\label{eq:Cbs_U1_5th-1}
\end{equation}
The effective coupling $\beta_{D_{5}\mu}$ is proportional to $s_{25}^{L}\approx0.1$,
which provides a further suppression of $\mathcal{O}((s_{25}^{L})^{3})$
that renders the loop negligible against the much larger tree-level
$Z'$-mediated contribution. The typical benchmark $s_{25}^{L}\approx0.1$
naturally suppresses the $\mu\mu Z'$ coupling, keeping the $Z'$
contribution to $\tau\rightarrow3\mu$ under control, and simultaneously
protects from $Z'\rightarrow\mu\mu$ at LHC (see Section \ref{subsec:Colliders}).
\begin{figure}[t]
\subfloat[\label{fig:tau_3mu}]{\includegraphics[scale=0.4]{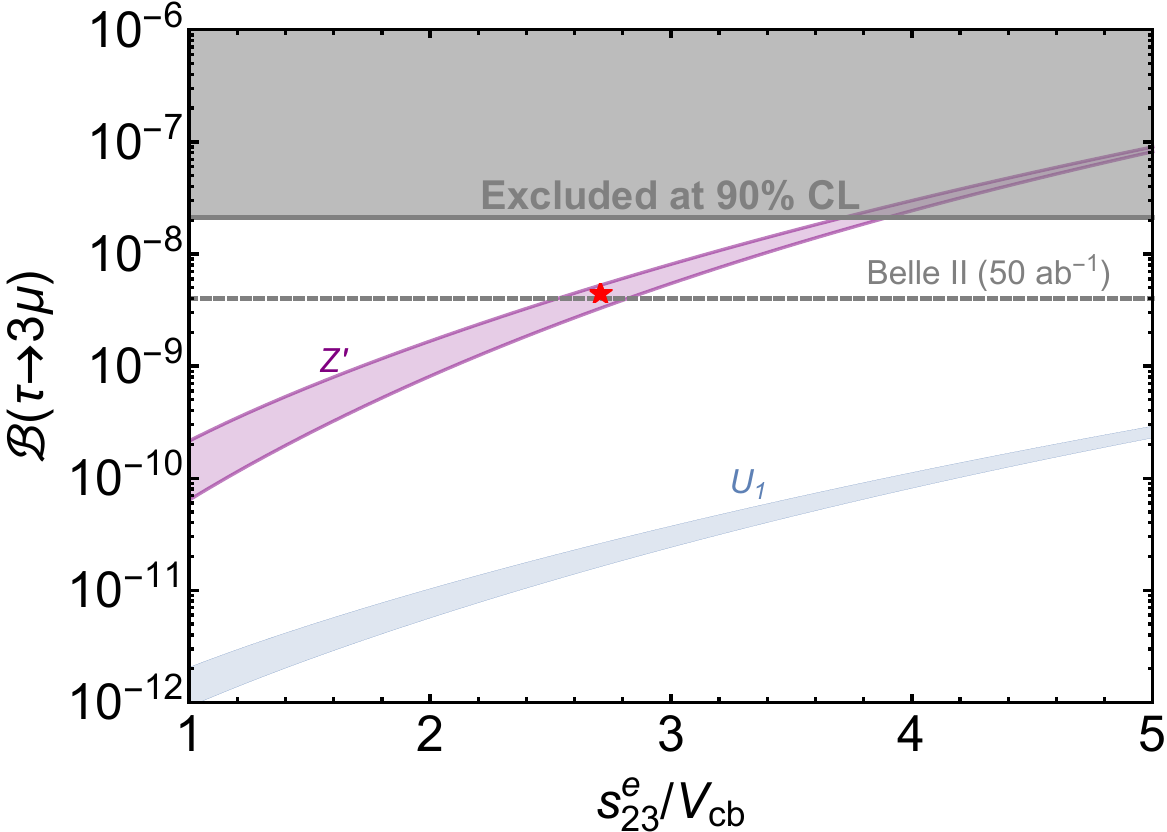}

}$\qquad$\subfloat[\label{fig:tau_muphoton}]{\centering{}\includegraphics[scale=0.4]{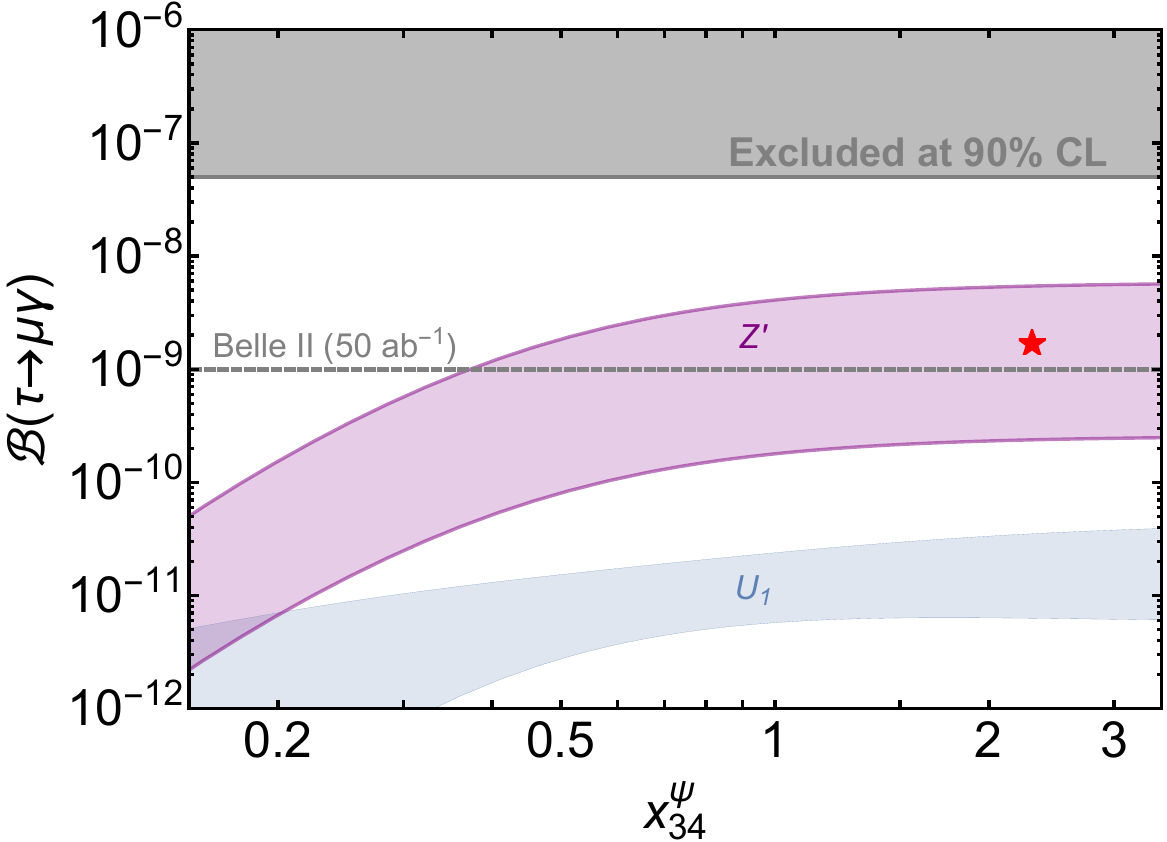}}

\caption{\textbf{\textit{(Left)}} $\mathcal{B}\left(\tau\rightarrow3\mu\right)$
as a function of the 2-3 charged lepton mixing sine $s_{23}^{e}$.
The purple region denotes the $Z'$ contribution while the blue region
denotes the $U_{1}$ contribution, for both we have varied $x_{25}^{\psi}=[0.3,0.35]$
which is compatible with $R_{K^{(*)}}^{2022}$. \textbf{\textit{(Right)}}
$\mathcal{B}\left(\tau\rightarrow\mu\gamma\right)$ as a function
of $x_{34}^{\psi}$. The purple region denotes the $Z'$ contribution
for which we have varied $s_{23}^{e}=[V_{cb},5V_{cb}]$. The blue
region denotes the $U_{1}$ contribution, for which we have varied
$x_{25}^{\psi}=[0.1,1]$. The gray regions are excluded by the experiment,
the dashed lines show the projected future bound. The red star shows
BP2.}
\end{figure}

As depicted in Fig.~\ref{fig:tau_3mu}, the $Z'$ contribution dominates
over the $U_{1}$ contribution, and the regions of the parameter space
with very large $s_{23}^{e}$ are already excluded by the experiment.
We have chosen to plot the results of the 2022 case only, since this
observable depends mostly on $s_{23}^{e}$ and there is little variation
with 2021 data. The Belle II collaboration will test a further region
of the parameter space \cite{Belle-II:2018jsg}, setting the bound
$s_{23}^{e}<2.8V_{cb}$ if no signal is detected. In all UV incomplete
4321 models (such as \cite{DiLuzio:2017vat,DiLuzio:2018zxy}) the
$\mu-\tau$ mixing is unspecified, so only the small $U_{1}$ signal
is predicted. Therefore, the large $Z'$ signal offers the opportunity
to disentangle the twin Pati-Salam model from other 4321 proposals.

As depicted in Fig.~\ref{fig:LFV_processes}, $\tau\rightarrow3\mu$
is the most constraining signal over the parameter space out of all
the LFV processes, provided that the 2-3 charged lepton mixing is
$\mathcal{O}(0.1)$.

\subsection*{$\boldsymbol{\tau\rightarrow\mu\gamma}$}

The process $\tau\rightarrow\mu\gamma$ receives 1-loop contributions
via both $Z'$ and $U_{1}$, as depicted in Fig.~\ref{fig:diagram_tau_mu_photon},
with formulae reported in Appendix \ref{subsec:Dipole-operators_Appendix}.
Provided that the 3-4 mixing is maximal, the $U_{1}$ loop is dominated
by the 5th vector-like quark, and in this situation the couplings
$\beta_{D_{5}\mu}^{*}\beta_{D_{5}\tau}$ are controlled by $x_{25}^{\psi}$.
The $Z'$ loop is dominated by chiral leptons, in particular by the
$\tau$ lepton, since the coupling $\xi_{\tau\tau}$ is maximal while
$\xi_{\mu\mu}$ is suppressed. In this scenario, the overall $Z'$
contribution is controlled by $\xi_{\tau\mu}$ which grows with the
$\mu-\tau$ mixing angle $s_{23}^{e}$, and the variation via $x_{25}^{\psi}$
is minimal.

In Fig.~\ref{fig:tau_muphoton} we can see that the $Z'$ contribution
dominates the branching fraction in the range of large $x_{34}^{\psi}$
motivated by $R_{D^{(*)}}$, leading to the predictions for $\mathcal{B}\left(\tau\rightarrow\mu\gamma\right)$
being one/three orders of magnitude below the current experimental
limit depending on the value of $s_{23}^{e}$. We have also included
the projected bound by Belle II (50 $\mathrm{ab^{-1}}$) \cite{Belle-II:2018jsg},
which will partially test the parameter space. In the 4321 models
of \cite{DiLuzio:2017vat,DiLuzio:2018zxy} the $\mu-\tau$ mixing
is unspecified, so only the blue $U_{1}$ signal is predicted. For
non-fermiophobic models, this signal is largely enhanced via a chirality
flip with the bottom quark running in the loop \cite{Cornella:2019hct,Cornella:2021sby,Barbieri:2022ikw,Fuentes-Martin:2020bnh,Fuentes-Martin:2020hvc},
predicting a larger signal $\mathcal{B}\left(\tau\rightarrow\mu\gamma\right)\approx10^{-8}$.
Instead, our $Z'$ signal lies below, offering the opportunity to
disentangle the twin Pati-Salam model from all other proposals.

\subsection*{$\boldsymbol{B_{s}\rightarrow\tau\mu}$, $\boldsymbol{B\rightarrow K\tau\mu}$
and $\boldsymbol{\tau\rightarrow\mu\phi}$ }

\begin{figure}[t]
\subfloat[\label{fig:Bs_to_taumu}]{\includegraphics[scale=0.4]{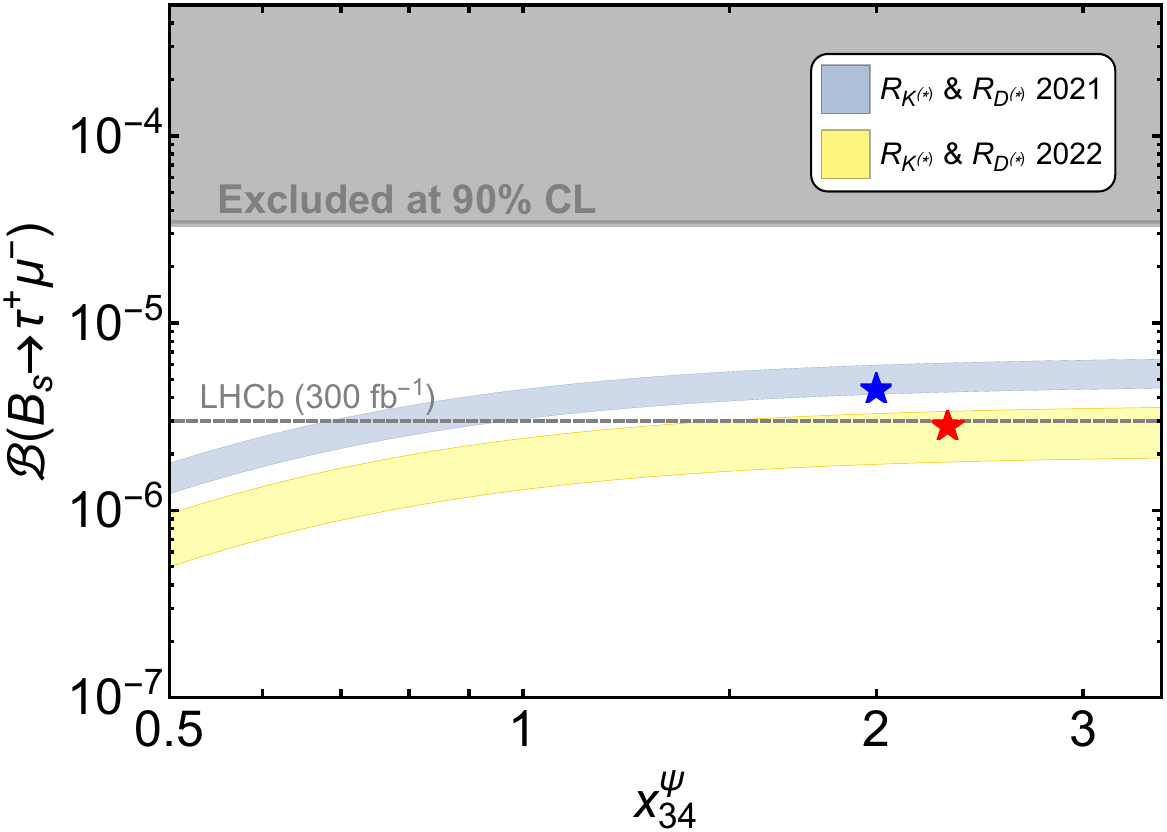}

}$\qquad$\subfloat[\label{fig:LFV_processes}]{\centering{}\includegraphics[scale=0.4]{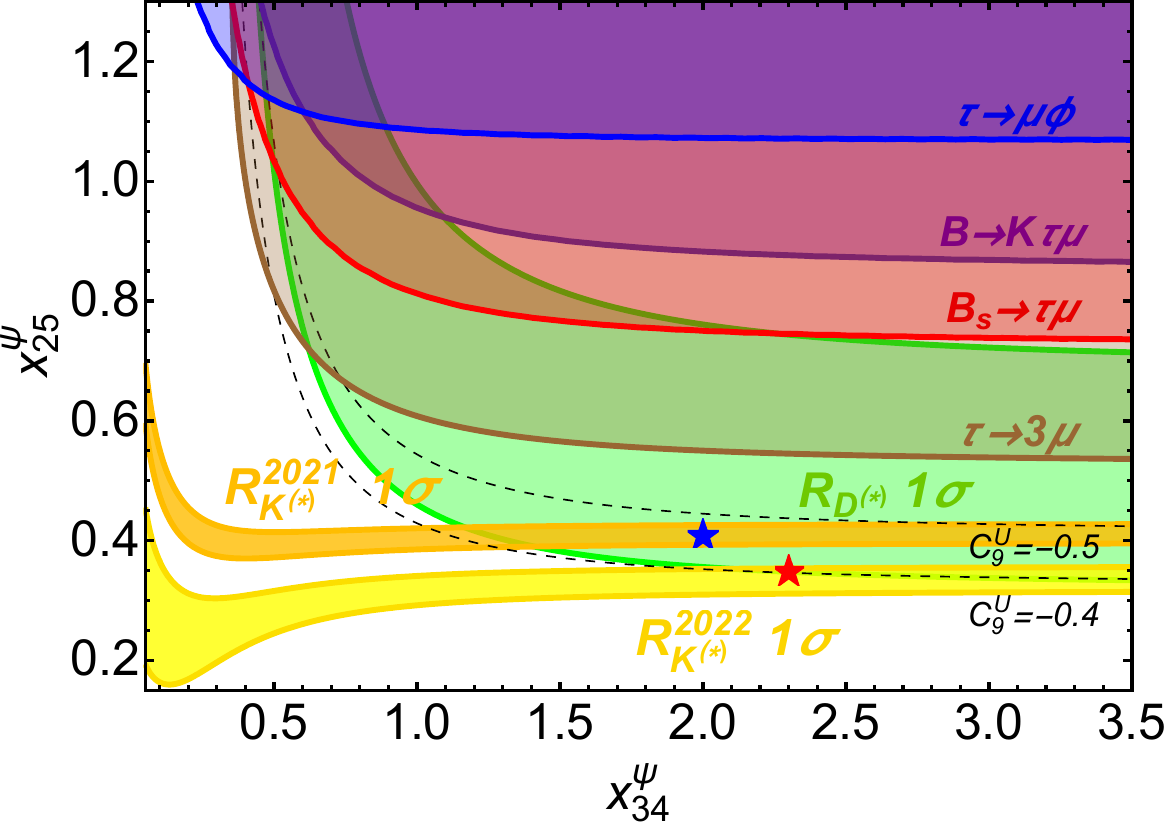}}

\caption{\textbf{\textit{(Left)}} $\mathcal{B}\left(B_{s}\rightarrow\tau^{+}\mu^{-}\right)$
as a function of $x_{34}^{\psi}$. The yellow (blue) band is obtained
by varying $x_{25}^{\psi}$ in the range $x_{25}^{\psi}=[0.3,\,0.35]$($[0.4,\,0.45]$)
preferred by $R_{K^{(*)}}^{2022}$ ($R_{K^{(*)}}^{2021}$ ). The gray
region is excluded by the experiment, the dashed line shows the projected
future bound. \textbf{\textit{(Right) }}Parameter space in the plane
($x_{34}^{\psi}$, $x_{25}^{\psi}$) compatible with $R_{D^{(*)}}$
and $R_{K^{(*)}}$ at 1$\sigma$. The remaining parameters are fixed
as in Table \ref{tab:BP}. The dashed lines show contours of constant
$C_{9}^{U}$. The regions excluded by LFV violating processes are
displayed. The blue (red) star shows BP1 (BP2). \label{fig:LFV_Btaumu}}
\end{figure}
The vector leptoquark $U_{1}$ mediates tree-level contributions to
flavour-violating (semi)leptonic $B$-decays to (kaons), taus and
muons. The experimental bound for $B_{s}\rightarrow\tau\mu$ was obtained
by LHCb \cite{LHCb:2019ujz}, while for $B\rightarrow K\tau\mu$ experimental
bounds are only available for the decays $B^{+}\rightarrow K^{+}\tau\mu$
\cite{BaBar:2012azg}. The process $\tau\rightarrow\mu\phi$ receives
tree-level contributions from both $U_{1}$ and $Z'$, see Appendix
\ref{subsec:b_s_ll_Appendix} and \ref{subsec:LFV_tau_Appendix}.
However, $\tau\rightarrow\mu\phi$ turns out to be suppressed by the
small effective couplings $\beta_{s\mu}\propto s_{25}^{Q}s_{25}^{L}$
and $\xi_{ss}\propto\left(s_{25}^{Q}\right)^{2}$ and we find $\mathcal{B}\left(\tau\rightarrow\mu\phi\right)\approx10^{-9}$,
roughly two orders of magnitude below the current experimental bounds,
and just below the future sensitivity of Belle II.

As can be seen in Fig.~\ref{fig:LFV_processes}, $B_{s}\rightarrow\tau^{+}\mu^{-}$
implies the largest constraint over the parameter space out of all
semileptonic LFV processes involving $\tau$ leptons, followed by
$B^{+}\rightarrow K^{+}\tau^{+}\mu^{-}$ and $\tau\rightarrow\mu\phi$.
The present experimental bounds lead to mild constraints over the
parameter space compatible with $R_{D^{(*)}}$. As depicted in Fig.~\ref{fig:Bs_to_taumu},
the 2021 region for $B_{s}\rightarrow\tau^{+}\mu^{-}$ was partially
within LHCb projected sensitivity, but the 2022 region will mostly
remain untested.

\subsection*{$\boldsymbol{K_{L}\rightarrow\mu e}$}

The LFV process $K_{L}\rightarrow\mu e$ sets a strong constraint
over all models featuring a vector leptoquark $U_{1}$ with first
and second family couplings \cite{Dolan:2020doe},

\begin{equation}
\mathcal{B}\left(K_{L}\rightarrow\mu e\right)=\frac{\tau_{K_{L}}G_{F}^{2}f_{K}^{2}m_{\mu}^{2}m_{K}}{8\pi}\left(1-\frac{m_{\mu}^{2}}{m_{K}^{2}}\right)^{2}C_{U}^{2}\left|\beta_{de}\beta_{s\mu}^{*}\right|^{2}\,.\label{eq:KL_mue}
\end{equation}
The first family coupling $\beta_{de}$ can be diluted via mixing
with vector-like fermions, which we parameterised via the effective
parameter $\epsilon$ in Eq.~\eqref{eq:LQ_couplings}, so that $\beta_{se}\approx s_{16}^{Q}s_{16}^{L}\epsilon$.
The mechanism to perform this and the definition of $\epsilon$ in
terms of fundamental parameters of the model is included in Appendix~\ref{sec:Epsilon_Dilution}.
\begin{figure}[t]
\subfloat[\label{fig:KL_mue}]{\includegraphics[scale=0.4]{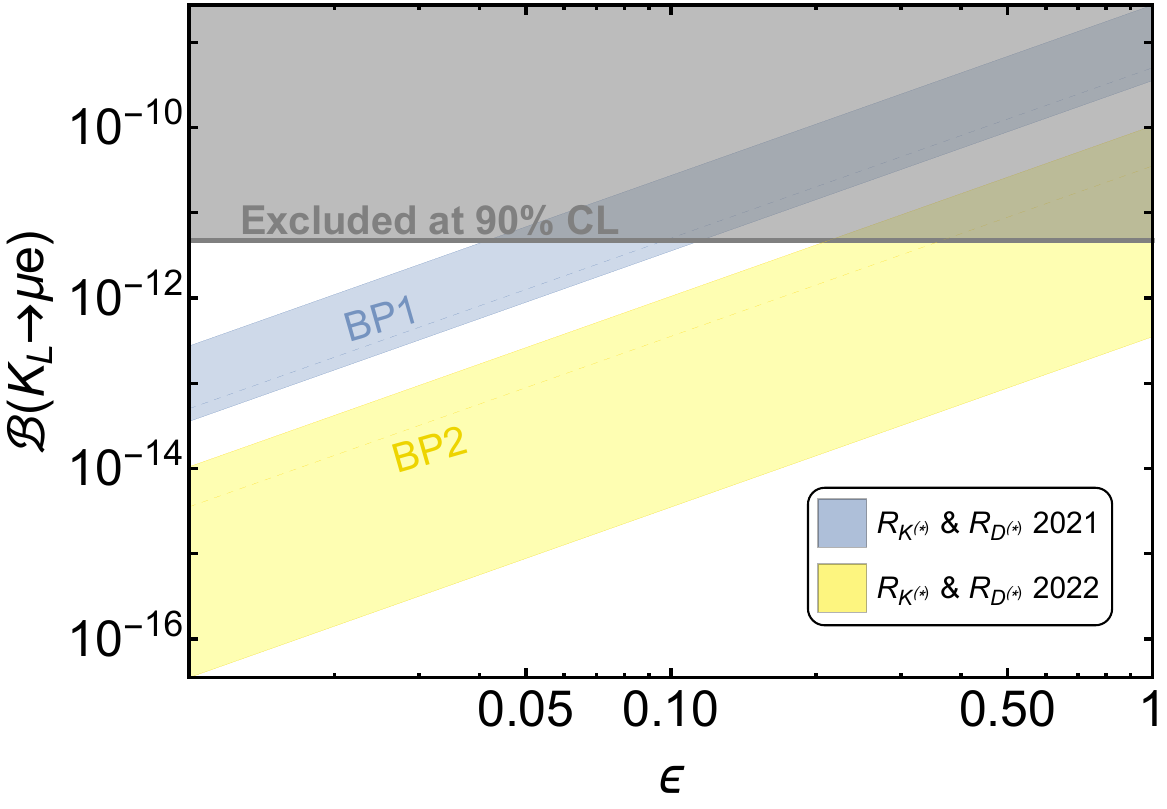}

}$\qquad$\subfloat[\label{fig:LFU_ratios_tau}]{\includegraphics[scale=0.4]{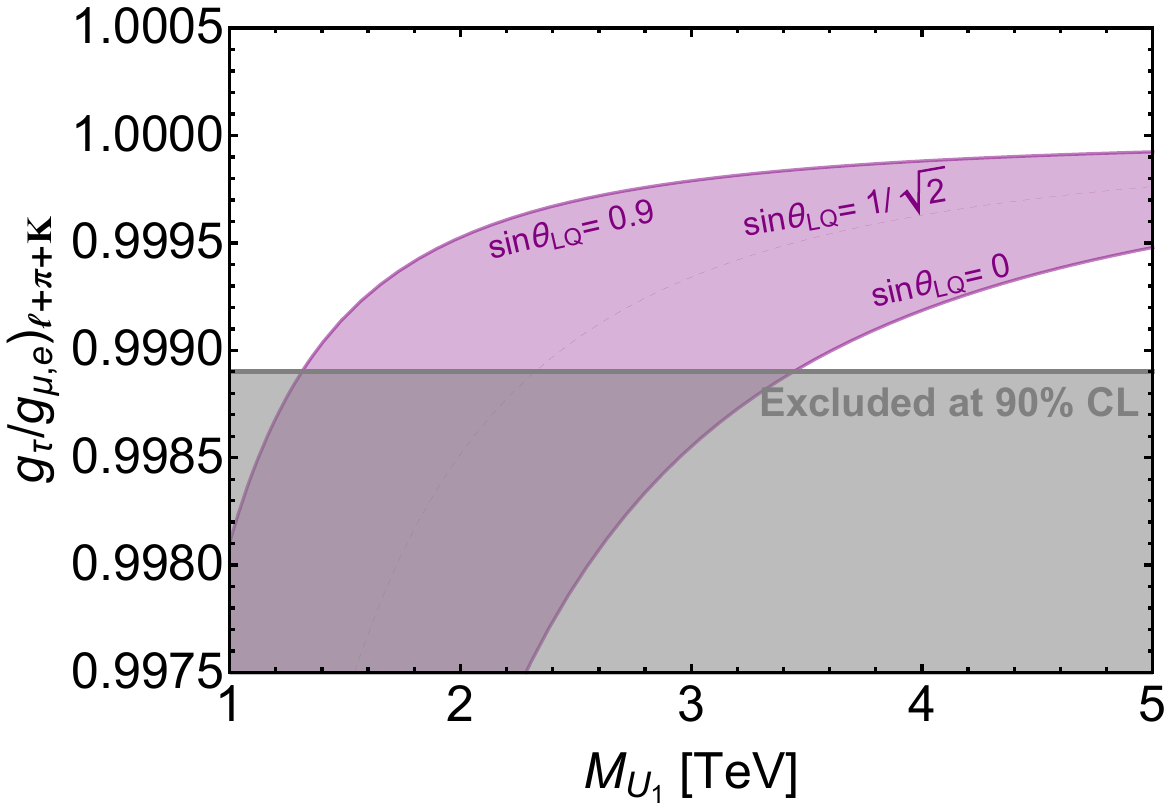}

}

\caption{\textbf{\textit{(Left)}}\textit{ }$\mathcal{B}\left(K_{L}\rightarrow\mu e\right)$
(Eq.~\eqref{eq:KL_mue}) as a function of $\epsilon$ (see main text
for details). $x_{25}^{\psi}$ is varied in the range $x_{25}^{\psi}=[0.3,\,0.35]$
($[0.4,\,0.45]$) preferred by $R_{K^{(*)}}^{2022}$ ($R_{K^{(*)}}^{2021}$
), obtaining the yellow (blue) band. \textbf{\textit{(Right)}} LFU
ratios originated from $\tau$ decays (Eq.~\eqref{eq:LFUratios_tau})
as a function of the mass of the vector leptoquark $M_{U_{1}}$, $\sin\theta_{LQ}$
is varied in the range $\sin\theta_{LQ}=[0,0.9]$ and $g_{4}=3.5$.
The remaining parameters are fixed as in Table~\ref{tab:BP} for
both panels, and current exclusion limits are shown.}
\end{figure}

In Fig.~\ref{fig:KL_mue} we can see that for the 2022 case, some
region of the parameter space is compatible with $K_{L}\rightarrow\mu e$
without the need of diluting the coupling. Instead, for the benchmark
values BP1 and BP2, a mild suppression is required. In Appendix~\ref{sec:Epsilon_Dilution},
a benchmark with the fundamental parameters of the model that provide
such a suppression is included. This signal is a direct consequence
of the underlying twin PS symmetry and the GIM-like mechanism, which
lead to quasi-degenerate mixing angles $s_{16}^{Q}\approx s_{16}^{L}$
that are equal to their 25 counterparts, and as a consequence $\beta_{de}\neq0$.
Therefore, it is not present in other 4321 models \cite{DiLuzio:2017vat,DiLuzio:2018zxy,Cornella:2019hct,Cornella:2021sby,Barbieri:2022ikw}.

\subsubsection{Tests of universality in leptonic $\tau$ decays}

NP contributions to $R_{D^{(*)}}$ commonly involve large couplings
to $\tau$ leptons, which can have an important effect over LFU ratios
originated from $\tau$ decays. Such tests are constructed by performing
ratios of the partial widths of a lepton decaying to lighter leptons
and/or hadrons. We find all ratios in our model to be well approximated
by (see Appendix~\ref{subsec:Tests-of-universality-tau-decays}),

\begin{equation}
\left(\frac{g_{\tau}}{g_{\mu,e}}\right)_{\ell+\pi+K}\approx1-0.079C_{U}\left|\beta_{b\tau}\right|^{2}\,,\label{eq:LFUratios_tau}
\end{equation}
where $\beta_{b\tau}\approx\cos\theta_{LQ}$ assuming maximal 3-4
mixing. Therefore, it can be seen as a constraint over the $\beta_{b\tau}$
coupling, and hence is not directly related to $R_{K^{(*)}}$ so we
do not plot two bands here. The high-precision measurements of these
effective ratios only allow for per mille modifications, see the HFLAV
average \cite{HFLAV:2022pwe} in Table~\ref{tab:Observables}. As
depicted in Fig.~\ref{fig:LFU_ratios_tau}, this constraint sets
the lower bound $M_{U_{1}}\gtrsim2.2\,\mathrm{TeV}$ for $\sin\theta_{LQ}=1/\sqrt{2}$
and $g_{4}=3.5$. This bound becomes more restrictive for $\cos\theta_{LQ}\approx1$,
or equivalently $\beta_{b\tau}\approx1$, for which we find $M_{U_{1}}\gtrsim3.3\,\mathrm{TeV}$
if $g_{4}=3.5$ and $M_{U_{1}}\gtrsim2.9\,\mathrm{TeV}$ if $g_{4}=3$.

\subsubsection{Signals in rare $B$-decays \label{subsec:Signals-in-rare-processes}}

\subsubsection*{$\boldsymbol{B_{s}\rightarrow\tau\tau}$ and $\boldsymbol{B\rightarrow K\tau\tau}$}

\begin{figure}[t]
\subfloat[\label{fig:Bs_tautau}]{\includegraphics[scale=0.4]{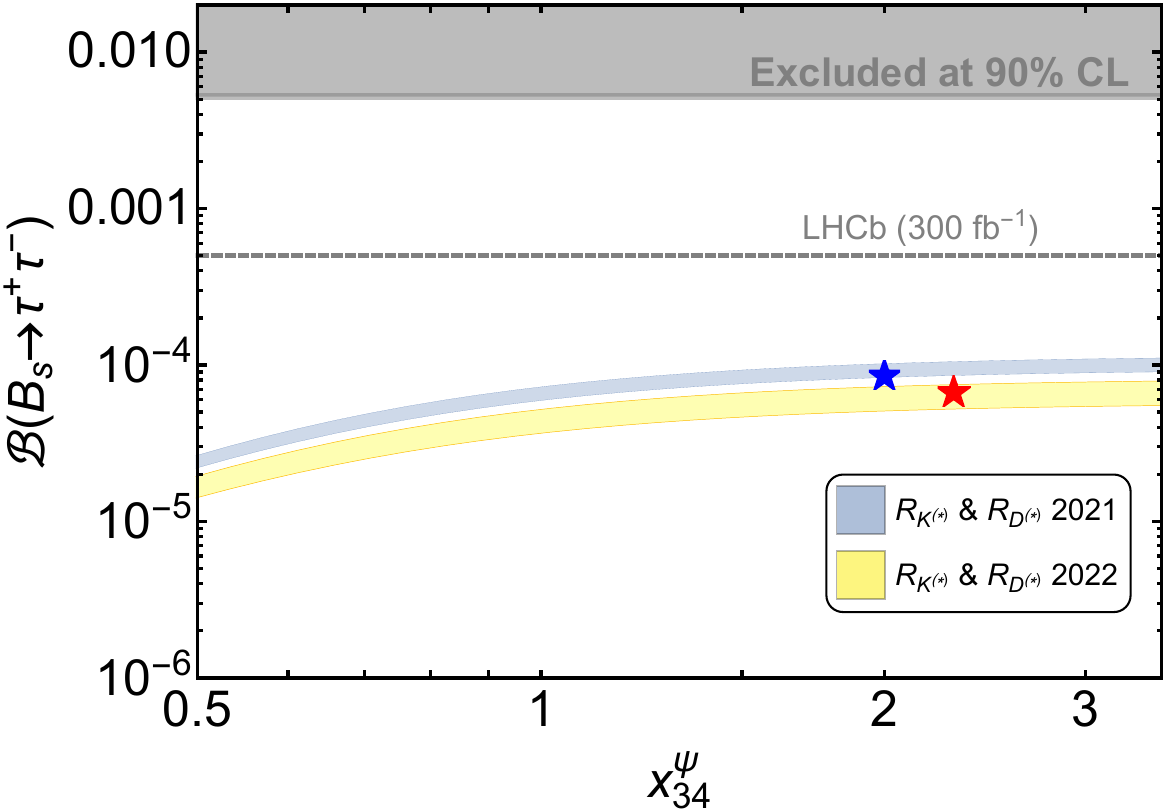}

}$\qquad$\subfloat[\label{fig:B_K_tautau}]{\includegraphics[scale=0.4]{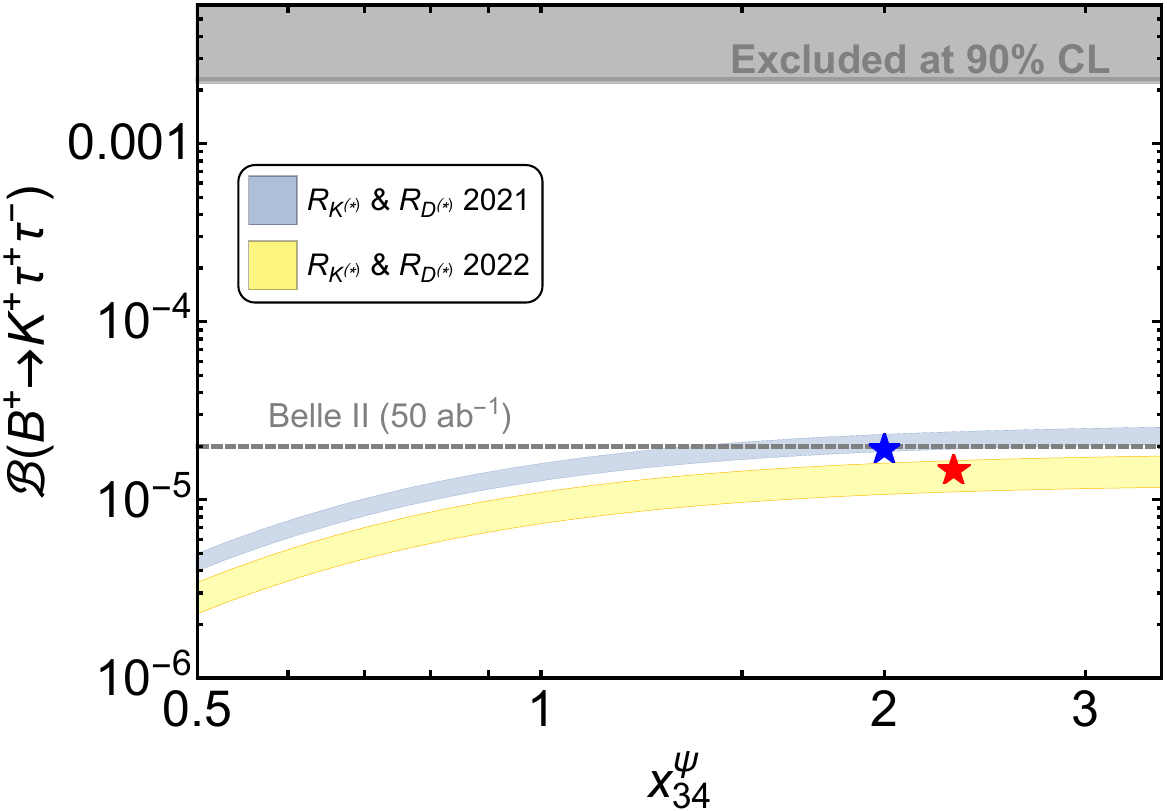}

}

\caption{The branching fractions $\mathcal{B}\left(B_{s}\rightarrow\tau^{+}\tau^{-}\right)$
(left) and $\mathcal{B}\left(B^{+}\rightarrow K^{+}\tau^{+}\tau^{-}\right)$
(right) as a function of $x_{34}^{\psi}$, with $x_{25}^{\psi}$ varied
in the range $x_{25}^{\psi}=[0.3,\,0.35]$ ($[0.4,\,0.45]$) preferred
by $R_{K^{(*)}}^{2022}$ ($R_{K^{(*)}}^{2021}$), obtaining the yellow
(blue) band. The rest of the parameters are fixed as in Table~\ref{tab:BP}.
Current exclusion limits are displayed, along with their future projections.
The blue (red) star shows BP1 (BP2).\label{fig:B_tautau}}
\end{figure}
The explanation of $R_{D^{(*)}}$ in our model requires a large $U_{1}$
contribution to the 4-fermion operator $(\bar{c}_{L}\gamma_{\mu}b_{L})(\bar{\tau}_{L}\gamma^{\mu}\nu_{\tau L})$.
Therefore, large contributions to the rare decays $B_{s}\rightarrow\tau\tau$
and $B\rightarrow K\tau\tau$ arise via $SU(2)_{L}$ invariance of
the $U_{1}$ couplings. The respective branching fractions are of
order $10^{-7}$ in the SM and mild upper bounds have been obtained
by LHCb \cite{LHCb:2017myy} and BaBar \cite{BaBar:2016wgb}, respectively. 

In Fig.~\ref{fig:B_tautau}, we plot the branching fractions as a
function of $x_{34}^{\psi}$, while $x_{25}^{\psi}$ is varied in
the ranges compatible with 2021 and 2022 $R_{K^{(*)}}$, respectively.
We find that the predictions are far below the current bounds, however
they lie closer to the expected future bounds from LHCb and Belle
II data \cite{LHCb:2018roe,Belle-II:2018jsg}. This prediction is
different in non-fermiophobic 4321 models \cite{Cornella:2019hct,Cornella:2021sby,Barbieri:2022ikw},
where these contributions are enhanced and all the parameter space
will be tested in $B^{+}\rightarrow K^{+}\tau^{+}\tau^{-}$ by Belle~II. 

\subsection*{$\boldsymbol{B\rightarrow K\nu\bar{\nu}}$}

The $U_{1}$ leptoquark does not contribute at tree-level to $b\rightarrow s\nu\nu$
transitions, and the tree-level exchange of the $Z'$ is suppressed
due to the down-aligned flavour structure. However, loop-level corrections
can lead to an important enhancement of the channel $B\rightarrow K\nu_{\tau}\bar{\nu}_{\tau}$
\cite{Cornella:2021sby}. We parameterise corrections to the SM branching
fraction as
\begin{figure}[t]
\begin{centering}
\includegraphics[scale=0.4]{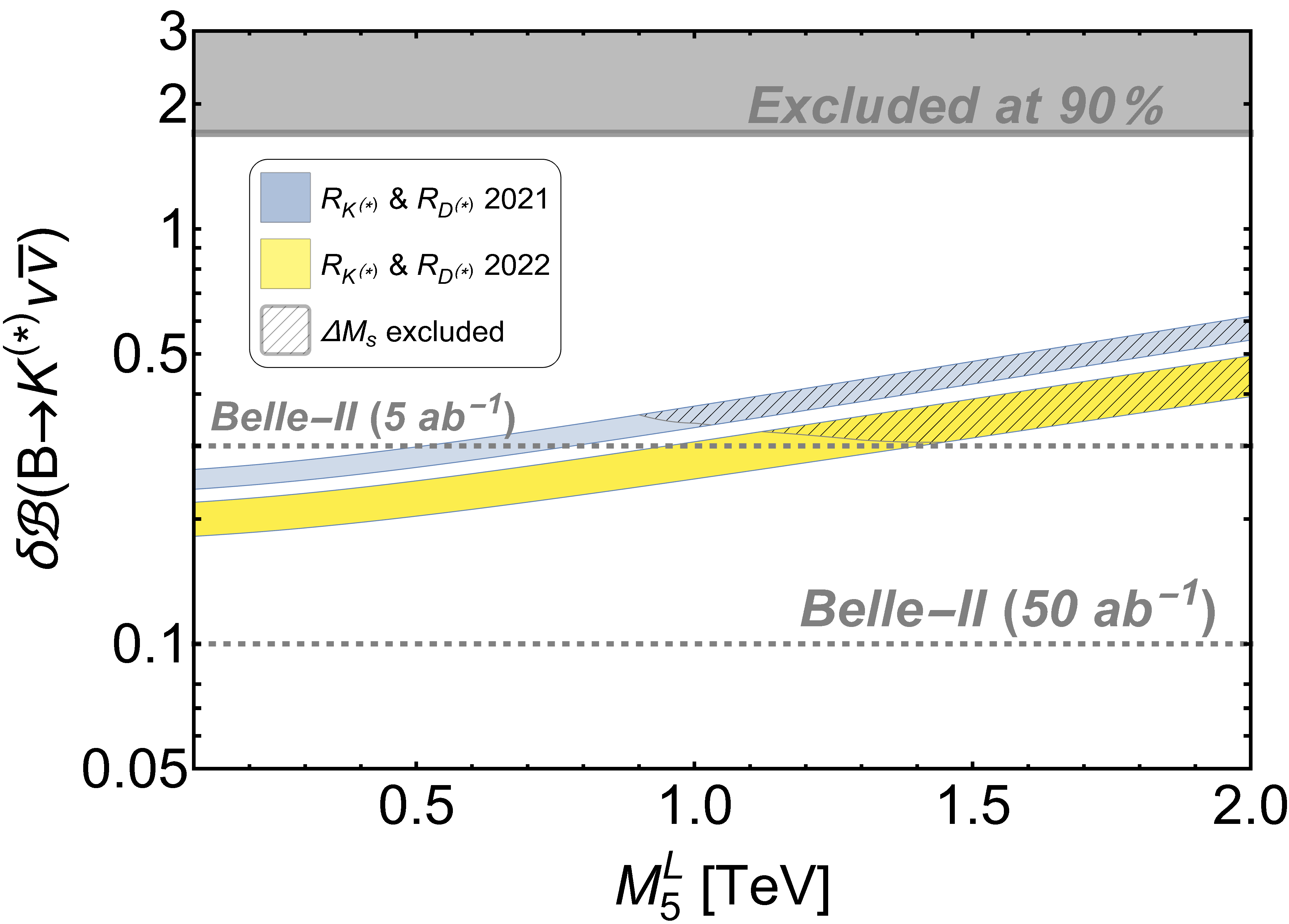}
\par\end{centering}
\caption{$\delta\mathcal{B}(B\rightarrow K^{(*)}\nu\bar{\nu})$ (Eq.~\eqref{eq:BtoK_nunu})
as a function of the 5th family vector-like mass term. $x_{25}^{\psi}$
is varied in the range $x_{25}^{\psi}=[0.3,\,0.35]$ ($[0.4,\,0.45]$)
preferred by $R_{K^{(*)}}^{2022}$ ($R_{K^{(*)}}^{2021}$ ), obtaining
the yellow (blue) band. The hatched region is excluded by the $\Delta M_{s}$
bound, see Eq.~\eqref{eq:DeltaMs_bound}. The gray region is excluded
by current experimental measurements, the dashed line indicates the
projected future bound. \label{fig:BtoK_nunu}}
\end{figure}

\begin{equation}
\delta\mathcal{B}(B\rightarrow K^{(*)}\nu\bar{\nu})=\frac{\mathcal{B}(B\rightarrow K^{(*)}\nu\bar{\nu})}{\mathcal{B}(B\rightarrow K^{(*)}\nu\bar{\nu})_{\mathrm{SM}}}-1\approx\frac{1}{3}\left|\frac{C_{\nu\nu}^{\mathrm{NP}}+C_{\nu\nu}^{\mathrm{SM}}}{C_{\nu\nu}^{\mathrm{SM}}}\right|^{2}-\frac{1}{3}\,,\label{eq:BtoK_nunu}
\end{equation}
where the EFT and the Wilson coefficients are defined in Appendix~\ref{subsec:BtoKnunu_Appendix}.

As depicted in Fig.~\ref{fig:Box-and-penguin_BtoKnunu}, the main
contributions are a semileptonic box diagram mediated by $U_{1}$
and a triangle diagram correction to the flavour-violating $bsZ'$
vertex, plus the RGE-induced contribution via the $U_{1}$-mediated
operator $(\overline{s}_{L}\gamma_{\mu}b_{L})$ $(\overline{\tau}_{L}\gamma^{\mu}\tau_{L})$,
denoted as $C_{\nu,U}^{\mathrm{RGE}}$. The former two 1-loop contributions
are dominated by the fifth VL charged lepton and grow with its bare
mass, $M_{5}^{L}$. This way, the overall contribution to $B\rightarrow K\nu\bar{\nu}$
can be sizable, yielding up to $\mathcal{O}(1)$ corrections with
respect to the SM value, as depicted in Fig.~\ref{fig:BtoK_nunu}.
The details of the calculation are found in Appendix~\ref{subsec:BtoKnunu_Appendix}.

For low $M_{5}^{L}$, the value of $\delta\mathcal{B}(B\rightarrow K^{(*)}\nu\bar{\nu})$
corresponds to $C_{\nu,U}^{\mathrm{RGE}}$. For large $M_{5}^{L}$,
however, we have seen that stringent constraints from $B_{s}-\bar{B_{s}}$
meson mixing play an important role, see Section~\ref{subsec:BsMixing_revisited}.
This constraint is depicted as the hatched region in Fig.~\ref{fig:BtoK_nunu},
correlating $B\rightarrow K\nu\bar{\nu}$ and $\Delta M_{s}$, a feature
which has not been highlighted in other analyses. In particular, $\Delta M_{s}$
rules out the region where $\delta\mathcal{B}(B\rightarrow K^{(*)}\nu\bar{\nu})$
can reach values close to current experimental limits. Nevertheless,
the Belle II collaboration will measure $\mathcal{B}(B\rightarrow K^{(*)}\nu\bar{\nu})$
up to 10\% of the SM value \cite{Belle-II:2018jsg}, hence testing
all the parameter space of the model. 

Our signal of $B\rightarrow K^{(*)}\nu\bar{\nu}$ also offers a great
opportunity to disentangle our twin PS framework from non-fermiophobic
4321 models and the $\mathrm{PS}^{3}$ model \cite{Cornella:2019hct,Cornella:2021sby,Barbieri:2022ikw,Bordone:2017bld},
as they predict a much smaller signal (see Fig. 4.4 of \cite{Cornella:2021sby}
and compare their purple region with our Fig.~\ref{fig:BtoK_nunu}).

\subsubsection{Perturbativity\label{subsec:Pertubativity}}

The explanation of the $R_{D^{(*)}}$ anomaly requires large mixing
angles $s_{34}^{Q}$ and $s_{34}^{L}$, which translate into a sizeable
Yukawa coupling $x_{34}^{\psi}$, thus pushing the model close to
the boundary of the perturbative domain. Perturbativity is a serious
constraint over our model, since we need the low-energy 4321 theory
to remain perturbative until the high scale of the twin Pati-Salam
symmetry. When assessing the issue of perturbativity, two conditions
must be satisfied:
\begin{itemize}
\item Firstly, the low-energy observables must be calculable in perturbation
theory. For Yukawa couplings, we consider the typical bound $x_{34}^{\psi}<\sqrt{4\pi}$.
Regarding the gauge coupling $g_{4}$, standard perturbativity criteria
imposes the beta function criterium \cite{Goertz:2015nkp} $\left|\beta_{g_{4}}/g_{4}\right|<1$,
which yields $g_{4}<4\pi\sqrt{3}/\sqrt{28}\approx4.11$.
\item Secondly, the couplings must remain perturbative up to the energy
scale of the UV completion, i.e.~we have to check that the couplings
of the model do not face a Landau pole below the energy scale of the
twin PS symmetry, namely $\mu\approx1\,\mathrm{PeV}$.
\end{itemize}
The phenomenologically convenient choice of large $g_{4}$ is not
a problem for the extrapolation in the UV, thanks to the asymptotic
freedom of the $SU(4)$ gauge factor (see Fig.~\ref{fig:Perturbativity_gauge}).
To investigate the running of the most problematic Yukawa $x_{34}^{\psi}$,
we use the one-loop renormalisation group equations of the 4321 model
(see Appendix~\ref{sec:RGE-equations}).

The running of the effective Yukawa couplings is protected, as the
top Yukawa is order 1 and all the other are smaller, SM-like (see
the discussion in Section~\ref{subsec:Effective_Yukawa_3VL}). This
feature is different from \cite{DiLuzio:2018zxy}, where the top mass
was accidentally suppressed by the equivalent of $c_{34}^{Q}$ in
our model, hence requiring a large, non-perturbative top Yukawa to
preserve the top mass. Instead, in our model the effective top Yukawa
arises proportional to the maximal angle $s_{34}^{Q}$, rendering
the top Yukawa natural and perturbative. The matrices of couplings
$x_{Q,L}$ and $\lambda_{15}$ are defined as (assuming small $x_{35}^{\psi}$
as discussed in Section \ref{subsec:BsMixing_revisited})

\begin{equation}
x_{\psi}=\left(\begin{array}{ccc}
x_{16}^{\psi} & 0 & 0\\
0 & x_{25}^{\psi} & 0\\
0 & 0 & x_{34}^{\psi}
\end{array}\right)\,,\quad\lambda_{15}=\left(\begin{array}{ccc}
\lambda_{15}^{6} & 0 & 0\\
0 & \lambda_{15}^{5} & 0\\
0 & 0 & \lambda_{15}^{4}
\end{array}\right)\,,\quad\psi=Q,\,L\,.
\end{equation}
The Yukawas $x_{25}^{\psi}$ and $x_{16}^{\psi}$ are not dangerous
as they are order 1 of smaller. The problematic Yukawa is $x_{34}^{\psi}$,
which is required to be large in order to both the LFU ratios, and
also it is connected with the physical mass of the fourth lepton as
per Eq.~\eqref{eq:Mass-_4th}. Large $\lambda_{15}^{5}$ is also
required to obtain a large splitting of VL masses, which leads to
a large $\theta_{LQ}$ as required by $R_{D^{(*)}}$.
\begin{figure}[t]
\begin{centering}
\subfloat[\label{fig:x_Yukawas}]{\includegraphics[scale=0.4]{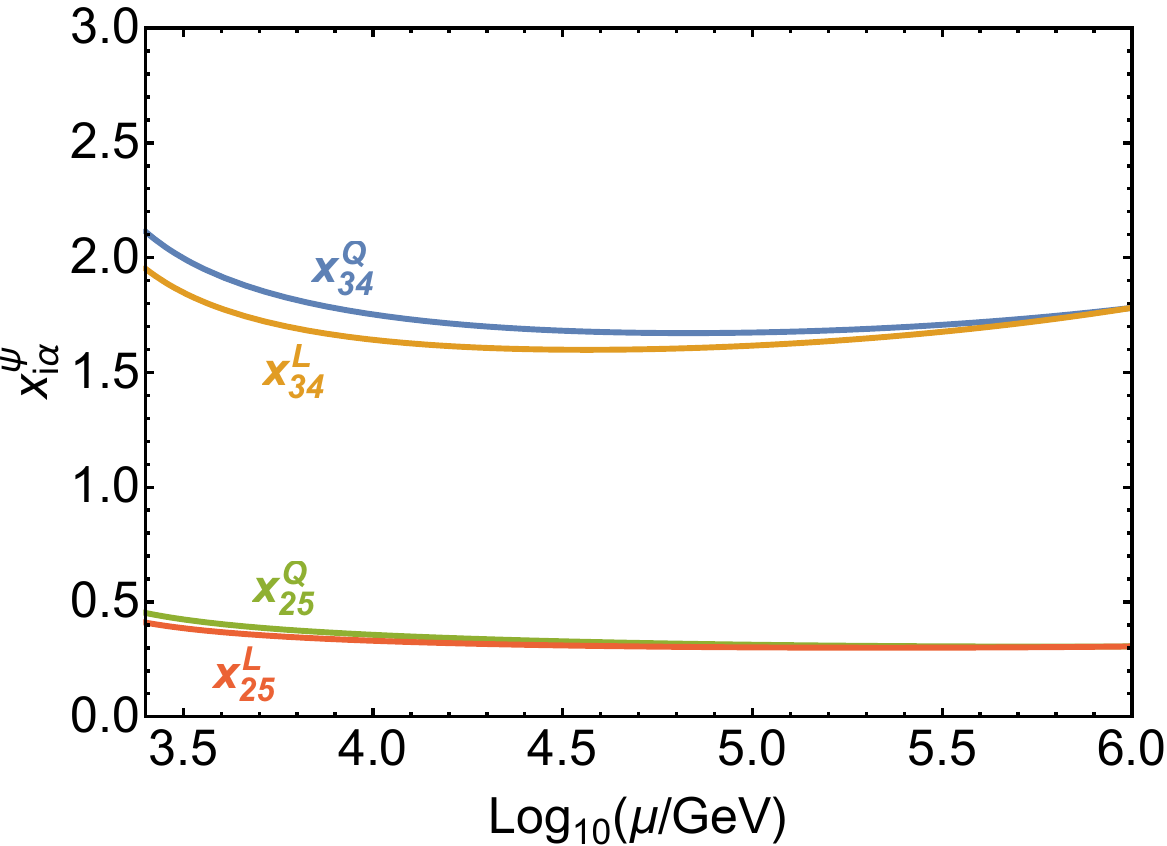}

}$\qquad$\subfloat[\label{fig:Lambda_15}]{\includegraphics[scale=0.4]{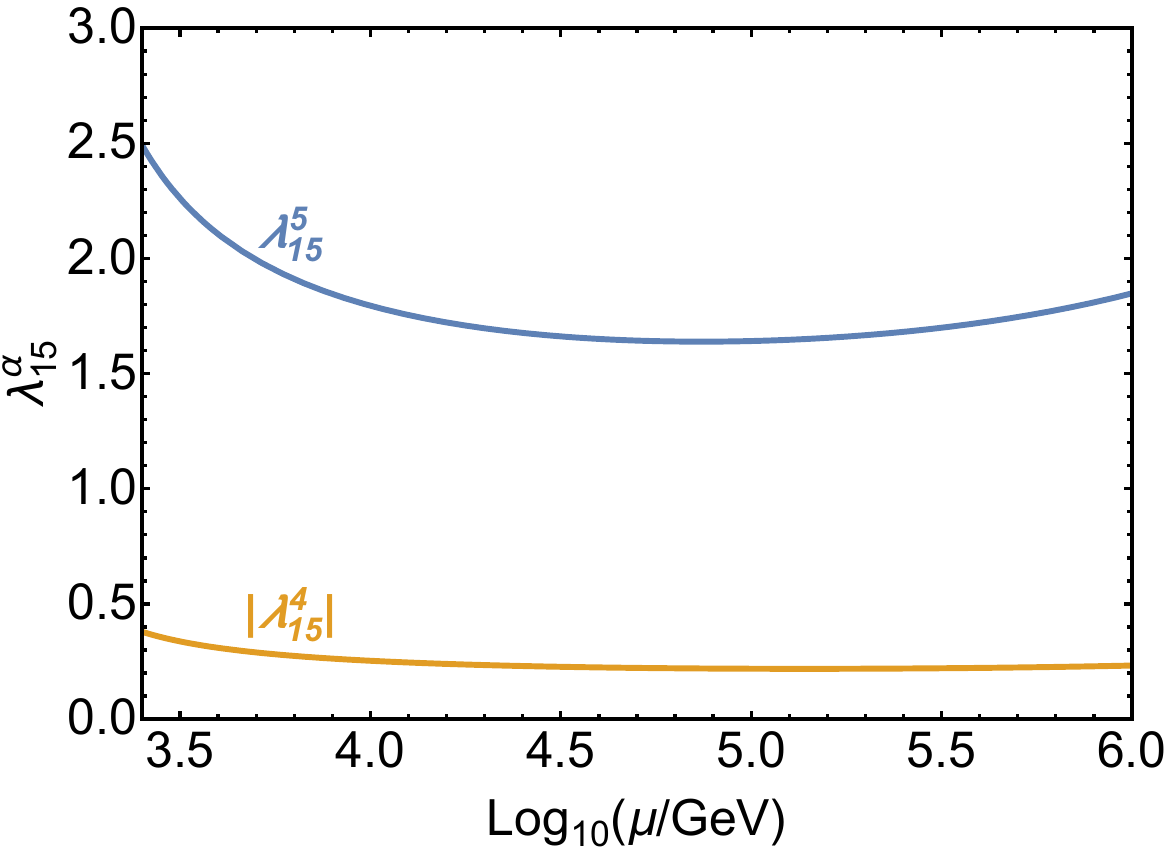}

}
\par\end{centering}
\caption{RGE of the fundamental Yukawa couplings in our benchmark scenario
(Table~\ref{tab:BP}) from the TeV scale to the scale of the twin
Pati-Salam symmetry $\mu\sim1\,\mathrm{PeV}$. The left panel shows
the $x_{i\alpha}^{\psi}$ Yukawas which lead to the mixing between
SM fermions and vector-like partners. The right panel shows the $\lambda_{15}$
Yukawas which split the vector-like masses of quarks and leptons.\label{fig:Perturbativity_Yukawas}}
\end{figure}

Fig.~\ref{fig:Perturbativity_Yukawas} shows that the Yukawas of
our benchmark scenario remain perturbative up to the high energy scale
$\mu\approx1\,\mathrm{PeV}$, thanks to the choice of a large $g_{4}=3.5$.
However, we have checked that the Landau pole is hit when $x_{34}^{\psi}>2.5$,
hence this region should be considered as disfavoured by perturbativity.

The small RGE effects that break the PS universality of the Yukawa
couplings are below 8\% in any case, hence the universality of the
couplings is preserved at the TeV scale in good approximation.

\subsubsection{High-$p_{T}$ signatures\label{subsec:Colliders}}

General 4321 models predict a plethora of high-$p_{T}$ signatures
involving the heavy gauge bosons and at least one family of vector-like
fermions, requiring dedicated analyses such as those in \cite{DiLuzio:2018zxy,Baker:2019sli,Cornella:2021sby}.
In particular, our model predicts a similar high-$p_{T}$ phenomenology
as that of \cite{DiLuzio:2018zxy}, which also considers effective
$U_{1}$ couplings via mixing with three families of vector-like fermions.
However, the bounds obtained in the high-$p_{T}$ analysis of \cite{DiLuzio:2018zxy}
are outdated. Moreover, certain differences arise due to the underlying
twin Pati-Salam symmetry in our model, plus the different implementation
of the scalar sector and VEV structure. Furthermore, we anticipate
that some bounds obtained in \cite{Cornella:2021sby,Baker:2019sli}
might be overestimated for our model, as they usually consider large
couplings to right-handed third family fermions, motivating a dedicated
analysis.
\begin{figure}[t]
\noindent \begin{centering}
\subfloat[\label{fig:Spectrum}]{\begin{raggedright}
\begin{tabular}{c}
\includegraphics[scale=0.35]{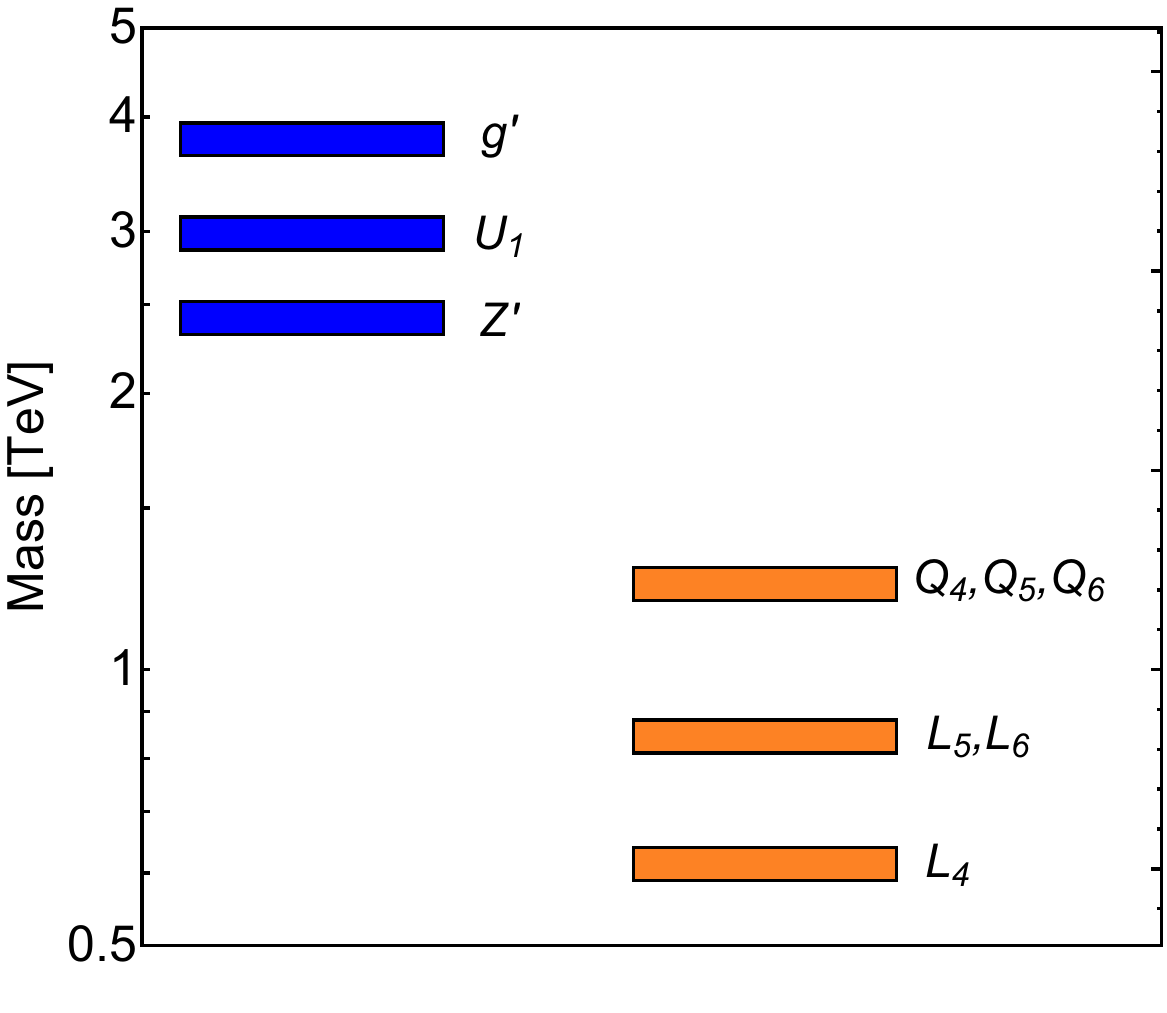}\tabularnewline
\end{tabular}
\par\end{raggedright}
}$\quad$\subfloat[\label{fig:Decay_Channels_Vectors}]{\begin{centering}
\begin{tabular}{cccc}
\toprule 
Particle & Decay mode & $\mathcal{B}(\mathrm{BP})$ & $\Gamma/M$\tabularnewline
\midrule
\midrule 
\multirow{4}{*}{$U_{1}$} & $Q_{3}L_{5}+Q_{5}L_{3}$ & $\sim0.47$ & \multirow{4}{*}{0.32}\tabularnewline
 & $Q_{3}L_{3}$ & $\sim0.22$ & \tabularnewline
 & $Q_{5}L_{5}$ & $\sim0.24$ & \tabularnewline
 & $Q_{i}L_{a}+Q_{a}L_{i}$ & $\sim0.07$ & \tabularnewline
\midrule 
\multirow{4}{*}{$g'$} & $Q_{3}Q_{3}$ & $\sim0.3$ & \multirow{4}{*}{0.5}\tabularnewline
 & $Q_{5}Q_{5}$ & $\sim0.3$ & \tabularnewline
 & $Q_{6}Q_{6}$ & $\sim0.3$ & \tabularnewline
 & $Q_{1}Q_{6}$+$Q_{2}Q_{5}+Q_{3}Q_{4}$ & $\sim0.1$ & \tabularnewline
\midrule 
\multirow{5}{*}{$Z'$} & $L_{5}L_{5}$ & $\sim0.29$ & \multirow{5}{*}{0.24}\tabularnewline
 & $L_{6}L_{6}$ & $\sim0.29$ & \tabularnewline
 & $L_{3}L_{3}$ & $\sim0.27$ & \tabularnewline
 & $Q_{3}Q_{3}$+$Q_{5}Q_{5}+Q_{6}Q_{6}$ & $\sim0.09$ & \tabularnewline
 & $L_{1}L_{6}+L_{2}L_{5}+L_{3}L_{4}$ & $\sim0.06$ & \tabularnewline
\end{tabular}
\par\end{centering}
}
\par\end{centering}
\caption{\textbf{\textit{(Left) }}Spectrum of new vector bosons and fermions
in our benchmark scenario (BP, Table~\ref{tab:BP}) around the TeV
scale. \textbf{\textit{(Right)}} Main decay channels of the new vectors
$U_{1}$, $g'$ and $Z'$ in BP. Addition (+) implies that the depicted
channels have been summed when computing the branching fraction $\mathcal{B}(\mathrm{BP})$.
$i=1,2$ and $a=5,6$.}
\end{figure}

We have included the particle spectrum of our benchmark scenario in
Fig.~\ref{fig:Spectrum}, as a typical configuration for the masses
of the new vectors and fermions. Table~\ref{fig:Decay_Channels_Vectors}
shows the main decay channels of the new vector bosons, which feature
large decay widths $\Gamma/M$ due to all the available decay channels
to vector-like fermions, plus the choice of large $g_{4}=3.5$ close
to perturbativity bounds.

In this section, we revisit some of the most simple collider signals,
such as coloron dijet searches and $Z'$ dilepton searches. We will
also comment on $U_{1}$ searches, coloron ditop searches and vector-like
fermions. We will point out the differences between our framework
and general 4321 models, motivating a future manuscript dedicated
to specific high-$p_{T}$ signals of the twin PS model.

\subsection*{Coloron signals}

The heavy colour octet has a large impact over collider searches for
4321 models, and its production usually sets the lower bound on the
scale of the model. In our case, the heavy coloron has a gauge origin,
hence the coloron couplings to two gluons are absent at tree-level,
reducing the coloron production at the LHC. Moreover, in the motivated
scenario $\left\langle \phi_{3}\right\rangle \gg\left\langle \phi_{1}\right\rangle $,
the coloron is slightly heavier than the vector leptoquark at roughly
$M_{g'}\approx\sqrt{2}M_{U_{1}}$, helping to suppress the impact
of the coloron over collider searches while preserving a slightly
lighter $U_{1}$ for $R_{D^{(*)}}$. In the scenario $g_{4}\gg g_{3,1}$,
the coupling strength of the coloron is roughly $g_{4}$, which receives
NLO corrections via the $K$-factor \cite{Fuentes-Martin:2019ign,Fuentes-Martin:2020luw}

\begin{equation}
K_{\mathrm{NLO}}\approx\left(1+2.65\frac{g_{4}^{2}}{16\pi^{2}}+8.92\frac{g_{s}^{2}}{16\pi^{2}}\right)^{-1/2}\,,\quad g_{g'}\approx K_{\mathrm{NLO}}g_{4}\,.
\end{equation}

\begin{figure}[t]
\subfloat[\label{fig:Coloron_Production}]{\includegraphics[scale=0.4]{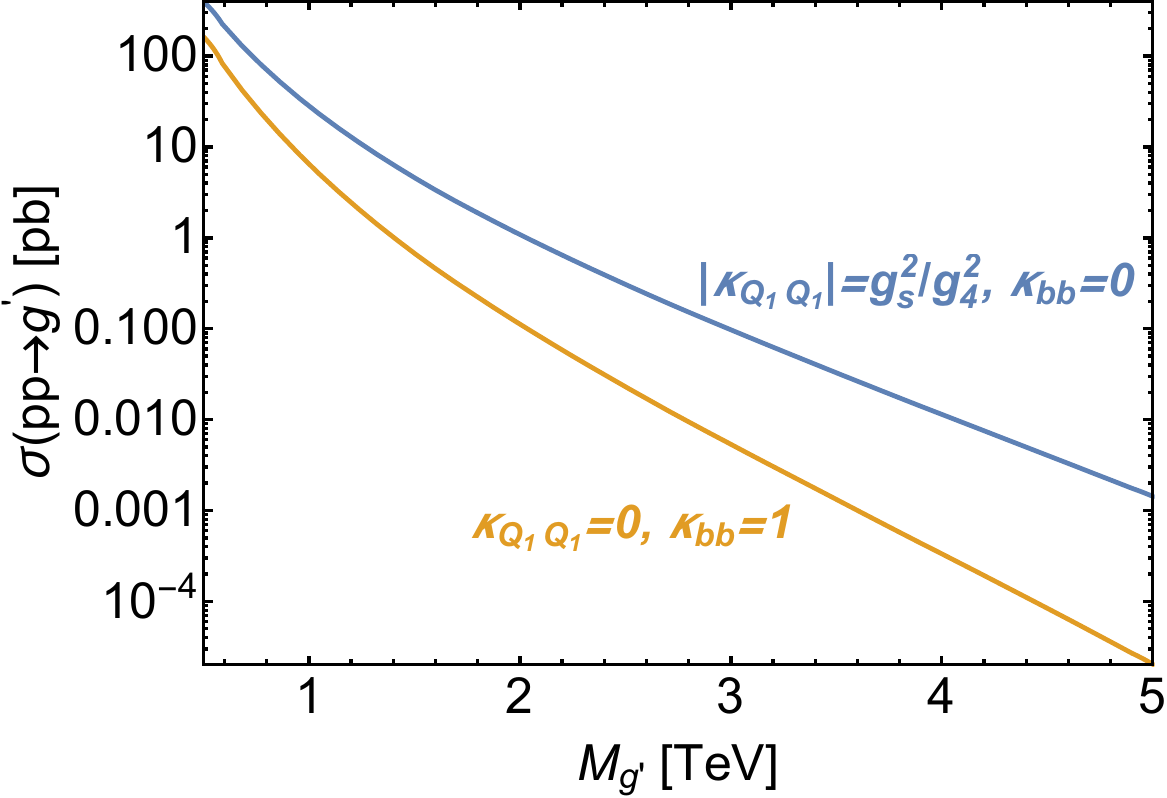}

}$\qquad$\subfloat[\label{fig:Zprime_Production}]{\includegraphics[scale=0.4]{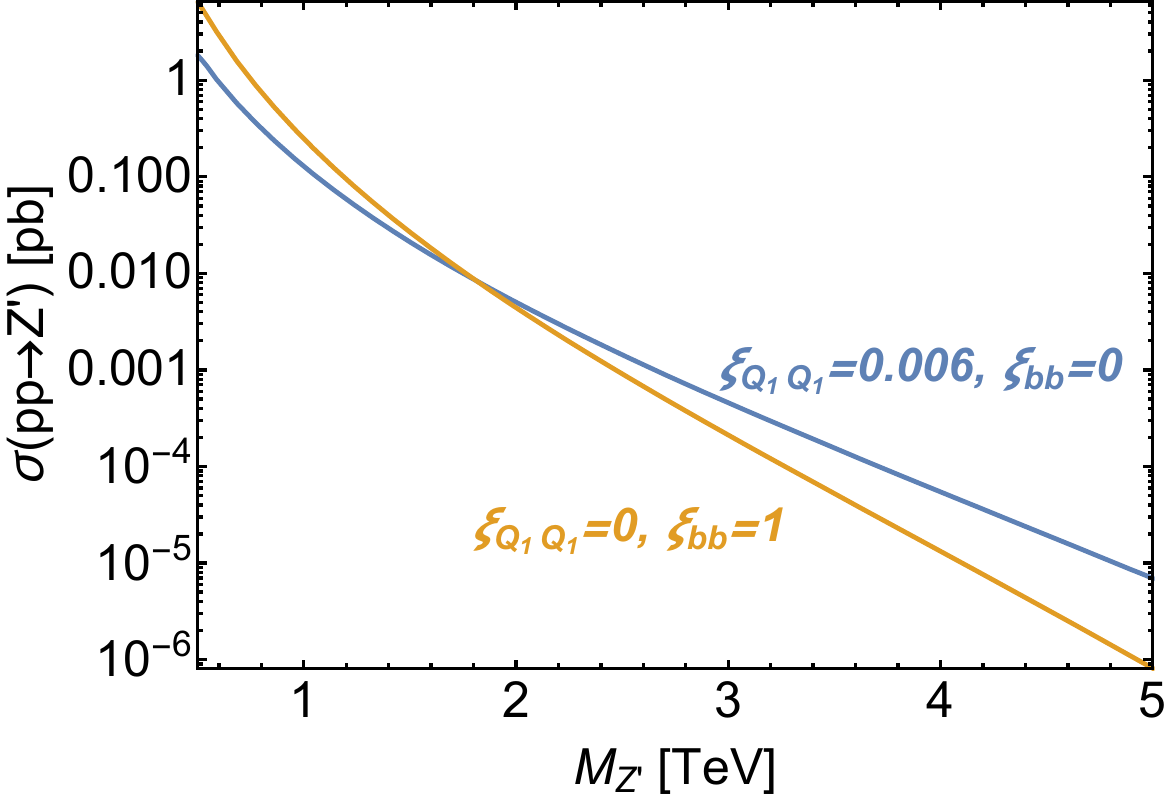}

}

\caption{Production cross sections via 13 TeV $pp$ collisions for the coloron
(left) and $Z'$ (right), via their typical couplings to valence quarks
(blue) and bottoms (orange). The choice of $\xi_{Q_{1}Q_{1}}=0.006$
corresponds to a mixing angle $s_{16}^{Q}\approx0.2$.\label{fig:B_tautau-1}}
\end{figure}

We have computed the coloron production cross section from 13 TeV
$pp$ collisions with \texttt{Madgraph5} \cite{Madgraph:2014hca}
using the default \texttt{NNPDF23LO} PDF set and the coloron UFO model,
publicly available in the \texttt{FeynRules} \cite{Feynrules:2013bka}
model database\footnote{\url{https://feynrules.irmp.ucl.ac.be/wiki/LeptoQuark}}.
We verify in Fig.~\ref{fig:Coloron_Production} that coloron production
is dominated by valence quarks, even though the coupling to left-handed
bottoms is maximal. The coloron couples to light left-handed quarks
(see Eq.~\eqref{eq:Coloron_couplings_3VL}) via the mixing $s_{25}^{Q}\approx s_{16}^{Q}$
of $\mathcal{O}(0.1)$, which interferes destructively with the flavour-universal
term, allowing for a certain cancellation of the left-handed couplings
to light quarks. However, the coloron is still produced via the flavour--universal
couplings to right-handed quarks.

\begin{figure}[t]
\subfloat[\label{fig:Coloron_dijet}]{\includegraphics[scale=0.4]{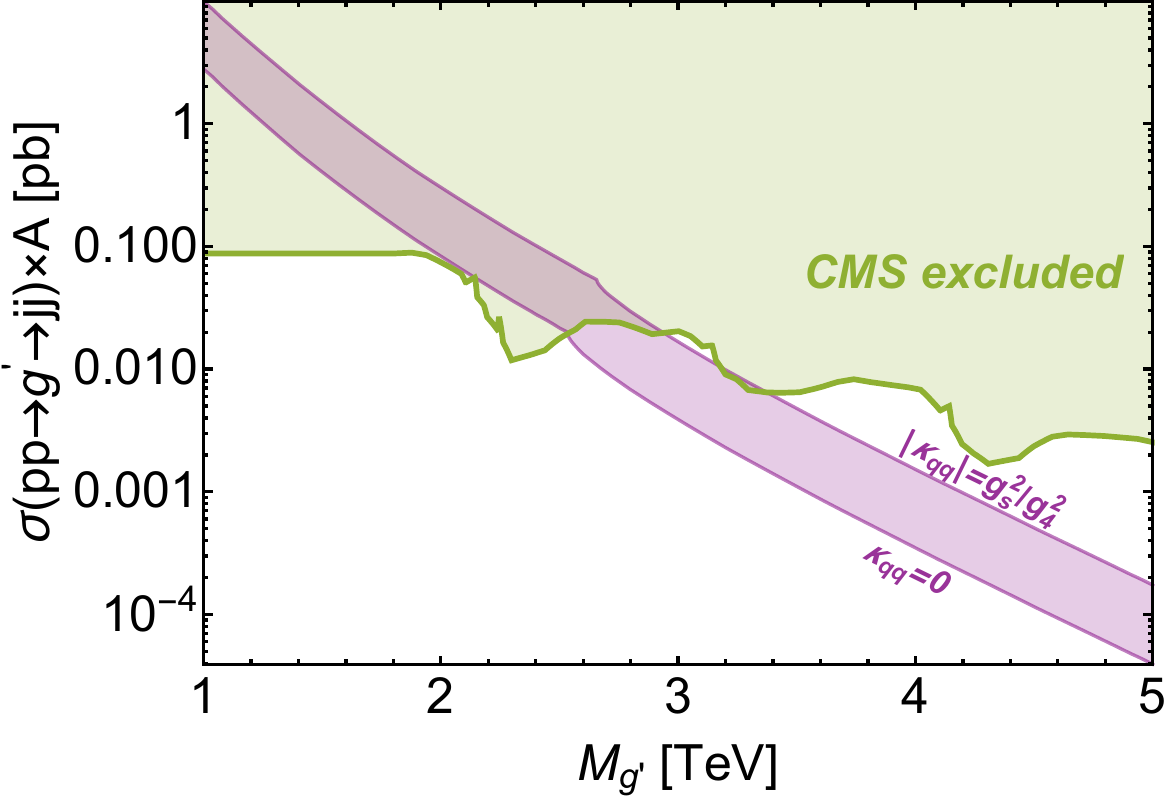}

}$\qquad$\subfloat[\label{fig:Colliders_All}]{\includegraphics[scale=0.39]{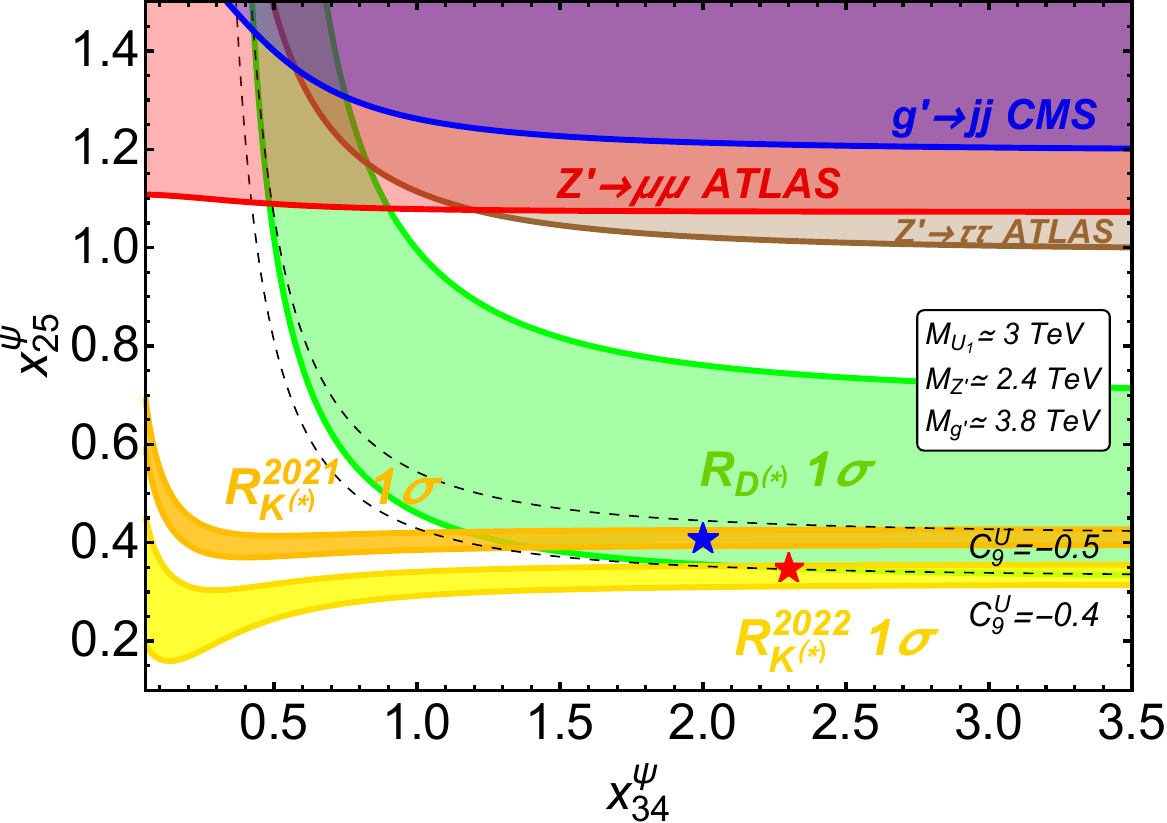}

}

\caption{\textbf{\textit{(Left)}} Total cross section for the coloron dijet
channel in the narrow width approximation, with $\left|\kappa_{qq}\right|$
varied in the range $\left|\kappa_{qq}\right|=[0,g_{s}^{2}/g_{4}^{2}]$,
where $q=Q_{1},Q_{2}$. The rest of parameters are fixed as in Table~\ref{tab:BP}
for both panels. The exclusion bound from CMS is shown in green. \textbf{\textit{(Right)}}
Parameter space in the plane ($x_{34}^{\psi}$, $x_{25}^{\psi}$)
compatible with the LFU ratios. The dashed lines show contours of
constant $C_{9}^{U}$. The regions excluded by the collider searches
considered are included. The blue (red) star shows BP1 (BP2).}
\end{figure}

We estimate analytically the branching fraction to all SM quarks excluding
tops, and then we compute the total cross section via the narrow width
approximation. Finally, we confront our results with the limits for
a $q\bar{q}$-initiated spin-1 resonance provided by CMS in Fig.~10
of \cite{CMS:2019gwf}. The results are displayed in Fig.~\ref{fig:Coloron_dijet},
where we have varied the coupling to light LH quarks $\kappa_{qq}$
and fixed the rest of parameters as in Table~\ref{tab:BP}. We find
bounds ranging from $M_{g'}\apprge2.5\,\mathrm{TeV}$ when $\kappa_{qq}\approx0$
and $M_{g'}\apprge3\,\mathrm{TeV}$ when $\kappa_{qq}\approx g_{s}^{2}/g_{4}^{2}$.
These bounds are slightly milder than those obtained in \cite{Cornella:2021sby},
the reason being that in \cite{Cornella:2021sby} right-handed bottom
quarks are assumed to couple maximally to the coloron, while in our
model this coupling is suppressed.

We expect to find more stringent bounds in resonant coloron production
with $t\bar{t}$ final states, due to the maximal couplings of the
coloron to the third generation EW quark doublet. According to the
recent analysis in \cite{Cornella:2021sby}, our benchmark scenario
would lie below current bounds, due to the large decay width $\Gamma_{g'}/M_{g'}\approx0.5$
provided by extra decay channels to TeV scale vector-like quarks.
The limit over the coloron mass is roughly 3.5 TeV, however this bound
might be overestimated again for our model due to the different description
of RH third family quarks. Reconstructing the $t\bar{t}$ channel
requires a dedicated analysis and a different methodology, which is
beyond the scope of this article, and we leave it for a future work.

\subsection*{$Z'$ signals}

For the $Z'$ boson, the flavour-universal couplings to valence quarks
are more heavily suppressed than those of the coloron, via the small
ratio $g_{Y}^{2}/g_{4}^{2}$. Therefore, cancellation between the
left-handed mixing term proportional to $s_{25}^{Q}\approx s_{16}^{Q}$
and the flavour-universal one is not possible here. In contrast with
the coloron, the large LH couplings to bottoms can play a role in
$Z'$ production. The production cross section is estimated via the
same methodology as for the coloron above. We do not consider any
NLO corrections in this case, following the methodology of \cite{Baker:2019sli}.
In Fig.~\ref{fig:Zprime_Production} we show that the production
via bottoms is larger than the production via valence quarks for a
light $Z'$, however the production via valence quarks is bigger for
$M_{Z'}\apprge2\,\mathrm{TeV}$, and shall not be neglected as it
commonly happens in the literature (see e.g.~\cite{Baker:2019sli}).

\begin{figure}[t]
\subfloat[\label{fig:Zprime_muons}]{\includegraphics[scale=0.41]{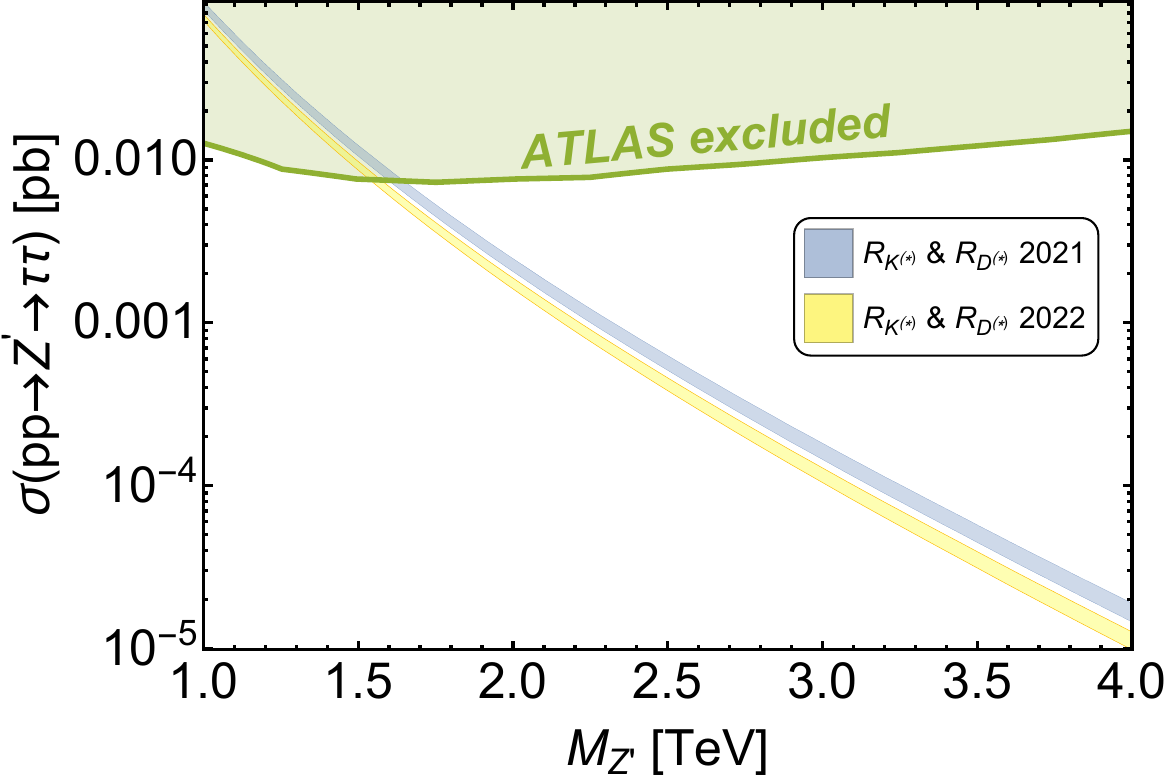}

}$\qquad$\subfloat[\label{fig:Zprime_taus}]{\includegraphics[scale=0.39]{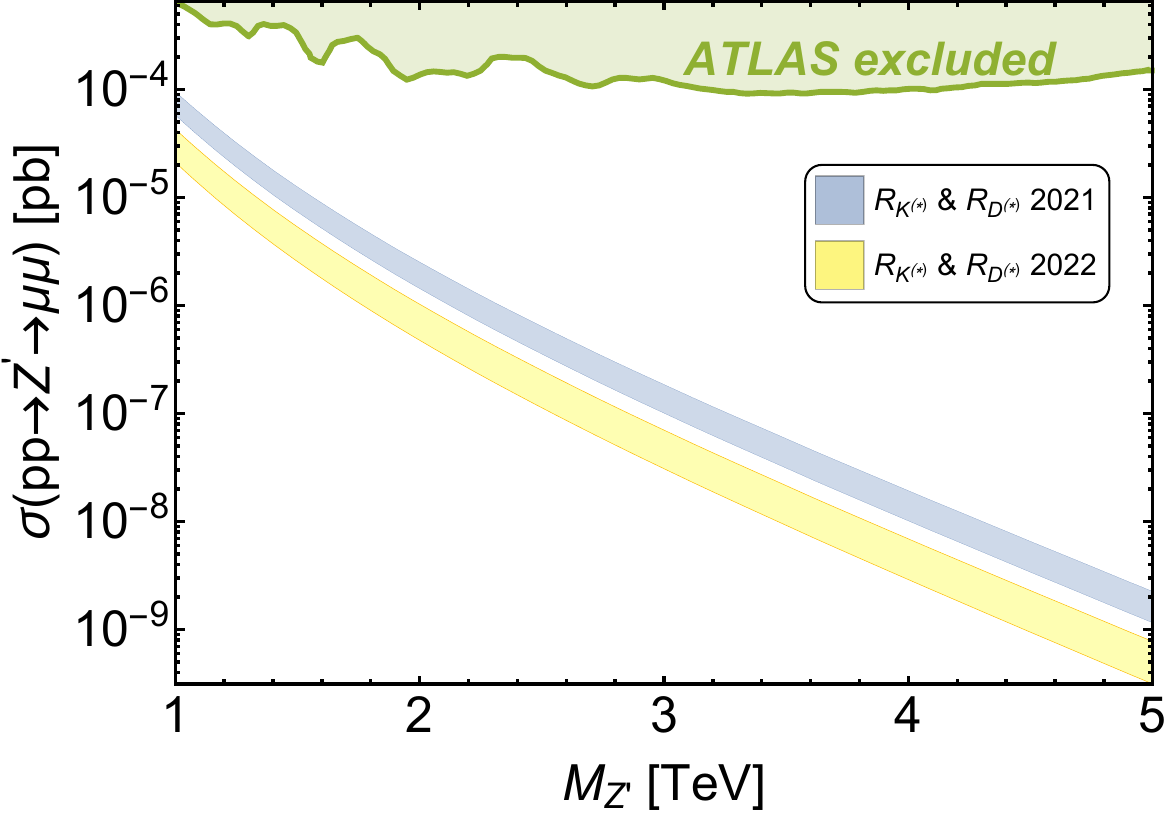}

}

\caption{Total cross section for ditau (left) and dimuon (right) production
via a heavy $Z'$ in the narrow width approximation, with $x_{25}^{\psi}$
varied in the range $x_{25}^{\psi}=[0.3,\,0.35]$ ($[0.4,\,0.45]$)
preferred by $R_{K^{(*)}}^{2022}$ ($R_{K^{(*)}}^{2021}$ ), obtaining
the yellow (blue) band. The exclusion bounds from ATLAS are shown
in green.}
\end{figure}

We estimate the branching fraction to muons and taus, and we compute
the total decay width via the narrow width approximation. We confront
our results with the limits from the dilepton resonance searches by
ATLAS, Fig.~4 of \cite{ATLAS:2017fih} for muons and Fig.~7 (c)
of \cite{ATLAS:2017eiz} for taus. We display the results in Figs.~\ref{fig:Zprime_muons}
and \ref{fig:Zprime_taus}. In Fig.~\ref{fig:Colliders_All} we see
that these processes, along with coloron dijet searches, mildly constrain
the region of large $x_{25}^{\psi}$. Ditau searches are more competitive
than dimuon searches or coloron dijet searches, due to the branching
fractions to muons and light quarks being suppressed by mixing angles
$s_{25}^{Q,L}\sim\mathcal{O}(0.1)$. Instead, the ditau channel is
enhanced by maximal 3-4 mixing, and sets bounds of roughly $M_{Z'}>1.5\,\mathrm{TeV}$,
see Fig.~\ref{fig:Zprime_taus}.

\subsection*{$U_{1}$ signals}

Leptoquark pair-production cross sections at the LHC are dominated
by QCD dynamics, and thus are largely independent of the leptoquark
couplings to fermions. Therefore, we are able to safely compare with
the analyses of Refs.~\cite{Cornella:2021sby,Baker:2019sli}. A certain
model dependence is present in the form of non-minimal couplings to
gluons, however these couplings are absent in models where $U_{1}$
has a gauge origin. According to Fig.~3.3 of \cite{Cornella:2021sby},
current bounds over direct production exclude $M_{U_{1}}<1.7\,\mathrm{TeV}$,
and the future bound is expected to exclude $M_{U_{1}}<2.1\,\mathrm{TeV}$
if no NP signal is found during the high-luminosity phase of the LHC.

An important constraint over $U_{1}$ arises from modifications of
the high-$p_{T}$ tail in the dilepton invariance mass distribution
of the Drell-Yan process $pp\rightarrow\tau^{+}\tau^{-}+X$, induced
by $t$-channel $U_{1}$ exchange \cite{Diaz:2017lit,Baker:2019sli,Angelescu:2021lln,Cornella:2021sby}.
This channel is well motivated by the $U_{1}$ explanation of $R_{D^{(*)}}$,
which unavoidably predicts a large $b\tau U_{1}$ coupling. The scenario
$\beta_{b\tau}^{R}=0$ considered in the study of \cite{Baker:2019sli,Cornella:2021sby}
fits well the twin Pati-Salam framework, up to a re-scaling of the
$U_{1}$ coupling strength as $g_{U}\rightarrow g_{U}\beta_{b\tau}^{L}$,
in order to account for the fact that our $\beta_{b\tau}^{L}$ is
not maximal but $\beta_{b\tau}^{L}\approx c_{\theta_{LQ}}\approx0.67$
in our benchmark scenario, where we obtain $g_{U}\approx2.3$. According
to the left panel of Fig.~3.3 in \cite{Cornella:2021sby}, the 3
TeV leptoquark of our benchmark lies well below the current bounds,
but within projected limits for the high luminosity phase of LHC.
Finding $U_{1}$ much below 3 TeV enters in tension with $pp\rightarrow g'\rightarrow t\bar{t}$
as explained before, due to the approximate relation $M_{g'}\approx\sqrt{2}M_{U_{1}}$
that entangles the masses of $U_{1}$ and the coloron (although the
bound is probably overestimated for our dominantly left-handed model). 

The twin Pati-Salam model could provide a good $U_{1}$ candidate
for the 3$\sigma$ excess at CMS \cite{CMS-PAS-EXO-19-016} pointing
to a 2 TeV $U_{1}$ leptoquark in the well motivated channel $pp\rightarrow U_{1}\rightarrow\tau\tau$,
once the extra decay channels to vector-like fermions are considered,
assuming that the bound from $pp\rightarrow g'\rightarrow t\bar{t}$
is indeed overestimated.

\subsection*{Vector-like fermions}

The presence of vector-like fermions is of fundamental importance
to discriminate between the different implementations of the 4321
model addressing the $B$-anomalies. A common constraint arises from
$\Delta F=2$ transitions at low energies, which require that the
vector-like charged lepton that mixes with muons is light (see Fig.~\ref{fig:ML5_deltaMs}).
The natural mass of the quark partner of $L_{5}$ should not lie far
away due to the approximate Pati-Salam universality, the small breaking
effects given by the VEV $\left\langle \Omega_{15}\right\rangle $.
In particular, in our benchmark scenario we obtained $M_{5}^{L}\approx0.8\,\mathrm{TeV}$
and $M_{5}^{Q}\approx1.2\,\mathrm{TeV}$. The flavour structure of
the model naturally predicts that both $Q_{5}$ and $L_{5}$ are strongly
coupled to the third generation of SM fermions.

The twin Pati-Salam model features also $L_{4}$ and $Q_{4}$ as a
relevant pair of vector-like fermions, which mix maximally with the
third generation in order to obtain the large couplings required for
$R_{D^{(*)}}$, and also to fit the top mass without perturbativity
issues. This implies that their bare mass terms in the original Lagrangian
are small, therefore their physical masses are dominated by $x_{34}^{\psi}\left\langle \phi_{3,1}\right\rangle $,
see Eq.~\eqref{eq:34_mixing_extended} and \eqref{eq:Mass-_4th}.
In the motivated scenario $\left\langle \phi_{3}\right\rangle \gg\left\langle \phi_{1}\right\rangle $
which slightly suppresses the production of the coloron, we found
$L_{4}$ to be very light, featuring roughly 600 GeV in our benchmark.
Instead, $Q_{4}$ can lie above 1 TeV, featuring roughly 1.2 TeV in
our benchmark. The couplings of $L_{4}$ to SM fermions are smaller
than those of $L_{5}$, but it is dominantly coupled to third generation
fermions.

Interestingly, CMS recently performed a search for the vector-like
leptons of the 4321 model \cite{CMS:2022cpe}, finding 2.8$\sigma$
preference for a vector-like lepton with 600 GeV, however the analysis
assumes EW production only and maximal couplings to the third generation.
If $Z'$-assisted production is included, $L_{5}$ a with mass at
800 GeV could be a good candidate for the anomaly. Furthermore, $L_{4}$
at 600 GeV could also provide a good fit once not maximal couplings
are considered, however this requires verification in a future dedicated
analysis. Non-fermiophobic 4321 models, such as \cite{Cornella:2019hct,Cornella:2021sby},
predict a heavier vector-like lepton, while \cite{DiLuzio:2018zxy}
also predicts $L_{5}$ at around 800 GeV but a heavier $L_{4}$. Regarding
the sixth vector-like fermions $L_{6}$ and $Q_{6}$, we expect them
to have similar masses as $L_{5}$ and $Q_{5}$ in order to preserve
the GIM-like suppression of 1-2 FCNCs, they are feebly coupled to
first generation fermions but not to the second or third generation.

Current bounds on vector-like quarks lie around 1 TeV, however they
are usually very model dependent. Our vector-like quarks are pair
produced through gluon fusion and through the decay of the coloron,
which is very likely to be kinematically allowed. Their decays leave
a large amount of third generation fermions in the final state, following
a similar mechanism as the one discussed in \cite{DiLuzio:2018zxy}.
The twin Pati-Salam model naturally predicts light vector-like quarks
with masses around 1 TeV, which is a feature not present in all other
4321 models and could motivate specific searches.

\section{Comparison with other models\label{sec:Comparison_models}}

Table~\ref{tab:Comparison_Models} includes a simplified set of observables
that allows to disentangle the twin PS model from the ones that are
already in the market. A further discussion can be found in the two
following sections. 

\subsection{Non-fermiophobic 4321 models (including the $\mathrm{PS}^{3}$ model)}

The twin PS model is built as a fermiophobic framework, where all
chiral fermions are singlets under the TeV scale $SU(4)$. This is
a crucial difference between our model and the non-fermiophobic 4321
models \cite{Cornella:2019hct,Cornella:2021sby,Barbieri:2022ikw,Fuentes-Martin:2020bnh,Fuentes-Martin:2020hvc}
and its UV completions (including the $\mathrm{PS}^{3}$ model \cite{Bordone:2017bld}),
where the third family of chiral fermions transforms under the TeV
scale $SU(4)$. This implies large left- and right-handed third family
couplings to $SU(4)$ gauge bosons. By contrast, in our theory of
flavour, the right-handed couplings of SM fermions to $U_{1}$ (and
also to $Z'$ and $g'$) arise via small mixing angles connected to
the origin of second family fermion masses, hence the twin PS model
predicts dominantly left-handed $U_{1}$ couplings. The low-energy
phenomenology between both approaches is radically different. 
\begin{table}[t]
\begin{tabular}{ccccc}
\toprule 
 & twin PS & fermiophobic 4321 & $\mathrm{PS}^{3}$ & non-fermiophobic 4321\tabularnewline
\midrule
\midrule 
Refs. & this paper & \cite{DiLuzio:2017vat,DiLuzio:2018zxy} & \cite{Bordone:2017bld,Bordone:2018nbg} & \cite{Cornella:2019hct,Cornella:2021sby,Barbieri:2022ikw,Greljo:2018tuh}\tabularnewline
\midrule 
Theory of flavour & Yes & No & Yes & No\tabularnewline
\midrule 
$R_{D^{(*)}}$ & $\Delta R_{D}=\Delta R_{D^{*}}$ & $\Delta R_{D}=\Delta R_{D^{*}}$ & $\Delta R_{D}>\Delta R_{D^{*}}$ & $\Delta R_{D}>\Delta R_{D^{*}}$\tabularnewline
\midrule 
$\mathcal{B}\left(\tau\rightarrow3\mu\right)$ & $10^{-8}$ & $\lesssim10^{-11}$ & $10^{-9}$ & -\tabularnewline
\midrule 
$\mathcal{B}\left(\tau\rightarrow\mu\gamma\right)$ & $10^{-9}$ & $\lesssim10^{-11}$ & $10^{-8}$ & $10^{-8}$\tabularnewline
\midrule 
$\mathcal{B}\left(\tau\rightarrow\mu\phi\right)$ & $10^{-9}$ & $10^{-11}$ & $10^{-10}$ & $10^{-10}$\tabularnewline
\midrule 
$\mathcal{B}\left(B_{s}\rightarrow\tau\mu\right)$ & $10^{-6}$ & \multicolumn{1}{c}{$10^{-7}$} & $10^{-5}$ & $10^{-5}$\tabularnewline
\midrule 
$\mathcal{B}\left(B_{s}\rightarrow\tau\tau\right)$ & $10^{-4}$ & - & $10^{-3}$ & $10^{-3}$\tabularnewline
\midrule 
$\mathcal{B}\left(B\rightarrow K\tau\tau\right)$ & $10^{-5}$ & - & $10^{-4}$ & $10^{-4}$\tabularnewline
\midrule 
$\delta\mathcal{B}(B\rightarrow K^{(*)}\nu\bar{\nu})$ \eqref{eq:BtoK_nunu} & $0.3$ & - & $0.2$ & $0.2$\tabularnewline
\midrule 
VL fermion families & 3 & 3 & 1 & 1\tabularnewline
\midrule 
High-$p_{T}$ constraints & Mild & Mild & Tight & Tight\tabularnewline
\bottomrule
\end{tabular}\caption{Main observables to distinguish the twin PS model from other proposals.
The numbers are only indicative, as these predictions may vary along
the parameter space of the different models. The dash (-) indicates
that the observable was not considered or numbers were not given in
the corresponding references. In the high-$p_{T}$ row we broadly
refer to how constrained is the model by high-$p_{T}$ searches.\label{tab:Comparison_Models}}
\end{table}

In terms of the charged current anomalies $R_{D^{(*)}}$, the twin
PS model only predicts the effective operator $(\bar{c}_{L}\gamma_{\mu}b_{L})(\bar{\text{\ensuremath{\tau}}}_{L}\gamma^{\mu}\nu_{\tau L})$,
and hence both $R_{D}$ and $R_{D^{*}}$ are corrected in the same
direction and with the same size. Instead, non-fermiophobic 4321 models
also predict the scalar operator $(\bar{c}_{L}b_{R})(\bar{\text{\ensuremath{\tau}}}_{R}\nu_{\tau L})$.
Due to the presence of this operator, the effect on $\Delta R_{D}$
is larger than on $\Delta R_{D^{*}}$ (about 5/2 larger for the $\mathrm{PS}^{3}$
model, see Eq.~(27) in \cite{Bordone:2017bld}).

Another key observable is $B\rightarrow K\nu\nu$, for which the twin
PS model predicts a larger branching fraction that will be fully tested
by Belle II. Instead, non-fermiophobic 4321 models predict a smaller
branching fraction, see Fig.~4.4 of \cite{Cornella:2021sby} and
compare their purple region with our Fig.~\ref{fig:BtoK_nunu}. Moreover,
in our analysis of $B\rightarrow K\nu\nu$ we have highlighted correlations
with $B_{s}-\bar{B}_{s}$ mixing due to the loops being dominated
by the same VL charged lepton, a feature which is missing in the analysis
of \cite{Cornella:2021sby}.

Regarding the rest of the observables, broadly speaking the twin PS
model predicts larger branching fractions for LFV processes. The exception
is $\tau\rightarrow\mu\gamma$, which is enhanced in non-fermiophobic
models via a chirality flip with the bottom quark running in the loop.
The rare decays $B_{s}\rightarrow\tau\tau$ and $B\rightarrow K\tau\tau$
are larger in non-fermiophobic models due to the presence of scalar
operators connected to the third family RH couplings. The LHCb and
Belle II collaborations will test regions of the parameter space and
allow to disentangle between the different 4321 approaches. 

High-$p_{T}$ searches also offer a window to disentangle both approaches,
since most of the constraints afflicting non-fermiophobic 4321 scenarios
are relaxed in the dominantly left-handed scenario of the twin PS
model. Particularly relevant are also the different implementations
of vector-like fermions.

Regarding the $\mathrm{PS}^{3}$ model, it predicts the same signals
as non-fermiophobic 4321 models plus enhanced LFV. It can be disentangled
from the twin PS model via the set of observables already discussed.

\subsection{Fermiophobic 4321 models}

Up to our knowledge, the only fermiophobic 4321 model proposed in
the literature is that of Ref.~\cite{DiLuzio:2017vat}, whose phenomenology
was studied in detail in \cite{DiLuzio:2018zxy}. This model presents
a simplified fermiophobic scenario with a rather assumed flavour structure
motivated by the phenomenology, including an ad-hoc alignment of SM-like
Yukawas and VL-chiral fermion mixing. Furthermore, \cite{DiLuzio:2017vat}
does not address the question of quark and lepton masses (is not a
theory of flavour), unlike the model proposed here. It lacks from
quark-lepton unification of SM fermions and leads to a less predictive
framework than the twin PS model.

The twin Pati-Salam model leads to an effective fermiophobic 4321
model. However, the underlying twin PS symmetry implies universality
of key parameters, leading to extra constraints and correlations between
observables, which are not present in the analyses of \cite{DiLuzio:2017vat,DiLuzio:2018zxy}.
For example, $R_{D^{(*)}}$ and $R_{K^{(*)}}$ are correlated here
due to the universality of $x_{25}^{\psi}$ and $x_{34}^{\psi}$,
leading to quasi-universal mixing angles $s_{25}^{Q,L}$ and $s_{34}^{Q,L}$.
By contrast, such mixing angles are free parameters in \cite{DiLuzio:2017vat,DiLuzio:2018zxy}
and one can explain $R_{D^{(*)}}$ without giving any contribution
to $R_{K^{(*)}}$. As a consequence, it was required a dedicated analysis
to show that $R_{D^{(*)}}$ can be explained in the twin PS model
while being compatible with the recent data on $R_{K^{(*)}}$, as
we did in this paper.

Since the twin PS model is a theory of flavour, while \cite{DiLuzio:2017vat,DiLuzio:2018zxy}
is not, new signals are predicted in LFV processes connected to the
origin of fermion masses and mixings. The twin PS model predicts non-vanishing
$\text{\ensuremath{\mu-\tau}}$ mixing, leading to striking signals
in $\tau\rightarrow3\mu$ and $\tau\rightarrow\mu\gamma$ close to
current experimental bounds, summarised in our Figs.~\ref{fig:tau_3mu}
and \ref{fig:tau_muphoton}. The large contributions to the branching
fractions of $\tau\rightarrow3\mu$ and $\tau\rightarrow\mu\gamma$
are mediated by the $Z'$ boson, see the purple region in Figs.~\ref{fig:tau_3mu}
and \ref{fig:tau_muphoton}. On the other hand, in general fermiophobic
4321 models \cite{DiLuzio:2017vat,DiLuzio:2018zxy} only a much smaller
1-loop $U_{1}$ mediated signal is predicted. This signal was not
computed in Refs.~\cite{DiLuzio:2017vat,DiLuzio:2018zxy} as it is
very small compared to the experimental bounds, but we have computed
it here for the sake of comparison, and it is depicted as the blue
region in Figs.~\ref{fig:tau_3mu} and \ref{fig:tau_muphoton}. In
a similar way, we obtain $\mathcal{B}(\tau\rightarrow\mu\phi)$ two
orders of magnitude larger than in \cite{DiLuzio:2017vat,DiLuzio:2018zxy}.

Finally, the fermion mixing predicted by the twin PS model avoids
current constraints coming from CKM unitarity, $\Delta F=2$ and EW
precision observables presented in \cite{Fajfer:2013wca}. The reasons
are the absence of SM-like Yukawa couplings for chiral fermions in
the original basis (as they will be generated indeed via this mixing),
along with the fact that VL quark EW doublets and SM quark EW singlets
do not mix, hence the VL quark doublet remains unsplitted. Remarkably,
this is different from \cite{DiLuzio:2017vat,DiLuzio:2018zxy}, where
mixing between the SM quark singlets and the VL (right-handed) quark
doublet was induced due to the presence of the SM-like Yukawa couplings
for chiral fermions, leading to possible splitting of the VL quark
doublet, which constrains the mixing angles for third generation quarks
according to the analysis in \cite{Fajfer:2013wca}.

\section{Conclusions\label{sec:Conclusions}}

We have performed a comprehensive phenomenological analysis of the
twin Pati-Salam model, which is capable of explaining anomalies in
LFU ratios of $B$-decays, while simultaneously accounting for the
fermion masses and mixings of the SM. The basic idea of this model
is that all three families of SM chiral fermions transform under one
PS group, while families of vector-like fermions transform under the
other one. Vector leptoquark couplings and SM Yukawa couplings emerge
together after mixing of the chiral fermions with the vector-like
fermions, thereby providing a direct link between $B$-physics and
fermion masses and mixings. The model was originally built to explain
the 2021 picture of LFU anomalies in the $R_{K^{(*)}}$ and $R_{D^{(*)}}$
ratios. In this updated version we have included an extended analysis
considering the new 2022 LHCb data on $R_{K^{(*)}}$, which has shifted
the preferred parameter space with respect to the 2021 case. The model
can still explain the $R_{D^{(*)}}$ anomalies at 1$\sigma$ in a
narrow window while being compatible with all data, however we expect
small deviations from the SM on the $R_{K^{(*)}}$ ratios, to be tested
in the future via more precise measurements of LFU by the LHCb collaboration.
We also predict $\Delta R_{D}=\Delta R_{D^{*}}$, with future measurements
shifting the world averages to slightly smaller central values.

Firstly, we presented a simplified version of the model that is able
to explain second and third family charged fermion masses and mixings
via effective Yukawa couplings, which arise naturally from mixing
effects with a fourth vector-like family of fermions. However, with
only a single vector-like family the model is unable to explain the
$B$-anomalies in a natural way, as it does not achieve the flavour
structure required by 4321 models, and hence is over constrained by
flavour-violating processes such as $B_{s}-\bar{B}_{s}$ meson mixing.
The latter are mediated by a heavy colour octet and a $Z'$ that also
acquire flavour-violating couplings with chiral fermions. 

We then extended the simplified model to include three vector-like
families, together with a $Z_{4}$ discrete symmetry to control the
flavour structure. This version of the model allows for larger flavour-violating
and dominantly left-handed $U_{1}$ couplings as required to address
$R_{D^{(*)}}$, thanks to mixing between a fourth and fifth vector-like
families which also mix with the second and third generations of SM
fermions. A sixth vector-like family is included to mix with the first
SM generation, for the sake of suppressing any FCNCs involving first
and second generation fermions. The mechanism resembles the GIM suppression
of FCNCs in the SM, featuring a similar Cabbibo-like matrix which
is present in leptoquark currents, but not in neutral currents mediated
by the coloron and $Z'$.

As in the simplified twin Pati-Salam model, the origin of second and
third generation charged fermion masses and mixings remains addressed
via effective Yukawa couplings, featuring now a down-aligned flavour
structure in the 2-3 sector (requiring a mild tuning as described
in the main text) that protects from the dangerous tree-level contributions
to $B_{s}-\bar{B}_{s}$ meson mixing. Non-zero 2-3 mixing in the charged
lepton sector is also predicted, leading to interesting signals in
$\tau\rightarrow3\mu$ and $\tau\rightarrow\mu\gamma$, mostly due
to $Z'$ exchange, which are close to present experimental bounds
in some region of the parameter space. Signals in LFV semileptonic
processes mediated by $U_{1}$ at tree-level are found to lie well
below current experimental limits, with the exception of $K_{L}\rightarrow e\mu$
which constrains a small region of the parameter space. However, this
tension can be alleviated if the first family $U_{1}$ coupling is
diluted via mixing with vector-like fermions. Tests of LFU in tau
decays set important bounds over the mass of $U_{1}$ depending on
its coupling to third generation fermions. Contributions of $U_{1}$
to the rare decays $B_{s}\rightarrow\tau\tau$ and $B\rightarrow K\tau\tau$
are broadly below current and projected experimental sensitivity.
Instead, the rare decay $B\rightarrow K^{(*)}\nu\bar{\nu}$ offers
the opportunity to fully test the model in the near future, since
Belle II is expected to cover all the parameter space compatible with
the LFU ratios. Remarkably, the model can be easily disentangled from
all other proposals via the previous set of observables, as discussed
in Section~\ref{sec:Comparison_models}.

Apart from the above low-energy predictions at LHCb and Belle II,
the model is also testable via high-$p_{T}$ searches at the LHC.
The study of the 1-loop contribution of vector leptoquark $U_{1}$
exchange to $B_{s}-\bar{B}_{s}$ mixing revealed that the fifth vector-like
lepton has to be light, around 1-2 TeV, to be compatible with the
stringent bound from $\Delta M_{s}$. This is easily achieved in the
twin Pati-Salam model, where light vector-like fermions are well motivated
in order to naturally obtain the large mixing to fit the $R_{D^{(*)}}$
anomaly, and also to fit the heavy top mass without perturbativity
issues. In particular, the fourth and fifth charged leptons are suggested
as good candidates to explain the CMS excess \cite{CMS:2022cpe},
but further study is required in this direction. Vector-like quarks
are found to lie not far above 1 TeV in the suggested benchmark, hence
motivating specific searches at LHC to be performed. Regarding the
heavy vectors, dijet searches and dilepton searches set mild bounds
over the mass of the coloron and $Z'$, respectively. The more stringent
bound over the scale of the model arises from the ditop searches in
\cite{Baker:2019sli,Cornella:2021sby}, which push the mass of the
coloron to lie above 3.5 TeV, however those bounds could be slightly
overestimated for our model as they only consider non-fermiophobic
4321 scenarios. Finally, the mass range for $U_{1}$ is compatible
with current bounds, and mostly lie within the projected sensitivity
of the high luminosity phase of LHC. A good fit for the $3\sigma$
CMS excess in $U_{1}$ searches \cite{CMS-PAS-EXO-19-016} could be
provided if the extra decay channels to vector-like fermions, assuming
that the bound from ditop searches is indeed overestimated, motivating
a future dedicated collider analysis.

The model proposed here features clear connections between the SM
fermion masses and the leptoquark couplings which can address anomalies
in LFU ratios, along with Pati-Salam universality of most of the parameters,
leading to a very predictive and testable framework. The masses and
mixings of first family fermion can arise along the lines of the original
twin Pati-Salam model \cite{King:2021jeo}, involving a new family
of vector-like fermions which does not couple to $U_{1}$. Thanks
to the GIM-like mechanism implemented, the phenomenology discussed
in this paper would remain unaltered. However, the GIM-like mechanism
introduced is accidental, and enforcing this mechanism via extra symmetries
would involve the first family. Therefore, we leave the origin of
first family masses, neutrino masses and the symmetry behind the GIM-like
mechanism for a future manuscript, as they are uncorrelated to $B$-physics
and the phenomenology discussed here.

\section*{Acknowledgements}

MFN would like to thank the Padova phenomenology group for hospitality
during an intermediate stage of this work, and in particular Luca
Di Luzio and Javier Fuentes-Martín for helpful discussions about the
4321 model. This project has received funding from the European Union's
Horizon 2020 Research and Innovation Programme under Marie Sk\l{}odowska-Curie
grant agreement HIDDeN European ITN project (H2020-MSCA-ITN-2019//860881-HIDDeN).
SFK acknowledges the STFC Consolidated Grant ST/T000775/1.

\appendix

\section{Mixing angle formalism\label{sec:Mixing-angle-formalism}}

The mixing between third family and fourth family fermions arises
from the following terms in the mass Lagrangian \cite{King:2020mau},

\begin{equation}
\mathcal{L}_{\mathrm{mass}}\supset x_{34}^{\psi}\phi\psi_{3}\overline{\psi}_{4}+M_{4}^{\psi}\psi_{4}\overline{\psi}_{4}+\mathrm{h.c.}
\end{equation}
After the scalar $\phi$ develops a VEV, we obtain

\begin{equation}
x_{34}^{\psi}\left\langle \phi\right\rangle \psi_{3}\overline{\psi}_{4}+M_{4}^{\psi}\psi_{4}\overline{\psi}_{4}=\left(x_{34}^{\psi}\left\langle \phi\right\rangle \psi_{3}+M_{4}^{\psi}\psi_{4}\right)\overline{\psi}_{4}=\tilde{M}_{4}^{\psi}\frac{x_{34}^{\psi}\left\langle \phi\right\rangle \psi_{3}+M_{4}^{\psi}\psi_{4}}{\sqrt{\left(x_{34}^{\psi}\left\langle \phi\right\rangle \right)^{2}+\left(M_{4}^{\psi}\right)^{2}}}\overline{\psi}_{4}\,,
\end{equation}
where we have defined

\begin{equation}
\tilde{M}_{4}^{\psi}=\sqrt{\left(x_{34}^{\psi}\left\langle \phi\right\rangle \right)^{2}+\left(M_{4}^{\psi}\right)^{2}}
\end{equation}
as the physical mass of the vector-like fermion. We can identify the
mixing angles as

\begin{equation}
s_{34}^{\psi}=\frac{x_{34}^{\psi}\left\langle \phi\right\rangle }{\sqrt{\left(x_{34}^{\psi}\left\langle \phi\right\rangle \right)^{2}+\left(M_{4}^{\psi}\right)^{2}}}\,,\quad c_{34}^{\psi}=\frac{M_{4}^{\psi}}{\sqrt{\left(x_{34}^{\psi}\left\langle \phi\right\rangle \right)^{2}+\left(M_{4}^{\psi}\right)^{2}}}\,.
\end{equation}
This way, the mass eigenstates are given by

\begin{equation}
\tilde{\psi}_{4}\equiv c_{34}^{\psi}\psi_{4}+s_{34}^{\psi}\psi_{3}\,,\qquad\tilde{\psi}_{3}\equiv c_{34}^{\psi}\psi_{4}-s_{34}^{\psi}\psi_{3}\,.
\end{equation}
We can follow the same procedure to obtain all the mixing angles and
physical masses of vector-like fermions,

{\small{}
\begin{flalign}
 & s_{34}^{Q}=\frac{x_{34}^{\psi}\left\langle \phi_{3}\right\rangle }{\sqrt{\left(x_{34}^{\psi}\left\langle \phi_{3}\right\rangle \right)^{2}+\left(M_{4}^{Q}\right)^{2}}}\,, &  & s_{34}^{L}=\frac{x_{34}^{\psi}\left\langle \phi_{1}\right\rangle }{\sqrt{\left(x_{34}^{\psi}\left\langle \phi_{1}\right\rangle \right)^{2}+\left(M_{4}^{L}\right)^{2}}}\,,\label{eq:34_mixing_extended}\\
 & s_{25}^{Q}=\frac{x_{25}^{\psi}\left\langle \phi_{3}\right\rangle }{\sqrt{\left(x_{25}^{\psi}\left\langle \phi_{3}\right\rangle \right)^{2}+\left(M_{5}^{Q}\right)^{2}}}\,, &  & s_{25}^{L}=\frac{x_{25}^{\psi}\left\langle \phi_{1}\right\rangle }{\sqrt{\left(x_{25}^{\psi}\left\langle \phi_{1}\right\rangle \right)^{2}+\left(M_{5}^{L}\right)^{2}}}\,,\label{eq:25_mixing_extended}\\
 & s_{35}^{Q}=\frac{c_{34}^{Q}x_{35}^{\psi}\left\langle \phi_{3}\right\rangle }{\sqrt{\left(c_{34}^{Q}x_{35}^{\psi}\left\langle \phi_{3}\right\rangle \right)^{2}+\left(x_{25}^{\psi}\left\langle \phi_{3}\right\rangle \right)^{2}+\left(M_{5}^{Q}\right)^{2}}}, &  & s_{35}^{L}=\frac{c_{34}^{L}x_{35}^{\psi}\left\langle \phi_{1}\right\rangle }{\sqrt{\left(c_{34}^{L}x_{35}^{\psi}\left\langle \phi_{1}\right\rangle \right)^{2}+\left(x_{25}^{\psi}\left\langle \phi_{1}\right\rangle \right)^{2}+\left(M_{5}^{L}\right)^{2}}}\,,\label{eq:35_mixing}\\
 & s_{16}^{Q}=\frac{x_{16}^{\psi}\left\langle \phi_{3}\right\rangle }{\sqrt{\left(x_{16}^{\psi}\left\langle \phi_{3}\right\rangle \right)^{2}+\left(M_{6}^{Q}\right)^{2}}}\,, &  & s_{16}^{L}=\frac{x_{16}^{\psi}\left\langle \phi_{1}\right\rangle }{\sqrt{\left(x_{16}^{\psi}\left\langle \phi_{1}\right\rangle \right)^{2}+\left(M_{6}^{L}\right)^{2}}}\,,\\
 & s_{24}^{q^{c}}=\frac{x_{42}^{\psi^{c}}\langle\overline{\phi_{3}}\rangle}{\sqrt{\left(x_{42}^{\psi^{c}}\langle\overline{\phi_{3}}\rangle\right)^{2}+\left(M_{4}^{\psi^{c}}\right)^{2}}}\,, &  & s_{24}^{e^{c}}=\frac{x_{42}^{\psi^{c}}\langle\overline{\phi_{1}}\rangle}{\sqrt{\left(x_{42}^{\psi^{c}}\langle\overline{\phi_{1}}\rangle\right)^{2}+\left(M_{4}^{\psi^{c}}\right)^{2}}}\,,\label{eq:sqc24_mixing}\\
 & s_{34}^{q^{c}}=\frac{x_{43}^{\psi^{c}}\langle\overline{\phi_{3}}\rangle}{\sqrt{\left(x_{42}^{\psi^{c}}\langle\overline{\phi_{3}}\rangle\right)^{2}+\left(x_{43}^{\psi^{c}}\langle\overline{\phi_{3}}\rangle\right)^{2}+\left(M_{4}^{\psi^{c}}\right)^{2}}}\,, &  & s_{34}^{e^{c}}=\frac{x_{43}^{\psi^{c}}\langle\overline{\phi}_{1}\rangle}{\sqrt{\left(x_{42}^{\psi^{c}}\langle\overline{\phi_{1}}\rangle\right)^{2}+\left(x_{43}^{\psi^{c}}\langle\overline{\phi_{1}}\rangle\right)^{2}+\left(M_{4}^{\psi^{c}}\right)^{2}}}\,,\label{eq:sqc34_mixing}\\
 & \tilde{M}_{4}^{Q}=\sqrt{\left(x_{34}^{\psi}\left\langle \phi_{3}\right\rangle \right)^{2}+\left(M_{4}^{Q}\right)^{2}}\,, &  & \tilde{M}_{4}^{L}=\sqrt{\left(x_{34}^{\psi}\left\langle \phi_{1}\right\rangle \right)^{2}+\left(M_{4}^{L}\right)^{2}}\,,\label{eq:Mass-_4th}\\
 & \tilde{M}_{5}^{Q}=\sqrt{\left(x_{25}^{\psi}\left\langle \phi_{3}\right\rangle \right)^{2}+\left(x_{35}^{\psi}\left\langle \phi_{3}\right\rangle \right)^{2}+\left(M_{5}^{Q}\right)^{2}}\,, &  & \tilde{M}_{5}^{L}=\sqrt{\left(x_{25}^{\psi}\left\langle \phi_{1}\right\rangle \right)^{2}+\left(x_{35}^{\psi}\left\langle \phi_{1}\right\rangle \right)^{2}+\left(M_{5}^{L}\right)^{2}}\,,\label{eq:Mass_5th}\\
 & \tilde{M}_{6}^{Q}=\sqrt{\left(x_{16}^{\psi}\left\langle \phi_{3}\right\rangle \right)^{2}+\left(M_{6}^{Q}\right)^{2}}\,, &  & \tilde{M}_{6}^{L}=\sqrt{\left(x_{16}^{\psi}\left\langle \phi_{1}\right\rangle \right)^{2}+\left(M_{6}^{L}\right)^{2}}\,,\label{eq:Mass_6th}\\
 & \widetilde{M}_{4}^{q^{c}}=\sqrt{\left(x_{42}^{\psi^{c}}\langle\overline{\phi_{3}}\rangle\right)^{2}+\left(x_{43}^{\psi^{c}}\langle\overline{\phi_{3}}\rangle\right)^{2}+\left(M_{4}^{\psi^{c}}\right)^{2}}\,, &  & \widetilde{M}_{4}^{e^{c}}=\sqrt{\left(x_{42}^{\psi^{c}}\langle\overline{\phi_{1}}\rangle\right)^{2}+\left(x_{43}^{\psi^{c}}\langle\overline{\phi_{1}}\rangle\right)^{2}+\left(M_{4}^{\psi^{c}}\right)^{2}}\,.
\end{flalign}
}{\small\par}

\section{Vector-fermion interactions\label{sec:Vector-fermion-interactions}}

\subsection{Simplified model}

For the $U_{1}$ couplings, in the basis of mass eigenstates we obtain

\begin{equation}
\frac{g_{4}}{\sqrt{2}}\hat{u}_{i}^{\dagger}\gamma^{\mu}\left(\begin{array}{ccc}
0 & 0 & 0\\
0 & 0 & s_{34}^{Q}s_{34}^{L}s_{23}^{u}\\
0 & 0 & s_{34}^{Q}s_{34}^{L}c_{23}^{u}
\end{array}\right)\hat{\nu}{}_{j}U_{1\mu}+\mathrm{h.c.}\,,\quad\frac{g_{4}}{\sqrt{2}}\hat{d}_{i}^{\dagger}\gamma^{\mu}\left(\begin{array}{ccc}
0 & 0 & 0\\
0 & s_{34}^{Q}s_{34}^{L}s_{23}^{d}s_{23}^{e} & s_{34}^{Q}s_{34}^{L}s_{23}^{d}c_{23}^{e}\\
0 & s_{34}^{Q}s_{34}^{L}c_{23}^{d}s_{23}^{e} & s_{34}^{Q}s_{34}^{L}c_{23}^{d}c_{23}^{e}
\end{array}\right)\hat{e}_{j}U_{1\mu}+\mathrm{h.c.}\label{eq:U1couplings_u_4th}
\end{equation}
For the coloron couplings, in the basis of mass eigenstates we obtain{\small{}
\begin{equation}
\mathcal{L}_{g'}^{\mathrm{gauge}}=\frac{g_{4}g_{s}}{g_{3}}\hat{d}_{i}^{\dagger}\gamma^{\mu}T^{a}\left(\begin{array}{ccc}
-\frac{g_{3}^{2}}{g_{4}^{2}} & 0 & 0\\
0 & -\left(c_{23}^{d}\right)^{2}\frac{g_{3}^{2}}{g_{4}^{2}}+\left(s_{34}^{Q}s_{23}^{d}\right)^{2} & \left(s_{34}^{Q}\right)^{2}s_{23}^{d}c_{23}^{d}\\
0 & \left(s_{34}^{Q}\right)^{2}s_{23}^{d}c_{23}^{d} & \left(s_{34}^{Q}c_{23}^{d}\right)^{2}-\left(c_{34}^{Q}c_{23}^{d}\right)^{2}\frac{g_{3}^{2}}{g_{4}^{2}}
\end{array}\right)\hat{d}_{j}g_{\mu}^{a'}+\left(d\rightarrow u\right)\,,\label{eq:coloron_d_couplings_4thVL}
\end{equation}
}and for the $Z'${\small{}
\begin{equation}
\mathcal{L}_{Z',q}^{\mathrm{gauge}}=\frac{\sqrt{3}}{\sqrt{2}}\frac{g_{4}g_{Y}}{g_{1}}\hat{d}_{i}^{\dagger}\gamma^{\mu}\left(\begin{array}{ccc}
-\frac{g_{1}^{2}}{9g_{4}^{2}} & 0 & 0\\
0 & -\left(c_{23}^{d}\right)^{2}\frac{g_{1}^{2}}{9g_{4}^{2}}+\frac{1}{6}\left(s_{34}^{Q}s_{23}^{d}\right)^{2} & \frac{1}{6}\left(s_{34}^{Q}\right)^{2}s_{23}^{d}c_{23}^{d}\\
0 & \frac{1}{6}\left(s_{34}^{Q}\right)^{2}s_{23}^{d}c_{23}^{d} & \frac{1}{6}\left(s_{34}^{Q}c_{23}^{d}\right)^{2}-\left(c_{34}^{Q}c_{23}^{d}\right)^{2}\frac{g_{1}^{2}}{9g_{4}^{2}}
\end{array}\right)\hat{d}_{j}Z'_{\mu}+\left(d\rightarrow u\right)\,,\label{eq:Z'_quark_couplings_4thVL}
\end{equation}
\begin{equation}
\mathcal{L}_{Z',e}^{\mathrm{gauge}}=\frac{\sqrt{3}}{\sqrt{2}}\frac{g_{4}g_{Y}}{g_{1}}\hat{e}_{i}^{\dagger}\gamma^{\mu}\left(\begin{array}{ccc}
\frac{g_{1}^{2}}{3g_{4}^{2}} & 0 & 0\\
0 & \left(c_{23}^{e}\right)^{2}\frac{g_{1}^{2}}{3g_{4}^{2}}-\frac{1}{2}\left(s_{34}^{L}s_{23}^{e}\right)^{2} & -\frac{1}{2}\left(s_{34}^{L}\right)^{2}s_{23}^{e}c_{23}^{e}\\
0 & -\frac{1}{2}\left(s_{34}^{L}\right)^{2}s_{23}^{e}c_{23}^{e} & -\frac{1}{2}\left(s_{34}^{L}c_{23}^{e}\right)^{2}+\left(c_{34}^{L}c_{23}^{e}\right)^{2}\frac{g_{1}^{2}}{3g_{4}^{2}}
\end{array}\right)\hat{e}_{j}Z'_{\mu}+\left(e\rightarrow\nu\right)\,.\label{eq:Z'_leptons_couplings_4thVL}
\end{equation}
}{\small\par}

\subsection{Extended model}

For the couplings to heavy gauge bosons we obtain,
\begin{equation}
\mathcal{L}_{U_{1}}^{\mathrm{gauge}}=\frac{g_{4}}{\sqrt{2}}Q'{}_{i}^{\dagger}\gamma_{\mu}\left(\begin{array}{ccc}
s_{16}^{Q}s_{16}^{L}\epsilon & 0 & 0\\
0 & c_{\theta_{LQ}}s_{25}^{Q}s_{25}^{L} & s_{\theta_{LQ}}s_{25}^{Q}s_{34}^{L}\\
0 & -s_{\theta_{LQ}}s_{34}^{Q}s_{25}^{L} & c_{\theta_{LQ}}s_{34}^{Q}s_{34}^{L}
\end{array}\right)L'_{j}U_{1}^{\mu}+\mathrm{h.c.}\,,\label{eq:LQ_couplings-1}
\end{equation}
{\small{}
\begin{equation}
\mathcal{L}_{g'}^{\mathrm{gauge}}=\frac{g_{4}g_{s}}{g_{3}}Q'{}_{i}^{\dagger}\gamma^{\mu}T^{a}\left(\begin{array}{ccc}
\left(s_{16}^{Q}\right)^{2}-\left(c_{16}^{Q}\right)^{2}\frac{g_{3}^{2}}{g_{4}^{2}} & 0 & 0\\
0 & \left(s_{25}^{Q}\right)^{2}-\left(c_{25}^{Q}\right)^{2}\frac{g_{3}^{2}}{g_{4}^{2}} & 0\\
0 & 0 & \left(s_{34}^{Q}\right)^{2}-\left(c_{34}^{Q}\right)^{2}\frac{g_{3}^{2}}{g_{4}^{2}}
\end{array}\right)Q'_{j}g'{}_{\mu}^{a}\,,\label{eq:Coloron_couplings_3VL-1}
\end{equation}
}{\small\par}

{\small{}
\begin{equation}
\mathcal{L}_{Z',q}^{\mathrm{gauge}}=\frac{\sqrt{3}}{\sqrt{2}}\frac{g_{4}g_{Y}}{g_{1}}Q'{}_{i}^{\dagger}\gamma^{\mu}\left(\begin{array}{ccc}
\frac{1}{6}\left(s_{16}^{Q}\right)^{2}-\left(c_{16}^{Q}\right)^{2}\frac{g_{1}^{2}}{9g_{4}^{2}} & 0 & 0\\
0 & \frac{1}{6}\left(s_{25}^{Q}\right)^{2}-\left(c_{25}^{Q}\right)^{2}\frac{g_{1}^{2}}{9g_{4}^{2}} & 0\\
0 & 0 & \frac{1}{6}\left(s_{34}^{Q}\right)^{2}-\left(c_{34}^{Q}\right)^{2}\frac{g_{1}^{2}}{9g_{4}^{2}}
\end{array}\right)Q'_{j}Z'_{\mu}\,,\label{eq:Z'_couplings_3VL_q}
\end{equation}
}{\small\par}

{\small{}
\begin{equation}
\mathcal{L}_{Z',\ell}^{\mathrm{gauge}}=-\frac{\sqrt{3}}{\sqrt{2}}\frac{g_{4}g_{Y}}{g_{1}}L'{}_{i}^{\dagger}\gamma^{\mu}\left(\begin{array}{ccc}
\frac{1}{2}\left(s_{16}^{L}\right)^{2}-\left(c_{16}^{L}\right)^{2}\frac{g_{1}^{2}}{3g_{4}^{2}} & 0 & 0\\
0 & \frac{1}{2}\left(s_{25}^{L}\right)^{2}-\left(c_{25}^{L}\right)^{2}\frac{g_{1}^{2}}{3g_{4}^{2}} & 0\\
0 & 0 & \frac{1}{2}\left(s_{34}^{L}\right)^{2}-\left(c_{34}^{L}\right)^{2}\frac{g_{1}^{2}}{3g_{4}^{2}}
\end{array}\right)L'_{j}Z'_{\mu}\,,\label{eq:Z'_couplings_3VL_l}
\end{equation}
}where the up-quark couplings above receive small corrections due
to 2-3 mixing arising after diagonalising the effective mass matrices
in Eqs.~\eqref{eq:MassMatrix_4thVL_effective_up-1},~\eqref{eq:MassMatrix_4thVL_effective_down-1}.
However, larger 2-3 charged lepton mixing is possible (see Section
\ref{subsec:Effective_Yukawa_3VL}), obtaining for charged leptons:{\footnotesize{}
\begin{equation}
\mathcal{L}_{Z',e}^{\mathrm{gauge}}\approx-\frac{\sqrt{3}}{\sqrt{2}}\frac{g_{4}g_{Y}}{g_{1}}\hat{e}{}_{i}^{\dagger}\gamma^{\mu}\left(\begin{array}{ccc}
\frac{1}{2}\left(s_{16}^{L}\right)^{2}-\left(c_{16}^{L}\right)^{2}\frac{g_{1}^{2}}{3g_{4}^{2}} & 0 & 0\\
0 & \frac{1}{2}\left(s_{25}^{L}\right)^{2}-\left(c_{25}^{L}\right)^{2}\frac{g_{1}^{2}}{3g_{4}^{2}} & \frac{1}{2}\left[\left(s_{34}^{L}\right)^{2}-\left(s_{25}^{L}\right)^{2}\right]s_{23}^{e}\\
0 & \frac{1}{2}\left[\left(s_{34}^{L}\right)^{2}-\left(s_{25}^{L}\right)^{2}\right]s_{23}^{e} & \frac{1}{2}\left(s_{34}^{L}\right)^{2}-\left(c_{34}^{L}\right)^{2}\frac{g_{1}^{2}}{3g_{4}^{2}}
\end{array}\right)\hat{e}_{j}Z'_{\mu}\,,\label{eq:Z'_couplings_3VL_e}
\end{equation}
}at first order in $s_{23}^{e}$ and taking $c_{23}^{e}\approx1$.
The flavour-violating couplings above can lead to interesting signals
in LFV processes such as $\tau\rightarrow3\mu$ and $\tau\rightarrow\mu\gamma$,
see more in Section~\ref{subsec:LFV_processes}. ..

\section{Benchmarks}

In Table~\ref{tab:BP} we include the benchmark points considered
in Section \ref{subsec:Low-energy-phenomenology}.
\begin{table}[t]
\noindent \begin{centering}
\begin{tabular}{lllllllll}
\toprule 
\multicolumn{4}{c}{Benchmark} &  & \multicolumn{4}{c}{Output}\tabularnewline
\midrule
\midrule 
$g_{4}$ & 3.5 & $\lambda_{15}^{44}$ & -0.5 &  & $s_{34}^{Q}$ & 0.978 & $M_{g'}$ & 3782.9 GeV\tabularnewline
\midrule 
$g_{3,2,1}$ & 1, 0.65, 0.36 & $\lambda_{15}^{55},\,\lambda_{15}^{66}$ & 2.5, 1.1 &  & $s_{34}^{L}$ & 0.977 & $M_{Z'}$ & 2414.3 GeV\tabularnewline
\midrule 
$x_{34}^{\psi}$ & 2 & $x_{42}^{\psi^{c}}$ & 0.4 &  & $s_{25}^{Q}=s_{16}^{Q}$ & $0.20^{*}$, $0.17^{**}$ & $s_{23}^{u}$ & 0.042556\tabularnewline
\midrule 
$x_{25}^{\psi}=x_{16}^{\psi}$ & $0.41^{*}$ , $0.35^{**}$ & $x_{43}^{\psi^{c}}$ & 1 &  & $s_{25}^{L}=s_{16}^{L}$ & 0.1455 & $s_{23}^{d}$ & 0.001497\tabularnewline
\midrule 
$M_{44}^{\psi}$ & 320 GeV & $M_{44}^{\psi^{c}}$ & 5 TeV &  & $s_{\theta_{LQ}}$ & 0.7097 & $s_{23}^{e}$ & -0.111\tabularnewline
\midrule 
$M_{55}^{\psi}$ & 780 GeV & $y_{53,43,34,24}^{\psi}$ & -0.3, 1, 1, 1 &  & $\widetilde{M}_{4}^{Q}$ & 1226.8 GeV & $V_{cb}$ & 0.04106\tabularnewline
\midrule 
$M_{66}^{\psi}$ & 1120 GeV & $\left\langle H_{t}\right\rangle $ & 177.2 GeV &  & $\widetilde{M}_{5}^{Q}$ & 1238.7 GeV & $m_{t}$ & 172.91 GeV\tabularnewline
\midrule 
$M_{45}^{\psi}$ & -700 GeV & $\left\langle H_{c}\right\rangle $ & 26.8 GeV &  & $\widetilde{M}_{4}^{L}$ & 614.04 GeV & $m_{c}$ & 1.270 GeV\tabularnewline
\midrule
$M_{54}^{\psi}$ & 50 GeV & $\left\langle H_{b}\right\rangle $ & 4.25 GeV &  & $\widetilde{M}_{5}^{L}$ & 845.26 GeV & $m_{b}$ & 4.180 GeV\tabularnewline
\midrule
$\left\langle \phi_{3}\right\rangle $ & 0.6 TeV & $\left\langle H_{s}\right\rangle $ & 2.1 GeV &  & $\widetilde{M}_{6}^{Q}$ & 1234.6 GeV & $m_{s}$ & 0.0987 GeV\tabularnewline
\midrule
$\left\langle \phi_{1}\right\rangle $ & 0.3 TeV & $\left\langle H_{\tau}\right\rangle $ & 1.75 GeV &  & $\widetilde{M}_{6}^{L}$ & 859.4 GeV & $m_{\tau}$ & 1.7765 GeV\tabularnewline
\midrule
$\left\langle \Omega_{15}\right\rangle $ & 0.4 TeV & $\left\langle H_{\mu}\right\rangle $ & 4.58 GeV &  & $M_{U_{1}}$ & 2987.1 GeV & $m_{\mu}$ & 105.65 MeV\tabularnewline
\bottomrule
\end{tabular}
\par\end{centering}
\caption{Input and output parameters for the benchmark points BP1 and BP2,
{*} indicates BP1 while {*}{*} indicates BP2, otherwise both benchmarks
share the same parameters. BP1 is compatible with 2021 data on $R_{K^{(*)}}$,
while BP2 is compatible with the 2022 updates by LHCb. \label{tab:BP}}
\end{table}

\section{Building the EFT of the model}

\subsection{4-fermion operators in the SMEFT\label{subsec:EFT_model_Appendix}}

Here we include the set of 4-fermion operators obtained at tree-level
after integrating out the heavy $U_{1}$, $Z'$ and $g'$ in the extended
model of Section \ref{sec:Twin-Pati-Salam-Theory_3VL},
\begin{align}
\mathcal{L}_{\text{4-\ensuremath{\mathrm{fermion}}}}= & -\frac{2}{v_{\mathrm{SM}}^{2}}\left[\left[C_{lq}^{(1)}\right]^{\alpha\beta ij}\left[\mathcal{Q}_{lq}^{(1)}\right]^{\alpha\beta ij}+\left[C_{lq}^{(3)}\right]^{\alpha\beta ij}\left[\mathcal{Q}_{lq}^{(3)}\right]^{\alpha\beta ij}\right.\nonumber \\
{} & \left.+\left[C_{qq}^{(1)}\right]^{ijkl}\left[\mathcal{Q}_{qq}^{(1)}\right]^{ijkl}+\left[C_{qq}^{(3)}\right]^{ijkl}\left[\mathcal{Q}_{qq}^{(3)}\right]^{ijkl}+\left[C_{ll}\right]^{\alpha\beta\delta\lambda}\left[\mathcal{Q}_{ll}\right]^{\alpha\beta\delta\lambda}\right]\,,\label{eq:SMEFT_4fermion}
\end{align}
where $v_{\mathrm{SM}}=\left(\sqrt{2}G_{F}\right)^{-1/2}\approx246\,\mathrm{GeV}$,
we choose latin indices for quark flavours and greek indices for lepton
flavours. The operators $\mathcal{Q}_{lq}^{(1,3)}$, $\mathcal{Q}_{qq}^{(1,3)}$
and $\mathcal{Q}_{ll}$ are defined as in the so-called Warsaw basis
\cite{Grzadkowski:2010es} of dim-6 operators built out of SM fields.
In our model, the Wilson coefficients are given by
\begin{align}
{} & \left[C_{lq}^{(1)}\right]^{\alpha\beta ij}=\frac{1}{2}C_{U}\beta_{i\alpha}\beta_{j\beta}^{*}-2C_{Z'}\xi_{ij}\xi_{\alpha\beta}\,, & {} & \left[C_{lq}^{(3)}\right]^{\alpha\beta ij}=\frac{1}{2}C_{U}\beta_{i\alpha}\beta_{j\beta}^{*}\,,\\
{} & \left[C_{qq}^{(1)}\right]^{ijkl}=\frac{1}{4}C_{g'}\kappa_{il}\kappa_{jk}-\frac{1}{6}C_{g'}\kappa_{ij}\kappa_{kl}+C_{Z'}\xi_{ij}\xi_{kl}\,, & {} & \left[C_{qq}^{(3)}\right]^{ijkl}=\frac{1}{4}C_{g'}\kappa_{il}\kappa_{jk}\,,\\
{} & \left[C_{ll}\right]^{\alpha\beta\delta\lambda}=C_{Z'}\xi_{\alpha\beta}\xi_{\delta\lambda}\,, & {} & {}
\end{align}
where we have defined
\begin{equation}
C_{U}=\frac{g_{4}^{2}v_{\mathrm{SM}}^{2}}{4M_{U_{1}}^{2}}\,,\qquad C_{g'}=\frac{g_{4}^{2}g_{s}^{2}}{2g_{3}^{2}}\frac{v_{\mathrm{SM}}^{2}}{M_{g'}^{2}}\,,\qquad C_{Z'}=\frac{3g_{4}^{2}g_{Y}^{2}}{4g_{1}^{2}}\frac{v_{\mathrm{SM}}^{2}}{M_{Z'}^{2}}\,.
\end{equation}

We consider all the fields in \eqref{eq:SMEFT_4fermion} to be mass
eigenstates, as the effects of fermion mixing are encoded into the
$U_{1}$ ($\beta_{i\alpha}$), $g'$ ($\kappa_{ij}$) and $Z'$ ($\xi_{ij}$,$\xi_{\alpha\beta}$)
couplings given in Eqs.~\eqref{eq:LQ_couplings-1}, \eqref{eq:Coloron_couplings_3VL-1}
\eqref{eq:Z'_couplings_3VL_q} and \eqref{eq:Z'_couplings_3VL_l}.

\subsection{$b\rightarrow c\tau\nu$\label{subsec:b_c_tau_nu_Appendix}}

The charged-current transition $b\rightarrow c\tau\nu_{\tau}$ is
described in our model by the effective Lagrangian,

\begin{equation}
\mathcal{L}_{b\rightarrow c\tau\nu_{\tau}}=-\frac{4G_{F}}{\sqrt{2}}V_{cb}\left(1+\left[C_{\nu edu}^{*}\right]^{3332}\right)(\bar{c}_{L}\gamma_{\mu}b_{L})(\bar{\tau}_{L}\gamma^{\mu}\nu_{\tau L})+\mathrm{h.c.}\,,
\end{equation}
where we have omitted all operators including right-handed fermions,
as they receive zero or negligible contributions in our model. The
matching with the SMEFT is

\begin{equation}
\left[C_{\nu edu}^{*}\right]^{3332}(m_{b})=\frac{2\eta_{V}^{\nu\tau}}{V_{cb}}\left[C_{lq}^{(3)}\right]^{\tau\tau23}(\Lambda)\,,\label{eq:R_D_WilsonCoefficient}
\end{equation}
where the negligible RGE effect is encoded as $\eta_{V}^{\nu\tau}\approx1.00144$
and has been computed with DsixTools 2.1 \cite{Fuentes-Martin:2020zaz}
for $\Lambda=1\,\mathrm{TeV}$. The Wilson coefficient $\left[C_{\nu edu}^{*}\right]^{3332}(m_{b})$
can provide a very good fit to $R_{D^{(*)}}$, here we take the $\text{1\ensuremath{\sigma}}$
interval from \cite{Angelescu:2021lln} (where $\mathcal{B}(B_{c}\rightarrow\tau\bar{\nu})<30\%$
was also imposed),

\begin{equation}
\left[C_{\nu edu}^{*}\right]^{3332}(m_{b})=0.07\pm0.02\,.
\end{equation}

\subsection{$b\rightarrow s\ell\ell$\label{subsec:b_s_ll_Appendix}}

The effective Lagrangian describing a generic $b\rightarrow s\ell\ell$
transition reads

\begin{equation}
\mathcal{L}_{b\rightarrow s\ell_{\alpha}\ell_{\beta}}=\frac{4G_{F}}{\sqrt{2}}\frac{\alpha_{\mathrm{EM}}}{4\pi}V_{tb}V_{ts}^{*}\left\{ \left(C_{9}^{\mathrm{SM}}\delta_{\alpha\beta}+C_{9}^{23\alpha\beta}\right)\mathcal{O}_{9}^{23\alpha\beta}+\left(C_{10}^{\mathrm{SM}}\delta_{\alpha\beta}+C_{10}^{23\alpha\beta}\right)\mathcal{O}_{10}^{23\alpha\beta}\right\} +\mathrm{h.c.}\,,
\end{equation}
where

\begin{align}
\mathcal{O}_{9}^{23\alpha\beta}=(\bar{s}_{L}\gamma^{\mu}b_{L})(\bar{\ell}_{\alpha}\gamma_{\mu}\ell_{\beta})\,, & \qquad\mathcal{O}_{10}^{23\alpha\beta}=(\bar{s}_{L}\gamma^{\mu}b_{L})(\bar{\ell}_{\alpha}\gamma_{\mu}\gamma_{5}\ell_{\beta})\,,\label{eq:O9_O10_operators}
\end{align}
we are interested in the matching to the SMEFT of the left-handed
operator,

\begin{equation}
C_{9}^{23\alpha\beta}=-C_{10}^{23\alpha\beta}=-\frac{2\pi}{\alpha_{\mathrm{EM}}V_{tb}V_{ts}^{*}}\eta_{V}^{\ell\ell}\left(\left[C_{lq}^{(3)}\right]^{\alpha\beta23}(\Lambda)+\left[C_{lq}^{(1)}\right]^{\alpha\beta23}(\Lambda)\right)\,,\label{eq:R_K_WilsonCoefficients}
\end{equation}
where the RGE is encoded as $\eta_{V}^{\ell\ell}\approx0.974$ and
has been computed with DsixTools 2.1 \cite{Fuentes-Martin:2020zaz}
for $\Lambda=1\,\mathrm{TeV}$. The expressions of the LFU ratios
$R_{K}$ and $R_{K^{*}}$ in terms of the Wilson coefficients $C_{9}^{23\mu\mu}$
and $C_{10}^{23\mu\mu}$ read

\begin{equation}
R_{K}^{[1.1,6]}=R_{K,\mathrm{SM}}^{[1.1,6]}\frac{1+0.24\mathrm{Re}(C_{9}^{23\mu\mu})-0.26\mathrm{Re}(C_{10}^{23\mu\mu})+0.03(\left|C_{9}^{23\mu\mu}\right|^{2}+\left|C_{10}^{23\mu\mu}\right|^{2})}{1+0.24\mathrm{Re}(C_{9}^{23ee})-0.26\mathrm{Re}(C_{10}^{23ee})+0.03(\left|C_{9}^{23ee}\right|^{2}+\left|C_{10}^{23ee}\right|^{2})}\,,
\end{equation}

\begin{equation}
R_{K^{*}}^{[1.1,6]}=R_{K^{*},\mathrm{SM}}^{[1.1,6]}\frac{1+0.18\mathrm{Re}(C_{9}^{23\mu\mu})-0.29\mathrm{Re}(C_{10}^{23\mu\mu})+0.03(\left|C_{9}^{23\mu\mu}\right|^{2}+\left|C_{10}^{23\mu\mu}\right|^{2})}{1+0.18\mathrm{Re}(C_{9}^{23ee})-0.29\mathrm{Re}(C_{10}^{23ee})+0.03(\left|C_{9}^{23ee}\right|^{2}+\left|C_{10}^{23ee}\right|^{2})}\,.
\end{equation}
We do not include expressions for the lower $q^{2}$ interval where
NP contributions are suppressed.

The theoretical expressions for the branching fractions of the relevant
leptonic and semileptonic $B$-decays are \cite{Becirevic:2016zri,Cornella:2021sby}

\begin{equation}
\mathcal{B}\left(B_{s}\rightarrow\ell^{+}\ell^{-}\right)=\mathcal{B}\left(B_{s}\rightarrow\ell^{+}\ell^{-}\right)_{\mathrm{SM}}\left|1+\frac{C_{10}^{23\ell\ell}}{C_{10}^{\mathrm{SM}}}\right|^{2}\,,\label{eq:Bs_tautau}
\end{equation}

\begin{equation}
\mathcal{B}\left(B^{+}\rightarrow K^{+}\tau^{+}\tau^{-}\right)=10^{-9}\left(2.2\left|C_{9}^{23\tau\tau}\right|^{2}+6.0\left|C_{10}^{23\tau\tau}\right|^{2}\right)\,,\label{eq:B_Ktautau}
\end{equation}

\begin{equation}
\mathcal{B}\left(B_{s}\rightarrow\tau^{-}\mu^{+}\right)=\frac{\tau_{B_{s}}m_{B_{s}}f_{B_{s}}^{2}}{64\pi^{3}}\alpha_{\mathrm{EM}}^{2}G_{F}^{2}m_{\tau}^{2}\left|V_{tb}V_{ts}^{*}\right|^{2}\left(1-\frac{m_{\tau}^{2}}{m_{B_{s}}^{2}}\right)^{2}\left(\left|C_{9}^{23\tau\mu}\right|^{2}+\left|C_{10}^{23\tau\mu}\right|^{2}\right)\,,
\end{equation}

\begin{equation}
\mathcal{B}\left(B_{s}\rightarrow\tau^{+}\mu^{-}\right)=\frac{\tau_{B_{s}}m_{B_{s}}f_{B_{s}}^{2}}{64\pi^{3}}\alpha_{\mathrm{EM}}^{2}G_{F}^{2}m_{\tau}^{2}\left|V_{tb}V_{ts}^{*}\right|^{2}\left(1-\frac{m_{\tau}^{2}}{m_{B_{s}}^{2}}\right)^{2}\left(\left|C_{9}^{23\mu\tau}\right|^{2}+\left|C_{10}^{23\mu\tau}\right|^{2}\right)\,,\label{eq:Bs_tau_mu}
\end{equation}

\begin{equation}
\mathcal{B}\left(B^{+}\rightarrow K^{+}\tau^{+}\mu^{-}\right)=10^{-9}\left(9.6\left|C_{9}^{23\tau\mu}\right|^{2}+10\left|C_{10}^{23\tau\mu}\right|^{2}\right)\,,\label{eq:B_Ktaumu}
\end{equation}

\begin{equation}
\mathcal{B}\left(B^{+}\rightarrow K^{+}\tau^{-}\mu^{+}\right)=10^{-9}\left(9.6\left|C_{9}^{23\mu\tau}\right|^{2}+10\left|C_{10}^{23\mu\tau}\right|^{2}\right)\,.
\end{equation}
where we use the numerical input $C_{10}^{\mathrm{SM}}=-4.17$ \cite{Bruggisser:2021duo},
$\mathcal{B}\left(B_{s}\rightarrow\tau^{+}\tau^{-}\right)_{\mathrm{SM}}=\left(7.73\pm0.49\right)\cdot10^{-7}$
\cite{Bobeth:2013uxa}, $f_{B_{s}}=230.3\pm1.3\,\mathrm{MeV}$ \cite{FlavourLatticeAveragingGroupFLAG:2021npn},
$m_{B_{s}}=5366.92\pm0.10\,\mathrm{MeV}$ \cite{PDG:2022ynf}, $\tau_{B_{s}}=1.515\pm0.005\,\mathrm{ps}$
\cite{PDG:2022ynf}, $\alpha_{\mathrm{EM}}=1/137.036$ and $G_{F}=1.166\cdot10^{-5}\mathrm{GeV}^{-2}$.

\subsection{Bounds from $B_{s}-\bar{B}_{s}$ mixing \label{subsec:Bs_Mixing_Appendix}}

We describe $B_{s}-\bar{B}_{s}$ mixing with the effective Lagrangian

\begin{equation}
\mathcal{L}_{\mathrm{eff}}^{bs}\supset-\frac{C_{bs}^{\mathrm{NP}}}{2}\left(\bar{s}_{L}\gamma_{\mu}b_{L}\right)^{2}\,,
\end{equation}
where in the simplified model the Wilson coefficient receives tree-level
NP contributions from the $Z'$ and $g'$ gauge bosons,
\begin{equation}
C_{bs}^{\mathrm{NP}}=C_{bs}^{g'}+C_{bs}^{Z'}=\left[\frac{1}{3M_{g'}^{2}}+\frac{1}{24M_{Z'}^{2}}\right]g_{4}^{2}\left(s_{34}^{Q}\right)^{4}\left(s_{23}^{d}c_{23}^{d}\right)^{2}\,,\label{eq:Cbs_4thVL}
\end{equation}
written in the phenomenological limit of interest $g_{4}\gg g_{3,1}$.
Here it is clear that the coloron contribution dominates over the
$Z'$ one. Even in the motivated scenario $\left\langle \phi_{3}\right\rangle \gg\left\langle \phi_{1}\right\rangle $,
where the coloron can be twice heavier than the $Z'$, the coloron
contribution is at least four times larger than the $Z'$ one.

Such a NP contribution is constrained by the results of the mass difference
$\Delta M_{s}$ of neutral $B_{s}$ mesons. The experimental value
is known very precisely, see for example the most recent HFLAV average
\cite{HFLAV:2022pwe}, which is dominated by the updated measurement
by LHCb \cite{LHCb:2021moh}. However, the SM prediction historically
suffered from larger uncertainties, and we need a precise knowledge
of the SM contribution in order to quantify the impact of possible
contributions from new physics. The theoretical determination of $\Delta M_{s}$
is limited by our understanding of non-perturbative matrix elements
of dimension six operators. The matrix elements can be determined
with lattice simulations or sum rules. As discussed in Ref.~\cite{DiLuzio:2019jyq},
the 2019 FLAG average \cite{FlavourLatticeAveragingGroup:2019iem}
is dominated by the lattice results \cite{FermilabLattice:2016ipl,Boyle:2018knm,Dowdall:2019bea},
and suffers from an uncertainty just below 10\% with the central value
being $1.8\sigma$ above the experiment,
\begin{equation}
\Delta M_{s}^{\mathrm{FLAG'19}}=\left(1.13_{-0.09}^{+0.07}\right)\Delta M_{s}^{\mathrm{exp}}\,.
\end{equation}
If one considers the value above as the SM prediction for $\Delta M_{s}$,
then NP models with positive contributions to $\Delta M_{s}$, such
as our coloron and $Z'$ contributions, have very small room to be
compatible with the experimental value at the $2\sigma$ level. Instead,

\begin{equation}
\Delta M_{s}^{\mathrm{Average'19}}=\left(1.04_{-0.07}^{+0.04}\right)\Delta M_{s}^{\mathrm{exp}}\,,
\end{equation}
was computed in \cite{DiLuzio:2019jyq} as a weighted average of both
the FLAG'19 average \cite{FlavourLatticeAveragingGroup:2019iem} and
sum rule results \cite{Kirk:2017juj,Grozin:2016uqy,King:2019lal}.
The weighted average shows better agreement with the experiment, and
a reduction of the total uncertainty (see the further discussion in
\cite{DiLuzio:2019jyq}). The Average'19 result for $\Delta M_{s}$
leaves some room for positive NP contributions at the $2\sigma$ level.
We extract an upper bound over the NP contribution by considering
the lower limit of the $2\sigma$ range, $\Delta M_{s}^{\mathrm{SM}}\approx0.9\Delta M_{s}^{\mathrm{exp}}$,
hence

\begin{equation}
\frac{\Delta M_{s}^{\mathrm{SM}}+\Delta M_{s}^{\mathrm{NP}}}{\Delta M_{s}^{\mathrm{exp}}}\approx0.9\frac{\Delta M_{s}^{\mathrm{SM}}+\Delta M_{s}^{\mathrm{NP}}}{\Delta M_{s}^{\mathrm{SM}}}\approx1\Rightarrow\Delta M_{s}^{\mathrm{NP}}\lesssim0.11\Delta M_{s}^{\mathrm{SM}}\,.\label{eq:DeltaMs_bound}
\end{equation}
In other words, $\Delta M_{s}^{\mathrm{Average'19}}$ allows for roughly
a 10\% positive NP correction over the SM value. This is in line with
the 10\% criteria considered in the analysis of \cite{Cornella:2021sby},
which was possibly motivated by $\Delta M_{s}^{\mathrm{Average'19}}$
as well. The bound in Eq.~\eqref{eq:DeltaMs_bound} translates directly
over the Wilson coefficient as

\begin{equation}
\delta(\Delta M_{s})\equiv\frac{\Delta M_{s}-\Delta M_{s}^{\mathrm{SM}}}{\Delta M_{s}^{\mathrm{SM}}}=\left|1+\frac{C_{bs}^{\mathrm{NP}}}{C_{bs}^{\mathrm{\mathrm{SM}}}}\right|-1=\frac{C_{bs}^{\mathrm{NP}}}{C_{bs}^{\mathrm{\mathrm{SM}}}}\apprle0.11\,,\label{eq:delta_DeltaMs}
\end{equation}
where in the second step we have assumed real and positive Wilson
coefficients. The SM contribution reads

\begin{equation}
C_{bs}^{\mathrm{SM}}=\frac{G_{F}^{2}m_{W}^{2}}{2\pi^{2}}\left(V_{tb}^{*}V_{ts}\right)^{2}S_{0}(x_{t})\,,
\end{equation}
with $S_{0}(x_{t})=2.37$ \cite{Buchalla:1995vs}. This way, we obtain
the numerical bound 

\begin{equation}
C_{bs}^{\mathrm{NP}}\lesssim\frac{1}{\left(225\,\mathrm{TeV}\right)^{2}}\,,\label{eq:bound_Bsmixing}
\end{equation}
which is in good agreement with the $(220\,\mathrm{TeV)^{-2}}$ obtained
in \cite{DiLuzio:2019jyq} from $\Delta M_{s}^{\mathrm{Average'19}}$.

\subsubsection{Loop functions for $U_{1}$-mediated $B_{s}-\bar{B}_{s}$ mixing
\label{subsec:Loop_functions_BsMixing}}

The loop function for the 1-loop $U_{1}$-mediated contribution to
$B_{s}-\bar{B}_{s}$ mixing reads
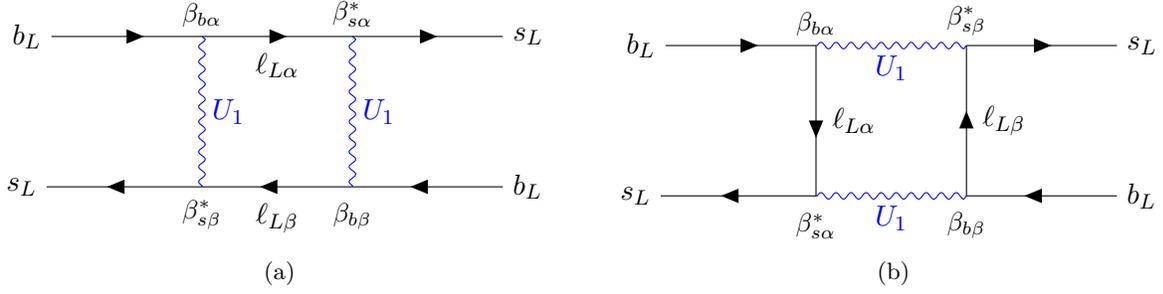
\begin{figure}[t]
\subfloat[]{\noindent \begin{centering}
\begin{tikzpicture}
	\begin{feynman}
		\vertex (a) {\(s_{L}\)};
		\vertex [right=24mm of a] (b) [label={ [yshift=-0.7cm] \small $\beta_{s\beta}^{*}$}];
		\vertex [right=20mm of b] (c) [label={ [yshift=-0.7cm] \small $\beta_{b\beta}$}];
		\vertex [right=20mm of c] (d) {\(b_{L}\)};
		\vertex [above=20mm of b] (f1) [label={ \small $\beta_{b\alpha}$}];
		\vertex [above=20mm of c] (f2) [label={ \small $\beta_{s\alpha}^{*}$}];
		\vertex [left=20mm of f1] (f3) {\(b_{L}\)};
		\vertex [right=20mm of f2] (f4) {\(s_{L}\)};
		\diagram* {
			(a) -- [anti fermion] (b) -- [boson, blue, edge label'=\(U_{1}\)] (f1) -- [anti fermion] (f3),
			(b) -- [anti fermion, edge label'=\(\ell_{L\beta}\), inner sep=6pt] (c) -- [boson, blue, edge label'=\(U_{1}\)] (f2) -- [fermion] (f4),
			(c) -- [anti fermion] (d),
			(f1) -- [fermion, edge label'=\(\ell_{L\alpha}\), inner sep=6pt] (f2),
	};
	\end{feynman}
\end{tikzpicture}
\par\end{centering}
}$\quad$\subfloat[]{\noindent \begin{centering}
\begin{tikzpicture}
	\begin{feynman}
		\vertex (a) {\(s_{L}\)};
		\vertex [right=24mm of a] (b) [label={ [yshift=-0.7cm] \small $\beta_{s\alpha}^{*}$}];
		\vertex [right=20mm of b] (c) [label={ [yshift=-0.7cm] \small $\beta_{b\beta}$}];
		\vertex [right=20mm of c] (d) {\(b_{L}\)};
		\vertex [above=20mm of b] (f1) [label={ \small $\beta_{b\alpha}$}];
		\vertex [above=20mm of c] (f2) [label={ \small $\beta_{s\beta}^{*}$}];
		\vertex [left=20mm of f1] (f3) {\(b_{L}\)};
		\vertex [right=20mm of f2] (f4) {\(s_{L}\)};
		\diagram* {
			(a) -- [anti fermion] (b) -- [anti fermion, edge label'=\(\ell_{L\alpha}\), inner sep=6pt)] (f1) -- [anti fermion] (f3),
			(b) -- [boson, blue, edge label'=\(U_{1}\)] (c) -- [fermion, edge label'=\(\ell_{L\beta}\), inner sep=6pt)] (f2) -- [fermion] (f4),
			(c) -- [anti fermion] (d),
			(f1) -- [boson, blue, edge label'=\(U_{1}\)] (f2),
	};
	\end{feynman}
\end{tikzpicture}
\par\end{centering}
}

\caption{Leptoquark-mediated one-loop diagrams contributing to $B_{s}-\bar{B}_{s}$
mixing. The indexes $\alpha,\beta$ run for all charged leptons including
vector-like, i.e.~$\ell_{L\alpha}=\left(\mu_{L},\tau_{L},E_{L4},E_{L5}\right)$.
\label{fig:Box_BsMixing}}
\end{figure}

\begin{equation}
F(x_{\alpha},x_{\beta})=\left(1+\frac{x_{\alpha}x_{\beta}}{4}\right)B(x_{\alpha},x_{\beta})\,,
\end{equation}
where

\begin{equation}
B(x_{\alpha},x_{\beta})=\frac{1}{\left(1-x_{\alpha}\right)\left(1-x_{\beta}\right)}+\frac{x_{\alpha}^{2}\log x_{\alpha}}{\left(x_{\beta}-x_{\alpha}\right)\left(1-x_{\alpha}^{2}\right)}+\frac{x_{\beta}^{2}\log x_{\beta}}{\left(x_{\alpha}-x_{\beta}\right)\left(1-x_{\beta}^{2}\right)}\,.\label{eq:loop_function}
\end{equation}
The loop function is dominated by the heavy vector-like partners.
In particular, in the motivated scenario with maximal $s_{34}^{L}$,
the couplings with the fourth family $\beta_{sE_{4}}^{*}\beta_{bE_{4}}$
are suppressed by the small cosine $c_{34}^{L}$. This way, the loop
is dominated by $E_{5}$ in good approximation. We obtain the effective
loop function in this scenario by removing all constants in $x_{\alpha,\beta}$,
which vanish due to the property \eqref{eq:unitarity_loop}, 

\begin{equation}
\tilde{F}(x)\approx F(x,x)-2F(x,0)+F(0,0)=\frac{x\left(x+4\right)\left(-1+x^{2}-2x\log x\right)}{4\left(x-1\right)^{3}}\,.\label{eq:F_tilde}
\end{equation}

\subsection{LFV $\tau$ decays \label{subsec:LFV_tau_Appendix}}

The model leads to contributions to LFV $\tau$ decays which we describe
via the effective Lagrangian

\begin{equation}
\mathcal{L}_{\tau\,\mathrm{LFV}}=-\frac{4G_{F}}{\sqrt{2}}\left(\left[C_{ee}^{V,LL}\right]^{2322}(\bar{\mu}_{L}\gamma_{\mu}\tau_{L})(\bar{\mu}_{L}\gamma^{\mu}\mu_{L})+\left[C_{ed}^{V,LL}\right]^{2322}(\bar{\mu}_{L}\gamma_{\mu}\tau_{L})(\bar{s}_{L}\gamma^{\mu}s_{L})\right)+\mathrm{h.c.}\,,
\end{equation}
where the matching to SMEFT Wilson coefficients reads

\begin{align}
\left[C_{ee}^{V,LL}\right]^{2322}(m_{b})=\left[C_{ll}\right]^{2322}(\Lambda)+C_{\tau\mu\mu\mu}^{U_{1}}\,,\quad & \left[C_{ed}^{V,LL}\right]^{2322}(m_{b})=\left[C_{lq}^{(1)}\right]^{2333}(\Lambda)+\left[C_{lq}^{(3)}\right]^{2333}(\Lambda)\,,
\end{align}
where we have neglected small RGE effects at the percent level. The
first contribution to $\left[C_{ee}^{V,LL}\right]^{2322}$ denotes
the tree-level matching to the SMEFT, while the second term describes
the 1-loop box diagram mediated by $U_{1}$, 

\begin{equation}
C_{\tau\mu\mu\mu}^{U_{1}}=-\frac{3g_{4}^{4}v_{\mathrm{SM}}^{2}}{256\pi^{2}M_{U_{1}}^{2}}\beta_{D_{5}\mu}^{*}\beta_{D_{5}\tau}\left(\beta_{D_{5}\mu}\right)^{2}\tilde{F}(x_{D_{5}})\,,
\end{equation}
where $\tilde{F}(x)$ is given in Eq.~\eqref{eq:F_tilde}. The branching
fraction is then given by

\begin{equation}
\mathcal{B}\left(\tau\rightarrow3\mu\right)=2\left(\left[C_{ee}^{V,LL}\right]^{2322}\right)^{2}<2.1\cdot10^{-8}\,,\label{eq:tau_3mu}
\end{equation}
with the bound given in \cite{Hayasaka:2010np}.

Finally, the formula for $\tau\rightarrow\mu\phi$ is given by \cite{Cornella:2021sby}
\begin{equation}
\mathcal{B}\left(\tau\rightarrow\mu\phi\right)=\frac{1}{\Gamma_{\tau}}\frac{G_{F}^{2}f_{\phi}^{2}m_{\tau}^{3}}{16\pi}\left(1-\frac{m_{\phi}^{2}}{m_{\tau}^{2}}\right)^{2}\left(1+2\frac{m_{\phi}^{2}}{m_{\tau}^{2}}\right)\left|\left[C_{ed}^{V,LL}\right]^{2322}\right|^{2}\,,\label{eq:tau_muphi}
\end{equation}
where $f_{\phi}=225\,\mathrm{MeV}$ and $m_{\phi}^{2}/m_{\tau}^{2}=0.33$
\cite{PDG:2022ynf}.

\subsubsection{Dipole operators for $\tau\rightarrow\mu\gamma$ \label{subsec:Dipole-operators_Appendix}}

The process $\tau\rightarrow\mu\gamma$ is described by the dipole
operator,

\begin{equation}
\mathcal{L}_{\mathrm{eff}}\supset-\frac{4G_{F}}{\sqrt{2}}eC_{\mu\tau}^{\mathrm{NP}}\left(\overline{L}_{L2}\sigma^{\mu\nu}\tau_{R}\right)HF_{\mu\nu}+\mathrm{h.c.}\,,
\end{equation}
which receives contributions via both $U_{1}$ and $Z'$,

\begin{equation}
\text{\ensuremath{C_{\mu\tau}^{\mathrm{NP}}}=\ensuremath{C_{\mu\tau}^{Z'}}+\ensuremath{C_{\mu\tau}^{U_{1}}}}\,,
\end{equation}
where
\begin{figure}[t]
\noindent \begin{centering}
\subfloat[]{\begin{centering}
\begin{tikzpicture}	
	\begin{feynman}
		\vertex (a) {\(\tau_{R}\)};
		\vertex [right=28mm of a] (b) [label={ [xshift=0.1cm, yshift=-0.55cm] \small $\beta_{i\tau}$}];
		\vertex [right=20mm of b] (c) [label={ [xshift=0.1cm, yshift=-0.6cm] \small $\beta_{i\mu}^{*}$}];
		\vertex [right=20mm of c] (d){\(\mu_{L}\)};
		\diagram* {
			(a) -- [edge label'=\(\tau_{L}\), near end, inner sep=6pt, insertion=0.5] (b) -- [boson, half left, edge label=$U_{1}$] (c),
			(b) --  [fermion, edge label'=\(d_{Li}\), inner sep=6pt] (c),
			(c) -- [fermion] (d),
	};
	\end{feynman}
\end{tikzpicture}
\par\end{centering}
}$\qquad$\subfloat[]{\begin{centering}
\begin{tikzpicture}	
	\begin{feynman}
		\vertex (a) {\(\tau_{R}\)};
		\vertex [right=28mm of a] (b) [label={ [xshift=0.1cm, yshift=-0.65cm] \small $\xi_{\tau\alpha}$}];
		\vertex [right=20mm of b] (c) [label={ [xshift=0.1cm, yshift=-0.7cm] \small $\xi_{\alpha\mu}$}];
		\vertex [right=20mm of c] (d){\(\mu_{L}\)};
		\diagram* {
			(a) -- [edge label'=\(\tau_{L}\), near end, inner sep=6pt, insertion=0.5] (b) -- [boson, half left, edge label=$Z'$] (c),
			(b) --  [fermion, edge label'=\(\ell_{L\alpha}\), inner sep=4pt] (c),
			(c) -- [fermion] (d),
	};
	\end{feynman}
\end{tikzpicture}
\par\end{centering}
}
\par\end{centering}
\caption{Leptoquark (left panel) and $Z'$ (right panel) 1-loop contributions
to $\tau\rightarrow\mu\gamma$. Photon lines are implicit. The index
$i$ runs for all down-quarks including vector-like, i.e~ $d_{Li}=\left(s_{L},b_{L},D_{L4},D_{L5}\right)$,
while $\alpha$ runs for all charged leptons including vector-like,
i.e.~$\ell_{L\alpha}=\left(\mu_{L},\tau_{L},E_{L4},E_{L5}\right)$.
\label{fig:diagram_tau_mu_photon}}
\end{figure}
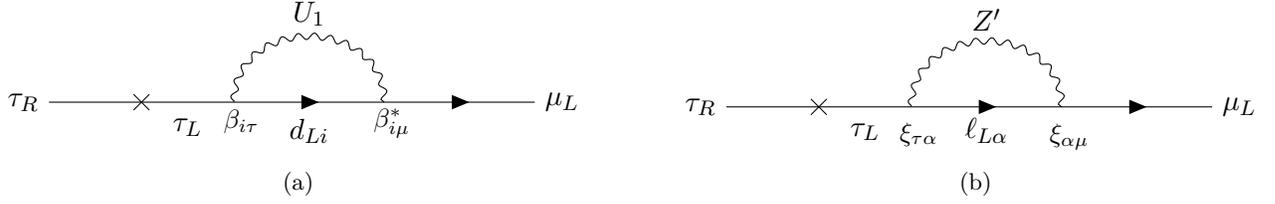

\begin{equation}
C_{\mu\tau}^{U_{1}}(\Lambda)=-\frac{C_{U}}{16\pi^{2}}\frac{y_{\tau}(\Lambda)}{2}\sum_{i}\beta_{i\mu}^{*}\beta_{i\tau}\left[G_{1}(x_{i})-2G_{2}(x_{i})\right]\,,
\end{equation}

\begin{equation}
C_{\mu\tau}^{Z'}(\Lambda)=-\frac{C_{Z'}}{16\pi^{2}}\frac{y_{\tau}(\Lambda)}{2}\sum_{\alpha}\xi_{\tau\alpha}\xi_{\alpha\mu}\widetilde{G}(x_{\alpha})\,,
\end{equation}
where $i=s,b,D_{4},D_{5}$ and $\alpha=\mu,\tau,E_{4},E_{5}$. The
effective tau Yukawa coupling $y_{\tau}$ in the Higgs basis is estimated
following the same procedure used in Eq.~\eqref{eq:Top_Effective Yukawa},
obtaining an effective SM-like Yukawa $y_{\tau}\approx0.01$. The
loop functions are defined as \cite{Cornella:2021sby,Fuentes-Martin:2020hvc,CarcamoHernandez:2019ydc}

\begin{equation}
G_{1}(x)=x\left[\frac{2-5x}{2\left(x-1\right)^{4}}\log x-\frac{4-13x+3x^{2}}{4\left(x-1\right)^{3}}\right]\,,\quad G_{2}(x)=x\left[\frac{4x-1}{2\left(x-1\right)^{4}}x\log x-\frac{2-5x-3x^{2}}{4\left(x-1\right)^{3}}\right]\,,
\end{equation}

\begin{equation}
\widetilde{G}(x)=\frac{5x^{4}-14x^{3}+39x^{2}-38x-18x^{2}\log x+8}{12(1-x)^{4}}\,.
\end{equation}
The running of the dipole operator from $\Lambda=2\,\mathrm{TeV}$
to the scale $\mu\sim m_{\tau}$ is given by $C_{\mu\tau}^{\mathrm{NP}}(m_{\tau})\approx0.92C_{\mu\tau}^{\mathrm{NP}}(\Lambda)$,
as estimated with DsixTools 2.1 \cite{Fuentes-Martin:2020zaz}. Neglecting
the muon mass, the branching ratio is given by

\begin{equation}
\mathcal{B}\left(\tau\rightarrow\mu\gamma\right)=\frac{8G_{F}^{2}\alpha_{\mathrm{EM}}m_{\tau}^{3}}{\Gamma_{\tau}}\left|C_{\mu\tau}^{\mathrm{NP}}(m_{\tau})\right|^{2}\,.\label{eq:tau_mu_photon}
\end{equation}

\subsection{Tests of universality in $\tau$ decays\label{subsec:Tests-of-universality-tau-decays}}

In our model, modifications to the ratios $(g_{\tau}/g_{\mu})$ and
$(g_{\tau}/g_{e})$ are given by

\begin{equation}
\left(\frac{g_{\tau}}{g_{\mu}}\right)_{\ell}=1+\frac{9}{12}C_{Z'}\left(\left|\xi_{\tau e}\right|^{2}-\left|\xi_{\mu e}\right|^{2}\right)-\eta^{\tau\mathrm{LFU}}C_{U}\left(\left|\beta_{b\tau}\right|^{2}-\left|\beta_{b\mu}\right|^{2}\right)\,,
\end{equation}

\begin{equation}
\left(\frac{g_{\tau}}{g_{e}}\right)_{\ell}=1+\frac{9}{12}C_{Z'}\left(\left|\xi_{\tau\mu}\right|^{2}-\left|\xi_{\mu e}\right|^{2}\right)-\eta^{\tau\mathrm{LFU}}C_{U}\left(\left|\beta_{b\tau}\right|^{2}-\left|\beta_{be}\right|^{2}\right)\,,
\end{equation}
where $\eta^{\tau\mathrm{LFU}}=0.079$ parameterises the running from
$\Lambda=2\,\mathrm{TeV}$, as computed in DsixTools 2.1 \cite{Fuentes-Martin:2020zaz}.
We find the $Z'$ contributions to be subleading due to the small
$Z'$ couplings being further suppressed by $T_{15}$ factors, see
Eq.~\eqref{eq:Z'_couplings_3VL_e}. Due to the hierarchy in leptoquark
couplings, we find $\beta_{b\tau}\gg\beta_{b\mu}$ and $\beta_{be}\approx0$,
hence in good approximation both ratios receive the same contribution
proportional to $\beta_{b\tau}$. Because of the same reason, tree-level
leptoquark contributions to the hadronic $\tau$ vs $\mu$ ratios
are found to be much smaller than the loop contribution, rendering
all the LFU ratios in $\tau$ to be well approximated by

\begin{equation}
\left(\frac{g_{\tau}}{g_{\mu,e}}\right)_{\ell+\pi+K}\approx1-\eta^{\tau\mathrm{LFU}}C_{U}\left|\beta_{b\tau}\right|^{2}\,,\label{eq:LFUratios_tau-2}
\end{equation}
where $\beta_{b\tau}\approx\cos\theta_{LQ}$ assuming maximal 3-4
mixing.

\subsection{$b\rightarrow s\nu\nu$ \label{subsec:BtoKnunu_Appendix}}

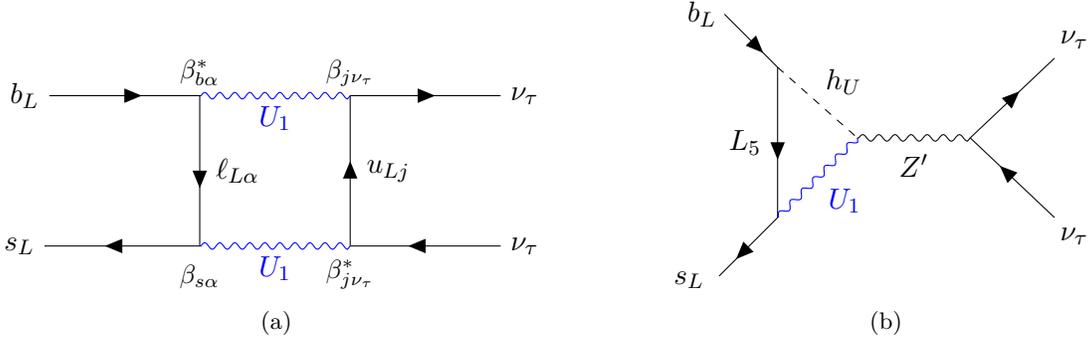
\begin{figure}[t]
\begin{centering}
\subfloat[\label{fig:Box_Diagram_BKnunu}]{\noindent \begin{centering}
\begin{tikzpicture}
	\begin{feynman}
		\vertex (a) {\(s_{L}\)};
		\vertex [right=24mm of a] (b) [label={ [yshift=-0.7cm] \small $\beta_{s\alpha}$}];
		\vertex [right=20mm of b] (c) [label={ [yshift=-0.7cm] \small $\beta_{j\nu_{\tau}}^{*}$}];
		\vertex [right=20mm of c] (d) {\(\nu_{\tau}\)};
		\vertex [above=20mm of b] (f1) [label={ \small $\beta_{b\alpha}^{*}$}];
		\vertex [above=20mm of c] (f2) [label={ \small $\beta_{j\nu_{\tau}}$}];
		\vertex [left=20mm of f1] (f3) {\(b_{L}\)};
		\vertex [right=20mm of f2] (f4) {\(\nu_{\tau}\)};
		\diagram* {
			(a) -- [anti fermion] (b) -- [anti fermion, edge label'=\(\ell_{L\alpha}\), inner sep=6pt)] (f1) -- [anti fermion] (f3),
			(b) -- [boson, blue, edge label'=\(U_{1}\)] (c) -- [fermion, edge label'=\(u_{Lj}\), inner sep=6pt)] (f2) -- [fermion] (f4),
			(c) -- [anti fermion] (d),
			(f1) -- [boson, blue, edge label'=\(U_{1}\)] (f2),
	};
	\end{feynman}
\end{tikzpicture}
\par\end{centering}
}$\quad\qquad$\subfloat[\label{fig:Radial_Diagram_BKnunu}]{\noindent \begin{centering}
\begin{tikzpicture}
	\begin{feynman}
		\vertex (a) [label={ [xshift=-0.4cm,yshift=-0.3cm] $s_{L}$}];
		\vertex [above right=11mm of a] (b);
		\vertex [above right=of b] (c);
		\vertex [above=20mm of b] (d);
		\vertex [above left=10mm of d] (e)  [label={ [xshift=-0.3cm,yshift=-0.3cm] $b_{L}$}];
		\vertex [right=of c] (f);
		\vertex [above right=of f] (f2) {\(\nu_{\tau}\)};
		\vertex [below right=of f] (f3) {\(\nu_{\tau}\)};
		\diagram* {
			(a) -- [anti fermion] (b) -- [anti fermion,  edge label=\(L_{5}\),  inner sep=6pt] (d) -- [anti fermion] (e),
			(b) -- [boson, blue, edge label'=\(U_{1}\), inner sep=4pt] (c),
			(d) -- [scalar,  edge label=\(h_{U}\), inner sep=3pt] (c),
			(c) -- [boson, edge label'=\(Z'\), inner sep=6pt] (f),
			(f) -- [fermion] (f2),
			(f) -- [anti fermion] (f3),
	};
	\end{feynman}
\end{tikzpicture}
\par\end{centering}
}
\par\end{centering}
\caption{Box and penguin diagrams contributing to $B\rightarrow K\nu\nu$.
The index $\alpha$ runs for all charged leptons including vector-like,
i.e.~$\ell_{L\alpha}=\left(\mu_{L},\tau_{L},E_{L4},E_{L5}\right)$,
and the index $j$ runs for all up-type quarks, including vector-like
$u_{Lj}=\left(c_{L},t_{L},U_{L4},U_{L5}\right)$. See more details
in the main text.\label{fig:Box-and-penguin_BtoKnunu}}
\end{figure}

We define the relevant Lagrangian to describe $b\rightarrow s\nu\nu$
transitions as

\begin{equation}
\mathcal{L}_{b\rightarrow s\nu\nu}=\frac{4G_{F}}{\sqrt{2}}V_{tb}V_{ts}^{*}\left(C_{\nu,\mathrm{NP}}^{\alpha\beta}+C_{\nu,\mathrm{SM}}\right)\left(\bar{s}_{L}\gamma_{\mu}b_{L}\right)\left(\bar{\nu}_{\alpha L}\gamma^{\mu}\nu_{\beta L}\right)+\mathrm{h.c.}\,.\label{eq:B_Knunu_Effective}
\end{equation}
The universal SM contribution reads

\begin{equation}
C_{\nu,\mathrm{SM}}=-\frac{\alpha_{W}}{2\pi}X_{t}\,,
\end{equation}
where $X_{t}=1.48\pm0.01$ \cite{Buchalla:1998ba}, and $\alpha_{W}=g_{2}^{2}/(4\pi)$
with $g_{2}\simeq0.65$ being the $SU(2)_{L}$ coupling. We further
split the NP effects into $Z'$-mediated and $U_{1}$-mediated contributions
as follows, and we only obtain sizable contributions in the $\tau\tau$
channel,
\begin{equation}
C_{\nu,\mathrm{NP}}^{\tau\tau}=-\frac{1}{V_{tb}V_{ts}^{*}}\frac{\sqrt{2}}{4G_{F}}\left(C_{\nu,Z'}^{\tau\tau}+C_{\nu,U}^{\tau\tau}\right)\,.
\end{equation}
The $U_{1}$ contribution at NLO accuracy reads \cite{Fuentes-Martin:2020hvc}
\begin{equation}
C_{\nu,U}^{\tau\tau}\approx C_{\nu,U}^{\mathrm{RGE}}+\frac{g_{4}^{4}}{32\pi^{2}M_{U_{1}}^{2}}\sum_{\alpha,j}\left(\beta_{s\alpha}^{*}\beta_{b\alpha}\right)\left(\beta_{j\nu_{\tau}}\right)^{2}F(x_{\alpha},x_{j})\,,
\end{equation}
where the second term arises from the semileptonic box diagram in
Fig.~\ref{fig:Box_Diagram_BKnunu}, and the first term encodes the
RGE-induced contribution from the tree-level leptoquark-mediated operator
$(\overline{s}_{L}\gamma_{\mu}b_{L})$ $(\overline{\tau}_{L}\gamma^{\mu}\tau_{L})$,
computed in DsixTools 2.1 \cite{Fuentes-Martin:2020zaz} as
\begin{equation}
C_{\nu,U}^{\mathrm{RGE}}=0.047\frac{g_{4}^{2}}{2M_{U_{1}}^{2}}\beta_{b\tau}\beta_{s\tau}\,.
\end{equation}
The $Z'$ contribution at NLO accuracy reads
\begin{equation}
C_{\nu,Z'}^{\tau\tau}\approx\frac{3g_{4}^{2}}{2M_{Z'}^{2}}\left[\xi_{bs}\xi_{\nu_{\tau}\nu_{\tau}}\left(1+\frac{3}{2}\frac{g_{4}^{2}}{16\pi^{2}}\xi_{\nu_{\tau}\nu_{\tau}}^{2}\right)+\frac{g_{4}^{2}}{16\pi^{2}}\beta_{sE_{5}}^{*}\beta_{bE_{5}}\xi_{\nu_{\tau}\nu_{\tau}}G_{\Delta Q=1}(x_{E_{5}},x_{Z'},x_{R})\right]\,,\label{eq:Z'_BKnunu}
\end{equation}
where $x_{E_{5}}\equiv(M_{5}^{L}/M_{U})^{2}$, $x_{Z'}\equiv M_{Z'}^{2}/M_{U}^{2}$
and $x_{R}\equiv M_{R}^{2}/M_{U}^{2}$ with $M_{R}$ being a scale
associated to the radial mode $h_{U}(3,1,2/3)$ arising from $\phi_{3,1}$.
The first term in Eq.~\eqref{eq:Z'_BKnunu} corresponds to the tree-level
contribution plus a 1-loop $Z'$ correction to the leptonic vertex.
The coupling $\xi_{bs}$ is suppressed by the small down mixing angle
$\theta_{23}^{d}$, leading to percent corrections to $\mathcal{B}(B\rightarrow K^{(*)}\nu\bar{\nu})$.
The second term in Eq.~\eqref{eq:Z'_BKnunu} corresponds to a 1-loop
correction to the flavour-violating $Z'$ vertex, with $U_{1}$, the
fifth vector-like lepton $E_{5}$ and $h_{U}$ running in the loop,
see Fig.~\ref{fig:Radial_Diagram_BKnunu}. The loop function is given
by \cite{Cornella:2021sby,Fuentes-Martin:2020hvc}
\begin{equation}
G_{\Delta Q=1}(x_{1},x_{2},x_{3})\approx\frac{5}{4}x_{1}+\frac{x_{1}}{2}\left(x_{2}-\frac{3}{2}\right)\left(\ln x_{3}-\frac{5}{2}\right)\,.
\end{equation}
In the twin PS framework, we expect extra radial modes associated
to $\bar{\phi}_{3,1}$ and $\bar{\phi'}_{3,1}$, however they only
couple to right-handed SM fermions and hence they cannot contribute
to the effective operator in Eq.~\eqref{eq:B_Knunu_Effective}.

\section{From CP-conjugated notation to left-right notation \label{sec:From-CP-conjugated-notation}}

Under the SM symmetry gauge group, the VL families decompose into
fermions with the usual SM quantum numbers of the chiral quarks and
leptons, but including partners in conjugate representations,
\begin{equation}
\psi_{a}\rightarrow\left(Q_{a},L_{a}\right)\equiv\left(Q_{La},L_{La}\right)\,,\quad\overline{\psi_{a}}\rightarrow\left(\overline{Q}_{a},\overline{L}_{a}\right)\overset{CP}{\rightarrow}\left(\tilde{Q}_{Ra},\widetilde{L}_{Ra}\right)\,,\label{eq:CP_LR_1}
\end{equation}
\begin{equation}
\psi_{a}^{c}\rightarrow\left(u_{a}^{c},d_{a}^{c},\nu_{a}^{c},e_{a}^{c}\right)\overset{CP}{\rightarrow}\left(u_{Ra},d_{Ra},\nu_{Ra},e_{Ra}\right)\,,\quad\overline{\psi_{a}^{c}}\rightarrow\left(\overline{u_{a}^{c}},\overline{d_{a}^{c}},\overline{\nu_{a}^{c}},\overline{e_{a}^{c}}\right)\equiv\left(\tilde{u}_{La},\tilde{d}_{La},\tilde{\nu}_{La},\tilde{e}_{La}\right)\,,\label{eq:CP_LR_2}
\end{equation}
where $a=4,5,6$. In the equations above we show the equivalence between
the CP-conjugated notation and the left (L) and right (R) convention,
by using a CP transformation where applicable. We use the tilde notation
to highlight the partners in conjugate representations. Similarly,
we can write the three chiral families of quarks and leptons in L,
R convention as,
\begin{equation}
\left(Q_{i},L_{i}\right)\equiv\left(Q_{Li},L_{Li}\right)\,,\quad\left(u_{i}^{c},d_{i}^{c},\nu_{i}^{c},e_{i}^{c}\right)\overset{CP}{\rightarrow}\left(u_{Ri},d_{Ri},\nu_{Ri},e_{Ri}\right)\,.\label{eq:CP_LR_3}
\end{equation}

\section{$\epsilon$ dilution of the first family $U_{1}$ coupling \label{sec:Epsilon_Dilution}}

In Eq.~\eqref{eq:LQ_couplings} we introduced a parameter $\epsilon$
which parameterises a possible suppression of the first family $U_{1}$
coupling via mixing with vector-like fermions. This idea of suppressing
leptoquark couplings via mixing with VL fermions is common in the
bibliography, as similar ideas are applied in \cite{Calibbi:2017qbu,Blanke:2018sro}
for the same purpose, and also to suppress right-handed couplings
in models were the third family is charged under the low-scale $SU(4)$
\cite{Cornella:2019hct,Cornella:2021sby,Barbieri:2022ikw}. The origin
of the first family $U_{1}$ coupling $\beta_{de}$ is mixing between
the sixth VL fermion family and the first chiral family, i.e.

\begin{equation}
\beta_{de}=s_{16}^{Q}s_{16}^{L}\,.
\end{equation}
However, prior to this mixing, the sixth VL fermion is allowed to
mix with another VL fermion family, following a mechanism similar
to the one that originated the Cabbibo-like matrix $W_{LQ}$. Let
us assume an extra sixth-primed VL family transforming in the same
way as the sixth family under the twin Pati-Salam symmetry, but discriminated
by a flavour symmetry which we assume as $Z_{2}$ for simplicity,
which forbids mixing between the sixth-primed family and any chiral
family. Instead, mixing between the sixth and sixth-primed fermion
families is allowed via a twin Pati-Salam singlet charged under the
new $Z_{2}$, i.e.

\begin{equation}
\mathcal{L}_{\mathrm{mix}}=x_{66}\chi\bar{\psi}_{6}\psi'_{6}+x'_{66}\chi^{*}\bar{\psi}'_{6}\psi{}_{6}\,+\mathrm{h.c.}
\end{equation}
The mass terms of the sixth and sixth-primed fields are splitted via
$\Omega_{15}$ in the usual way,

\begin{equation}
\mathcal{L}_{\mathrm{mass}}=(M_{66}^{\psi}+\lambda_{15}^{66}T_{15}\Omega_{15})\bar{\psi}_{6}\psi_{6}+(M_{66'}^{\psi}+\lambda_{15}^{66'}T_{15}\Omega_{15})\bar{\psi}'_{6}\psi'_{6}+\mathrm{h.c.}
\end{equation}
After $\Omega_{15}$ and the singlet $\chi$ develop VEVs, we obtain
the following mass matrices for quarks and leptons

\begin{equation}
\mathcal{L}_{\mathrm{mass}}+\mathcal{L}_{\mathrm{mix}}=\left(\begin{array}{@{}lcc@{}}
 & Q_{6} & Q'_{6}\\
\cmidrule(l){2-3}\left.\bar{Q}{}_{6}\right| & M_{66}^{Q} & x_{66}\left\langle \chi\right\rangle \\
\left.\bar{Q}'_{6}\right| & x'_{66}\left\langle \chi\right\rangle  & M_{66'}^{Q}
\end{array}\right)+\left(\begin{array}{@{}lcc@{}}
 & L_{6} & L'_{6}\\
\cmidrule(l){2-3}\left.\bar{L}{}_{6}\right| & M_{66}^{L} & x_{66}\left\langle \chi\right\rangle \\
\left.\bar{L}'_{6}\right| & x'_{66}\left\langle \chi\right\rangle  & M_{66'}^{L}
\end{array}\right)+\mathrm{h.c.}\,,\label{eq:MassMatrix_4thVL_effective_up-1-1}
\end{equation}
where we have defined

\begin{equation}
M_{66}^{Q}=M_{66}^{\psi}+\frac{\lambda_{15}^{66}}{2\sqrt{6}}\left\langle \Omega_{15}\right\rangle \,,\quad M_{66}^{L}=M_{66}^{\psi}-3\frac{\lambda_{15}^{66}}{2\sqrt{6}}\left\langle \Omega_{15}\right\rangle \,,
\end{equation}

\begin{equation}
M_{66'}^{Q}=M_{66'}^{\psi}+\frac{\lambda_{15}^{66'}}{2\sqrt{6}}\left\langle \Omega_{15}\right\rangle \,,\quad M_{66'}^{L}=M_{66'}^{\psi}-3\frac{\lambda_{15}^{66'}}{2\sqrt{6}}\left\langle \Omega_{15}\right\rangle \,.
\end{equation}
The mass matrices in Eq.~\eqref{eq:MassMatrix_4thVL_effective_up-1-1}
are diagonalised by different unitary transformations in the quark
and lepton sector, $V_{66'}^{Q}$ and $V_{66'}^{L}$, in such a way
that the $U_{1}$ couplings are given by
\begin{table}[t]
\begin{centering}
\begin{tabular}{cc}
\toprule 
field & $Z_{2}$\tabularnewline
\midrule
\midrule 
$\bar{\psi}_{6}$, $\psi_{6}$ & 1, 1\tabularnewline
\midrule 
$\bar{\psi'}_{6}$, $\psi'_{6}$ & -1, -1\tabularnewline
\midrule 
$\chi$ & -1\tabularnewline
\bottomrule
\end{tabular}$\qquad\qquad$%
\begin{tabular}{cccc}
\toprule 
\multicolumn{2}{c}{Input} & \multicolumn{2}{c}{Output}\tabularnewline
\midrule
\midrule 
$M_{66}^{\psi}$ & 900 GeV & $\tilde{M}_{66}^{Q}$ & 1211 GeV\tabularnewline
\midrule 
$M_{66'}^{\psi}$ & 1100 GeV & $\tilde{M}_{66}^{L}$ & 834 GeV\tabularnewline
\midrule 
$x_{66}\left\langle \chi\right\rangle $ & -700 & $s_{66}^{Q}$ & 0.298\tabularnewline
\midrule 
$x'_{66}\left\langle \chi\right\rangle $ & 680 & $s_{66}^{L}$ & 0.967\tabularnewline
\midrule 
$\lambda_{15}^{66}$, $\lambda_{15}^{66'}$ & 1.5, 2.5 & $\cos\theta_{6}$ & 0.045\tabularnewline
\bottomrule
\end{tabular}
\par\end{centering}
\caption{\textbf{\textit{(Left)}} Charge assignments under $Z_{2}$ that allow
the desired mixing. \textbf{\textit{(Right)}} Benchmark parameters
which lead to a dilution $\epsilon<0.1$. \label{tab:benchmark_suppression}}
\end{table}

\begin{equation}
\mathcal{L}_{U_{1}}=\frac{g_{4}}{\sqrt{2}}\left(\begin{array}{cc}
Q_{6}^{\dagger} & Q_{6}^{\dagger'}\end{array}\right)\gamma_{\mu}V_{66'}^{Q}\mathrm{diag}(1,1)V_{66'}^{L\dagger}\left(\begin{array}{c}
L_{6}\\
L'_{6}
\end{array}\right)U_{1}^{\mu}+\mathrm{h.c.}
\end{equation}
If we define 

\begin{equation}
V_{66'}^{Q}V_{66'}^{L\dagger}\equiv\left(\begin{array}{cc}
\cos\theta_{6} & \sin\theta_{6}\\
-\sin\theta_{6} & \cos\theta_{6}
\end{array}\right)\,,
\end{equation}
then the $Q_{6}^{\dagger}L_{6}U_{1}$ coupling receives a suppression
via $\cos\theta_{6}$ as

\begin{equation}
\beta_{de}=s_{16}^{Q}s_{16}^{L}\cos\theta_{6}\,.
\end{equation}
which is identified with the suppression parameter $\epsilon$ in
Eq.~\eqref{eq:LQ_couplings},

\begin{equation}
\epsilon\equiv\cos\theta_{6}.
\end{equation}
We can achieve values of $\cos\theta_{6}$ smaller than 0.1 without
any aggressive tuning of the parameters, obtaining the mild suppression
desired for $K_{L}\rightarrow\mu e$ as per Fig.~\ref{fig:KL_mue}.
A suitable benchmark can be found in Table~\ref{tab:benchmark_suppression}.
Interestingly, this mechanism does not affect the $Z'$ and $g'$
interactions, as the unitary matrices $V_{66'}^{Q}$ and $V_{66'}^{L}$
cancel in neutral currents. This allows the GIM-like suppression of
1-2 FCNCs to remain in place for both the quark and lepton sector
via $s_{16}^{Q}=s_{25}^{Q}$ and $s_{16}^{L}=s_{25}^{L}$, without
entering in conflict with $K_{L}\rightarrow\mu e$ nor with $B$-physics.

\section{RGE equations \label{sec:RGE-equations}}

To investigate the perturbativity of the model we use the RGE equations
of the 4321 model. For the gauge couplings beta functions $\beta_{g_{i}}=(dg_{i}/d\mu)/\mu$
we have \cite{DiLuzio:2018zxy}

\begin{equation}
\left(4\pi\right)^{2}\beta_{g_{1}}=\frac{131}{18}g_{1}^{3}\,,\quad\left(4\pi\right)^{2}\beta_{g_{2}}=\left(-\frac{19}{6}+\frac{8n_{\Psi}}{3}\right)g_{2}^{3}\,,
\end{equation}

\begin{equation}
\left(4\pi\right)^{2}\beta_{g_{3}}=-\frac{19}{3}g_{3}^{3}\,,\quad\left(4\pi\right)^{2}\beta_{g_{4}}=\left(-\frac{40}{3}+\frac{4n_{\Psi}}{3}\right)g_{4}^{3}\,,
\end{equation}
where $n_{\Psi}=3$ is the number of vector-like fermion families.
The Pati-Salam universality of the Yukawas $x_{i\alpha}^{\psi}$ is
broken by RGE effects which we quantify through the equations

\begin{align}
\begin{aligned}\left(4\pi\right)^{2}\beta_{x_{Q}} & =\frac{7}{2}x_{Q}x_{Q}^{\dagger}x_{Q}+\frac{1}{2}x_{Q}x_{L}^{\dagger}x_{L}+\frac{15}{8}x_{Q}\lambda_{15}\lambda_{15}^{\dagger}+2\mathrm{Tr}\left(x_{Q}x_{Q}^{\dagger}\right)x_{Q}\\
{} & -\frac{1}{12}g_{1}^{2}x_{Q}-\frac{9}{2}g_{2}^{2}x_{Q}-4g_{3}^{2}x_{Q}-\frac{45}{8}g_{4}^{2}x_{Q}\,,
\end{aligned}
\label{eq: full_lagrangian-1-2}
\end{align}

\begin{align}
\begin{aligned}\left(4\pi\right)^{2}\beta_{x_{L}} & =\frac{5}{2}x_{L}x_{L}^{\dagger}x_{L}+\frac{3}{2}x_{L}x_{Q}^{\dagger}x_{Q}+\frac{15}{8}x_{L}\lambda_{15}\lambda_{15}^{\dagger}+2\mathrm{Tr}\left(x_{L}x_{L}^{\dagger}\right)x_{L}\\
{} & -\frac{3}{4}g_{1}^{2}x_{L}-\frac{9}{2}g_{2}^{2}x_{L}-\frac{45}{8}g_{4}^{2}x_{L}\,,
\end{aligned}
\label{eq: full_lagrangian-1-1-2}
\end{align}

\begin{align}
\begin{aligned}\left(4\pi\right)^{2}\beta_{\lambda_{15}} & =\frac{21}{4}\lambda_{15}\lambda_{15}\lambda_{15}^{\dagger}+\frac{3}{2}\lambda_{15}x_{Q}^{\dagger}x_{Q}+\frac{1}{2}\lambda_{15}x_{L}^{\dagger}x_{L}+4\mathrm{Tr}\left(\lambda_{15}\lambda_{15}^{\dagger}\right)\lambda_{15}\\
{} & -\frac{9}{2}g_{2}^{2}\lambda_{15}-\frac{45}{4}g_{4}^{2}\lambda_{15}\,,
\end{aligned}
\label{eq: full_lagrangian-1-1-1-1}
\end{align}
where any contributions from the Yukawas of the personal Higgs, $y_{i\alpha}^{\psi}$,
are negligible as they are all 1 or smaller. 
\begin{figure}
\begin{centering}
\includegraphics[scale=0.4]{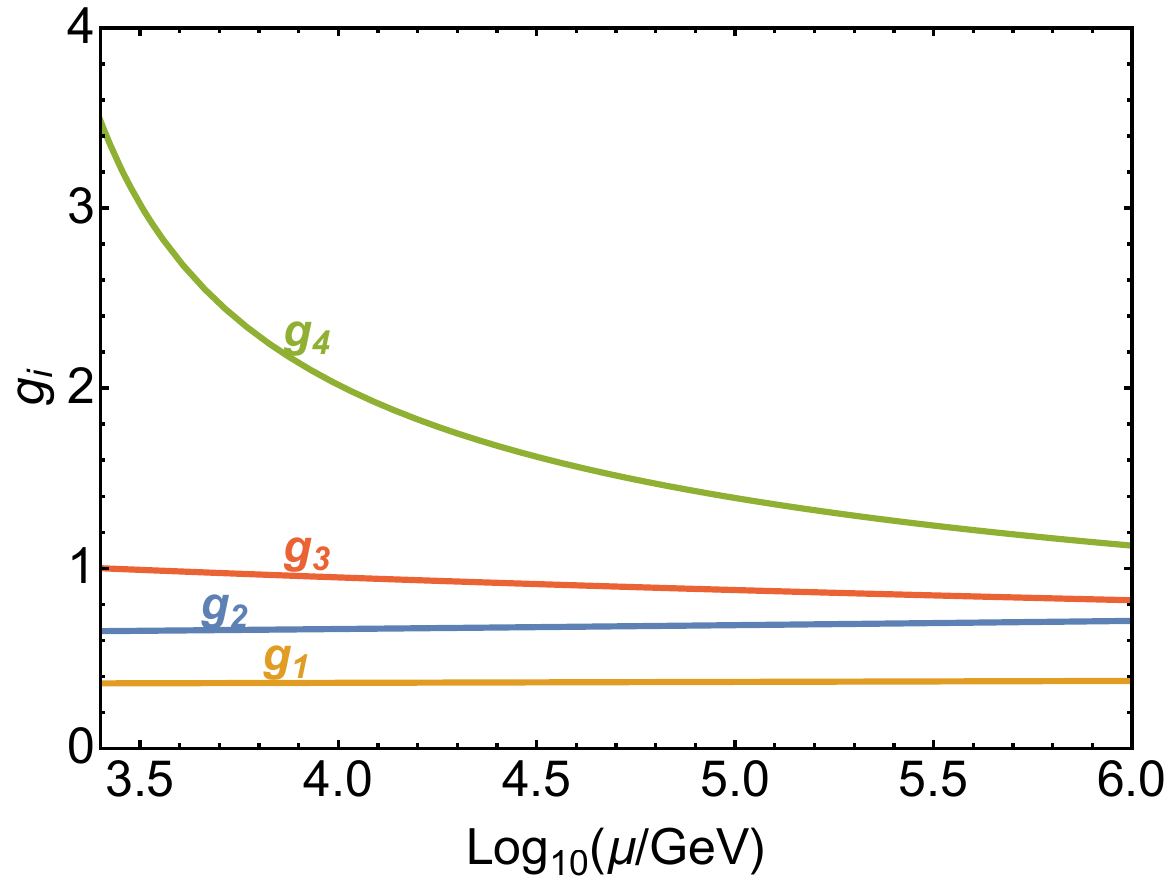}
\par\end{centering}
\caption{RGE of the gauge couplings in our benchmark scenario from the TeV
scale to the scale of the twin Pati-Salam symmetry $\mu\sim1\,\mathrm{PeV}$.\label{fig:Perturbativity_gauge}}
\end{figure}

\providecommand{\href}[2]{#2}\begingroup\raggedright\endgroup

\end{document}